\documentclass[]{emulateapj} 
\usepackage{apjfonts}

\newcommand{\acrylonitrile}{C$_2$H$_3$CN} 
\newcommand{\formaldehyde}{H$_2$CO}
\newcommand{\form}{H$_2$CO}
\newcommand{\formamide}{NH$_2$CHO}
\newcommand{\ntwohplus}{N$_2$H$^+$} 
\newcommand{\dimethylether}{CH$_3$OCH$_3$}
\newcommand{\ethylcyanide}{C$_2$H$_5$CN}
\newcommand{\methylformate}{HCOOCH$_3$}
\newcommand{\meth}{HCOOCH$_3$}
\newcommand{\methylcyanide}{CH$_3$CN}
\newcommand{\ethanol}{C$_2$H$_5$OH}
\newcommand{\htwos}{H$_2$S}
\newcommand{\hto}{H$_2$O}

\newcommand{\water}{H$_2$O} 
\newcommand{\methanol}{CH$_3$OH}
\newcommand{\thirteenmethanol}{$^{13}$CH$_3$OH}
\newcommand{\ammonia}{NH$_3$}
\newcommand{\hcccn}{HC$_3$N}
\newcommand{\lsun}{L$_\odot$}
\newcommand{\msun}{M$_\odot$}
\newcommand{\ngc}{NGC~6334}

\newcommand{\ngci}{NGC~6334~I}
\newcommand{\ngcin}{NGC~6334~I(N)}
\newcommand{\HII}{H\,{\sc ii}}
 
\newcommand{\mjb}{mJy~beam$^{-1}$} 
\newcommand{\percm}{cm$^{-1}$} 
\newcommand{\jykms}{Jy~km~s$^{-1}$} 
\newcommand{\jyb}{Jy~beam$^{-1}$} 
\newcommand{\mjyb}{mJy~beam$^{-1}$}
\newcommand{\mujb}{$\mu$Jy~beam$^{-1}$}
\newcommand{\kms}{km~s$^{-1}$}  
\newcommand{\mum}{$\mu$m}
\newcommand{\Tr}{$T_{rot}$}
\newcommand{\cc}{cm$^{-3}$}
\newcommand{\ct}{cm$^{-2}$}
\usepackage{natbib}
\usepackage{apjfonts}

\begin{document}

\shortauthors{Brogan et al.}

\shorttitle{Imaging of a protostellar cluster}

\title{Digging into \ngcin: Multiwavelength Imaging of a 
Massive Protostellar Cluster}
\author{C.L. Brogan\altaffilmark{1}, 
T.R. Hunter\altaffilmark{1}, 
C.J. Cyganowski\altaffilmark{2},
R.Indebetouw\ \altaffilmark{1,3},
H. Beuther\altaffilmark{4}, 
K.M. Menten\altaffilmark{5}, 
S. Thorwirth\altaffilmark{5,6}
}

\email{cbrogan@nrao.edu}

\altaffiltext{1}{NRAO, 520 Edgemont Rd, Charlottesville, VA, 22903, USA} 
\altaffiltext{2}{University of Wisconsin, Madison, WI, 53706, USA} 
\altaffiltext{3}{University of Virginia, Charlottesville, VA, 22903, USA} 
\altaffiltext{4}{MPIA Heidelberg, Germany} 
\altaffiltext{5}{MPIfR Bonn, Germany}  
\altaffiltext{6}{I. Physikalisches Institut, Universit\"at zu K\"oln, Germany}  


\begin{abstract}

We present a high-resolution, multi-wavelength study of the massive
protostellar cluster \ngcin\ that combines new spectral line data from
the Submillimeter Array (SMA) and VLA with a reanalysis of archival
VLA continuum data, 2MASS and {\em Spitzer} images.  As shown previously, the
brightest 1.3~mm source SMA1 contains substructure at subarcsecond
resolution, and we report the first detection of SMA1b at 3.6~cm along
with a new spatial component at 7~mm (SMA1d).  We find SMA1 (aggregate
of sources a, b, c, and d) and SMA4 to be comprised of free-free and
dust components, while SMA6 shows only dust emission. Our $1\farcs5$
resolution 1.3~mm molecular line images reveal substantial hot-core
line emission toward SMA1 and to a lesser degree SMA2.  We find
\methanol\ rotation temperatures of $165\pm 9$~K and $145\pm 12$~K for
SMA1 and SMA2, respectively.  We estimate a diameter of 1400~AU for
the SMA1 hot core emission, encompassing both SMA1b and SMA1d, and
speculate that these sources comprise a $\gtrsim 800$~AU separation
binary that may explain the previously-suggested precession of the
outflow emanating from the SMA1 region.  Compact line emission from
SMA4 is weak, and none is seen toward SMA6.  The LSR velocities of
SMA1, SMA2, and SMA4 all differ by 1-2~\kms.  Outflow activity from
SMA1, SMA2, SMA4, and SMA6 is observed in several molecules including
SiO(5--4) and IRAC $4.5~\mu$m emission; 24~\mum\ emission from SMA4 is
also detected. Eleven water maser groups are detected, eight of which
coincide with SMA1, SMA2, SMA4, and SMA6, while two others are
associated with the Sandell et al. (2000) source SM2.  We also detect
a total of 83 Class~I \methanol\/ 44~GHz maser spots which likely
result from the combined activity of many outflows.  Our observations
paint the portrait of multiple young hot cores in a protocluster prior
to the stage where its members become visible in the near-infrared.

\end{abstract}

\keywords{stars: formation --- infrared: stars --- ISM: individual
  (\ngcin) --- masers --- techniques: interferometric }

\section{INTRODUCTION}

The formation process of massive stars is an important but
poorly-understood phenomenon.  Because most regions that are forming
massive stars lie at distances of several kiloparsecs, and most
massive stars form in clusters, the identification of high mass
protostars requires both good sensitivity and high angular resolution
to resolve the continuum and line emission within young protoclusters.
The advent of (sub)millimeter interferometers allows the study of
deeply-embedded thermal gas emission at arcsecond angular resolution,
a realm that has historically been accessible only via maser and
cm~$\lambda$ continuum emission.  Research that combines observations
of dust, thermal gas and maser gas offers a more complete picture of
the star formation process, as these phenomena potentially trace
objects at different evolutionary stages.

\ngc\ is a luminous ($>10^6$ L$_{\sun}$) and relatively nearby
\citep[1.7~kpc;][]{Neckel78} region of massive star formation
containing sites at various stages of evolution \citep{Straw89a}.
\citet{Matthews08} have recently completed a comprehensive single dish
study of the submillimeter dust properties of the whole complex with
$\sim 14\arcsec$ resolution and find a total dust mass of $\sim
1.7\times 10^4$ \msun\/.  Recent work by \citet{Persi08} suggests that
the distance to NGC~6334 may be as close as 1.6~kpc, though we adopt
1.7~kpc for ease of comparison with earlier work.  The dust core 
``I(N)'' was first detected at 1~mm by \citet{Cheung78} and later at
$400~\mu$m by \citet{Gezari82} toward the northern end of the complex.

In the past decade, direct evidence of protostars in the I(N) region
were found in the form of Class I and II methanol masers
\citep{Kogan98,Walsh98} and a faint 3.6~cm source \citep{Carral02}.
In a previous paper, we presented the first millimeter interferometric
observations of I(N) with $\sim 1\farcs5$ resolution which revealed a
cluster of compact dust continuum cores likely to be a protocluster
\citep[][see also Figure~1]{Hunter06}.  Subsequent subarcsecond
resolution 7~mm continuum observations by \citet{Rodriguez07} resolved
the principal source SMA1 into several components, and detected
possible counterparts to SMA4 and SMA6 \citep[also see][]{Carral02}.
Our recent observations of the ammonia inversion transitions up to
(6,6) demonstrated the presence of compact clumps of hot thermal gas
within this protocluster \citep{Beuther07}.  Although water masers are
excellent tracers of high mass star formation, previous observations
of this region have been limited to single dish data \citep{Moran80}.
Likewise, most of the millimeter line studies of this region have been
performed with single-dish telescopes.  For example, \citet{Pirogov03}
imaged the \ntwohplus\ (1--0) line with a beamsize of 40\arcsec\ and
found significantly stronger emission toward I(N) than toward the more
luminous massive protocluster (source I) that lies $\sim 1\arcmin$
south of I(N).  In addition, the SiO (2--1) and (5--4) transitions
show strong wing emission from $-50$ to +40 \kms\/ when observed with
beamsizes of $58''$ and $23''$, respectively \citep{Megeath99}.

Recently, we reported 3~mm interferometric imaging of I(N), which
revealed compact \methylcyanide\ (5--4) emission toward SMA1 along
with a bipolar outflow seen in HCN(1--0) \citep{Beuther08}.  In this
paper, we present the details of our 1.3~mm Submillimeter
Array\footnote{The Submillimeter Array (SMA) is a collaborative
  project between the Smithsonian Astrophysical Observatory and the
  Academia Sinica Institute of Astronomy \& Astrophysics of Taiwan.}
(SMA) spectral line dataset, along with new NRAO Very Large
Array\footnote{The National Radio Astronomy Observatory is a facility
  of the National Science Foundation operated under agreement by the
  Associated Universities, Inc.} (VLA) \water\ maser and 44 GHz Class
I \methanol\ maser data. We have also re-analyzed archival VLA
continuum data from 3.6~cm to 7~mm, 2MASS near-IR and {\em Spitzer}
mid-IR IRAC and MIPS images. Together these data confirm the
identification of \ngcin\ as an actively forming massive protostellar
cluster.

\begin{figure*} 
\includegraphics[scale=0.75,angle=0]{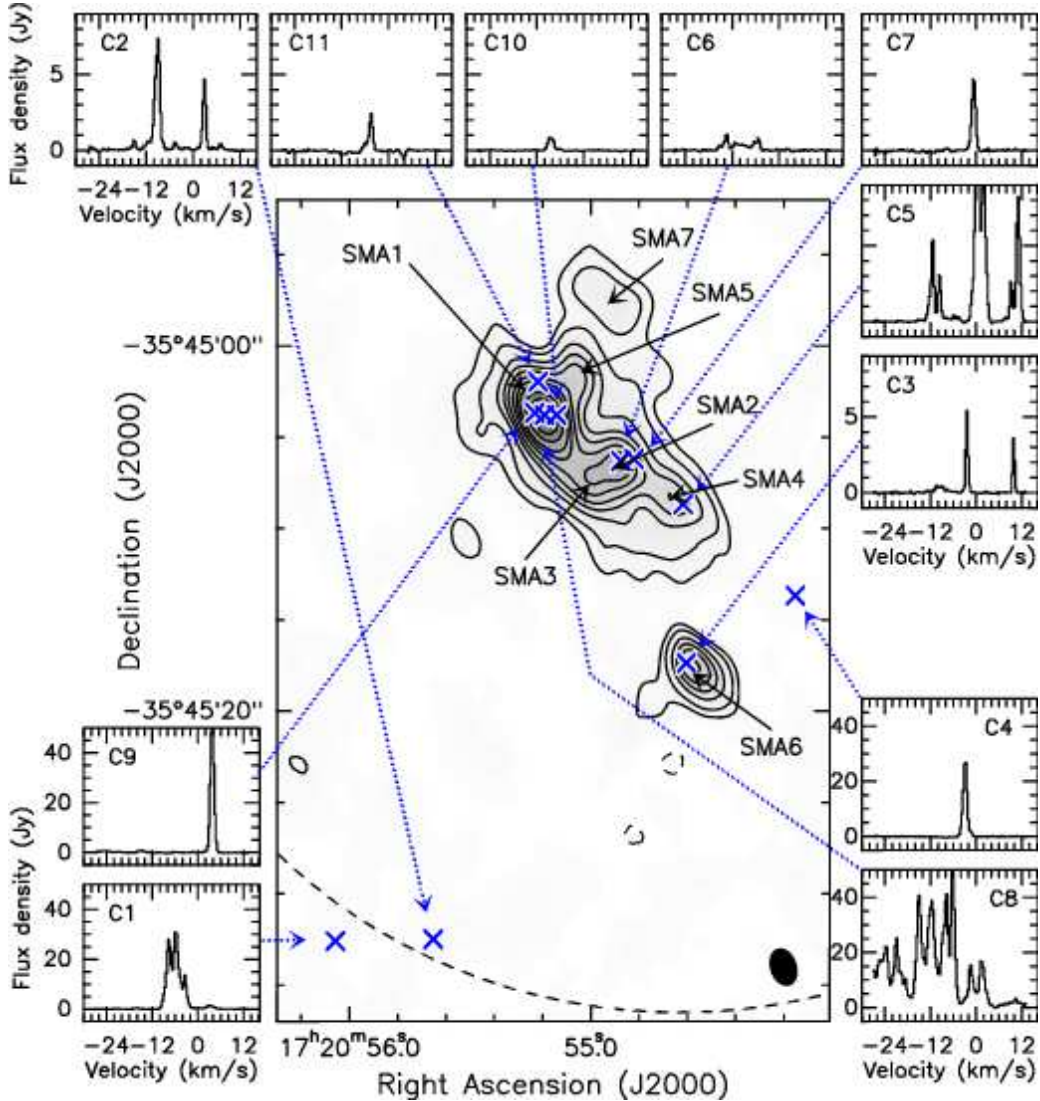}
\caption{Composite image showing the naturally weighted SMA 1.3~mm
  continuum emission from I(N) with surrounding insets showing the VLA
  \water\/  (22.235 GHz) maser spectra. The 1.3~mm continuum image has a 
resolution of $2\farcs 2 \times 1\farcs 3$ (P.A.=+15\arcdeg) and the contour
  levels are 30 ($3.5\sigma$), 60, 120, 180, 240, 320, 400, 640, 800,
  960 \mjb\/. The blue crosses show the intensity weighted centroid
  positions of the water maser groups listed in Table~\ref{watertable}
  (derived from Table~\ref{bigwatertable}). The water maser spectra
  show the average emission from each group. The dashed arc indicates the
  SMA primary beam, and the dashed arc indicates the VLA primary beam
  at 22~GHz.  The 1.3~mm image has not been corrected for primary beam
  attenuation.
  \label{contwithmasers}}
\end{figure*}

\section{OBSERVATIONS}

\subsection{Submillimeter Array (SMA)}

Our SMA observations were made with six antennas -- in May 2004 in the
compact configuration, and in May 2005 in the extended configuration.
In both tracks, two pointings were observed: source I at 17$^{\rm
  h}$20$^{\rm m}$53.44$^{\rm s}$, $-35\arcdeg 47\arcmin 02.2\arcsec$
and source I(N) at 17$^{\rm h}$20$^{\rm m}$54.63$^{\rm s}$,
$-35\arcdeg 45\arcmin 08.5\arcsec$ (J2000). Only the data for source
I(N) are presented in this paper. The SMA receivers are double
sideband SIS mixers with 2~GHz bandwidth \citep{Blundell98}. The local
oscillator was tuned to provide 216.6-218.6~GHz in LSB and 226.6-228.6 GHz
in USB, yielding a FWHP primary beam size of $\sim 55\arcsec$.  The
correlator was configured for 3072 channels per sideband with a
uniform spectral resolution of 0.8125 MHz. The correlator sideband
separation was about 18~dB at the time of these observations.  Typical
system temperatures were 200~K.  The projected baseline lengths ranged
from 10 to 180~m. A continuum image produced from these data was
presented in \citet{Hunter06}, however in order to correct for a
half-channel error in SMA velocity labeling discovered in November
2007, the data have been completely re-reduced.

The gain calibrators were the quasars NRAO~530 ($23^o$ distant) and
J1924$-$292 ($27^o$ distant).  Bandpass calibration was performed with
observations of Uranus (2004 data) and 3C279 (2005 data).  Flux
calibration is based on observations of Jovian satellites and regular
SMA monitoring of the quasar flux densities. 
The data were calibrated in Miriad, then exported to AIPS for imaging;
the two sidebands were reduced independently.  The AIPS task UVLSF was used
to separate the line and continuum emission. Self-calibration was
derived from the continuum, and solutions were transferred to the
spectral line data.  After self-calibration, the continuum data from
both sidebands were combined and imaged with both natural and uniform
weighting in order to achieve the best surface brightness sensitivity
(the former) and best angular resolution (the latter). The naturally
weighted image has an angular resolution of $2\farcs2\times 1\farcs3$
at P.A. $+15^\circ$ and the uniformly weighted image has an angular
resolution of $1\farcs8 \times 0\farcs9$ at P.A. $+8^\circ$. The rms
noise levels are 8.4 and 6.4 \mjb\/ for the natural and uniform
weighted images, respectively. The peak signal-to-noise of the
naturally weighted continuum image has increased to 126 compared to
114 for the image presented in \citet{Hunter06}. The spectral line
data were resampled to a spectral resolution of 1.1 \kms\/ and imaged
with natural weighting; the rms noise per channel is 180~\mjb\/.  The
absolute position uncertainty of these data is estimated to be
0\farcs3.

\subsection{Very Large Array (VLA)}

\subsubsection{Water Maser Observations}

The 1.3~cm (22.235 GHz) \water\ maser data were taken with the
 VLA on 2006 July 03 (project AH915)
with online Hanning smoothing and one polarization (RR). The full
bandwidth of the data is 3.125 MHz (40 \kms\/ usable) and the channel
separation and resolution is 24.4 kHz (0.33 \kms\/). The absolute flux
scale was set assuming a 22.235 GHz flux density of 2.54 Jy for 3C286.
Fast-switching was employed using J1717-337 and a cycle time of 2
minutes.  The calibrator J1924-292 was used for bandpass
calibration. The pointing model was updated once per hour during the
observation. After flagging periods of poor phase coherence (due to
weather), the on-source time is approximately 60 minutes. The FWHM
primary beam of these data is $120''$ and the synthesized beam is
$0\farcs 79 \times 0\farcs 25$ at P.A.=+7\arcdeg.

It was necessary to remove the remarkably strong masers detected
toward \ngci\ in the sidelobes ($112\arcsec$ from the phase center of
I(N)) in order to achieve high dynamic range on the target masers in
source I(N); source I masers were detected with peak intensities
(uncorrected for primary beam attenuation) as high as 112 \jyb\/
despite their location near the first null. The removal was
accomplished by self-calibrating the data on the strongest source I
channel, applying this calibration to the full line dataset, imaging
both fields, subtracting the clean components from source I (only) in
the $u-v$ plane (UVSUB), inverting the calibration (CLINV),
re-self-calibrating on the strongest I(N) maser channel, applying
these new corrections and making the final I(N) line cube (this
process is often called ``peeling''). This procedure improved the
dynamic range in the I(N) channels at the same velocities as the
strong source I masers by a factor of $\sim 3$.  The rms noise in a
single channel is 6~\mjb. The absolute position uncertainty is
approximately $0\farcs01$. Flux density measurements were made on an
image corrected for primary beam attenuation.

\subsubsection{44 GHz Methanol Maser Observations} 

The 7~mm (44.06941~GHz) Class~I \methanol\ ($7_{0,7}-6_{1,6}$)~A$^{+}$
maser transition was observed on 2008 February 18 with the VLA in
CnB-configuration (project code AC904).  Correlator mode 2AB was used
with a velocity channel width of 0.17 \kms\/.  NGC6334I(N) was
observed for 7.5 minutes (on-source) as a check source for a larger
survey of massive protostellar candidates \citep{Cyganowski09}. The
gain, bandpass, and flux calibrators were J1717-337, J2253+161, and
3C286, respectively.  Fast switching was employed with a cycle time of
2 minutes. The absolute positions are expected to be better than
$0\farcs2$. The synthesized beam is $1\farcs{30} \times 0\farcs{40}$
at P.A. = $-44^\circ$, and the velocity resolution is 0.33 \kms\/
after Hanning smoothing. Due to the snapshot nature of these data, and
consequent poor $u-v$ coverage, the noise is dynamic range limited in
channels with strong maser emission. Channels with relatively weak
maser emission have rms noise levels of $\sim 100$ \mjb\/, while it is
up to $5\times$ poorer in channels with strong maser emission
(i.e. peak $> 50$ Jy).

\subsubsection{Archival 7~mm, 1.3~cm, and 3.6~cm Continuum Observations}

We have re-reduced the archival VLA 7~mm and 1.3~cm continuum
observations originally presented by \citet[][project codes AZ159 and
  AZ152, further details of these observations can be found in that
  paper]{Rodriguez07}.  We achieved a significantly better rms noise
in the 7~mm image (210 \mujb\/) than those authors report (320
\mujb\/).  This improvement is likely due to the fact that we
simultaneously imaged a field toward source I which contains a strong
ultracompact \HII\/ region \citep[$2\arcmin$ south of I(N), see for
  example][]{Hunter06}; images made without cleaning this \HII\/
region show significant sidelobes at the location of I(N).  We also
used the AIPS task VLANT to retrieve and apply post-observation
antenna position corrections to the data, and used a model for the
brightness distribution of the flux calibrator 3C286. The flux
densities derived for the phase calibrator (J1720-355) are within
$2\%$ of those reported by \citet{Rodriguez07}. The angular resolution
of the 7~mm data is $0\farcs65\times 0\farcs44$,
P.A.=$+25\arcdeg$. The 1.3~cm data were reduced in a similar manner
and the rms noise is 64 \mujb\/, similar to that obtained by
\citet{Rodriguez07}; the angular resolution is $0\farcs38\times
0\farcs27$, P.A.=$+19\arcdeg$. The 1.3~cm and 7~mm data show
remarkable position agreement given that they were observed on
different days -- a testament to the accuracy of the VLA when
reference pointing and fast switching are employed. We estimate that
the absolute position accuracies for these two datasets are better
than $0\farcs05$.

We have also re-reduced the archival VLA 3.6~cm continuum observations
originally presented by \citet[][project code AM495, further details
of the observing parameters can be found in that paper]{Carral02}. The
pointing center for these data is located $1.5\arcmin$ south of source
I(N) (the 3.6~cm FWHP is $5\farcm4$). These data were re-reduced as
described above for the 7~mm and 1.3~cm data. In particular, a model
for the brightness distribution of the flux calibrator (3C286) was
employed -- a method not available at the time of the original
\citet{Carral02} reduction. The average flux densities (between the
two IFs) derived for the two phase calibrators, J1733-130 and
J1751-253 are $6.67\pm 0.03$ and $0.267\pm 0.001$, respectively. In
contrast \citet{Carral02} obtained $7.6\pm 0.2$ for J1733-130, $14\%$
higher than our more accurate result (a flux density for J1751-253 was
not reported by these authors). After self-calibration, the final
image was made with baselines longer than 100k$\lambda$ to minimize
artifacts from the nearby (within $2\arcmin$) extended \HII\/ regions
''I'' and ''E'', which fall within the VLA 3.6~cm FWHP primary beam.
The resolution of the final 3.6~cm image of the I(N) region is
$0\farcs51\times 0\farcs20$, P.A.=$-1\arcdeg$ and the rms noise is
41~\mujb\/ after primary beam correction \citep[][report an rms noise
of 60~\mujb\/ but it is unclear whether they applied primary beam
correction]{Carral02}.  The 3.6~cm data did not employ reference
pointing or fast switching, and we estimate an absolute position
accuracy for these data of $0\farcs3$. We note that there is a
consistent $0\farcs2$ offset (to the SE) between the 3.6~cm data and
the 1.3~cm and 7~mm data; we have not shifted the images to match.

\subsection{Archival Australia Telescope Compact Array (ATCA) 3.4~mm Observations}

We have re-imaged the 3.4~mm ATCA continuum data reported by
\citet{Beuther08}.  The rms noise is 3.3~\mjb\/ (after primary beam
correction) and the resolution is $2\farcs3\times 1\farcs8$,
P.A.=$83\arcdeg$. Due to the small primary beam of the ATCA at this
wavelength (FWHP $33\arcsec$), primary beam correction for this image
is critical. This accounts for the apparent increase in rms noise
compared to that reported by \citet[][2.6~\mjb\/; our rms noise is
  2.2~\mjb\/ before primary beam correction]{Beuther08}; the
peak signal-to-noise has increased by a factor of 1.3. We estimate an 
absolute position accuracy for these data of $0\farcs3$.

\subsection{Archival Infrared Data}

For near-infrared photometry, we analyzed archival 2MASS\footnote{Two
  Micron All Sky Survey, a joint project of the University of
  Massachusetts and the Infrared Processing and Analysis
  Center/California Institute of Technology, funded by the National
  Aeronautics and Space Administration and the National Science
  Foundation} survey images \citep{skrutskie}.  We have also examined
all of the {\em Spitzer} archival data for the I(N) region. Of the
available IRAC data \citep{Fazio04}, the data obtained in project 46
on 2004, August 2 by PI Fazio are by far the deepest (666 s). In this
paper we use the PBCD images produced by pipeline version 14.0
\citet[a different version of these data were also presented
  in][]{Hunter06}.  Since saturation in the I(N) region was not an
issue, only the long integration (10.4 s) data are
used. Unfortunately, of the available MIPS data, only the 24 \mum\/
data are usable -- the 70 and 160 \mum\/ data show tremendous striping
and other artifacts. In this paper we utilize the 24 \mum\/ data from
the MIPSGAL Legacy survey \citep{Carey09}.

We performed aperture photometry on all infrared images by defining
irregular polygonal shapes (the ``source'' shape) surrounding each
extended knot of emission (presumed to be shock fronts from one or
more outflows, not point sources). A background is taken as the mode
of pixel values in an scaled annulus, between 1.5 and 2 times radially
expanded from the centroid of the source shape.  These polygons are
translated to each infrared image by their locations on the sky, so
sample the identical regions even though the images have different
pixel sizes and world coordinate systems.  Fluxes were converted to Jy
using the photometric information for each observatory.

\subsection{Caltech Submillimeter Observatory (CSO)}

Our CSO\footnote{The Caltech Submillimeter Observatory 10.4 m is
operated by Caltech under a contract from the National Science
Foundation.}  observations were obtained on 13 April 1993 using the
facility 345~GHz SIS receiver with lead junctions \citep{Ellison89}
and a 500~MHz AOS backend.  
Several on-the-fly maps of the CO 3--2 line were obtained on a
10\arcsec\ grid, providing near-Nyquist sampling of the
20\arcsec\ beam of the telescope at this frequency.  At the end of
each row, a designated off position (17$^{\rm h}$24$^{\rm
  m}$56.06$^{\rm s}$, $-36\arcdeg 21\arcmin 40.1\arcsec$ J2000)
relatively free of CO emission was observed to provide the spectral
baseline.  After combining the maps, the total area covered was
$4\arcmin\times 4\arcmin$ ($25\times 25$ grid), with 30 seconds of
on-source integration per point in the central portion containing
source I and I(N), and 15 seconds per point on the periphery.  The DSB
receiver temperature was measured to be 320~K using the y-factor
method, and the typical SSB system temperature was 1800~K at an
elevation of 30\arcdeg.  The spectra have been corrected to the main
beam brightness temperature scale using an efficiency of 68\% as
measured on Jupiter on the same night.

\section{RESULTS} 

\begin{figure*}
\includegraphics[scale=0.9,angle=0]{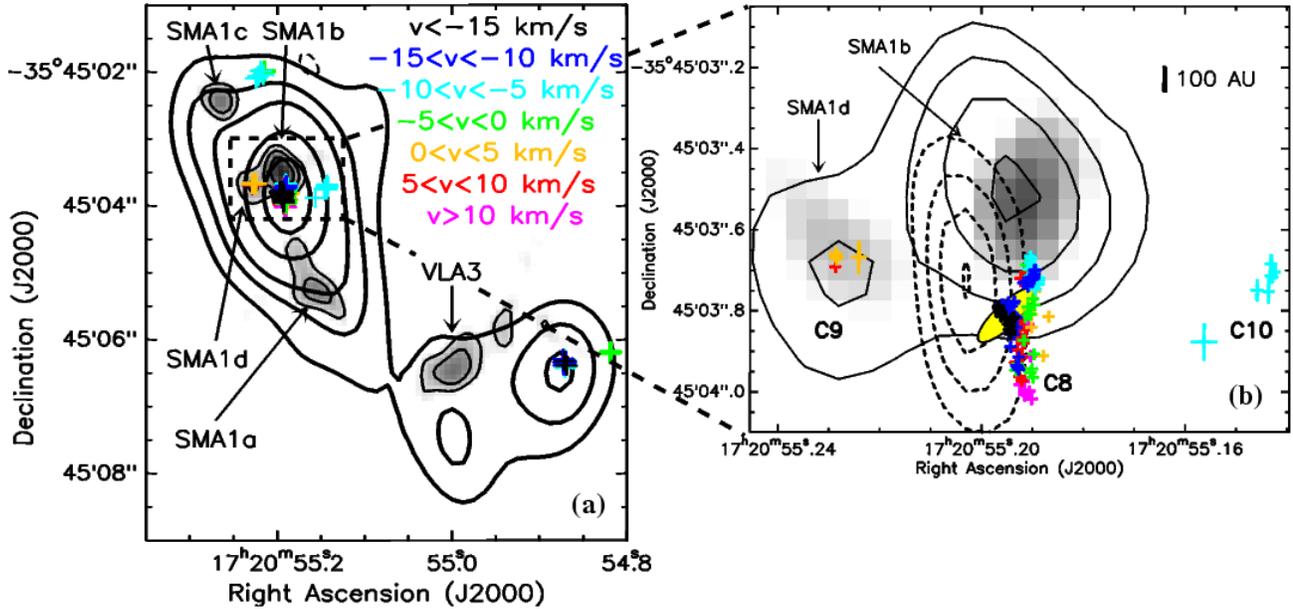}
\caption{Zoomed multiwavelength images of I(N). (a) Zoom shows the
  SMA1, SMA2, and SMA3 regions with thick contours showing the
  uniformly weighted SMA 1.3~mm continuum image with contour levels of
  170, 250, 330, 490, and 650 \mjb\/.  Thin contours and greyscale
  show the VLA 7~mm continuum emission (contour levels are -0.6, 0.6,
  0.8, 1.0, 1.2 \mjyb).  The water maser positions are shown as
  colored crosses with a corresponding color coded velocity legend;
  note that the color scale was set using an assumed $-3$ \kms\/
  systemic velocity for SMA1. (b) Zoom of the SMA1 region at 3.6~cm
  (dashed contours), 1.3~cm (greyscale),and 7~mm (solid contours). The
  3.6~cm contour levels are 0.12, 0.20, 0.28, and 0.36 \mjb\/,
  and the 7~mm contour levels are same as in (a). The water maser
  positions are also indicated with the same velocity color coding as
  (a). The position of the 44 GHz \methanol\ maser is indicated by the
  yellow ellipse and it has a velocity of $-0.9$ \kms\/. For both
  types of maser symbol size indicates position uncertainty.
  \label{qbandabsolute}}
\end{figure*} 

Our multiwavelength images reveal several sites of star formation
activity within the I(N) region.  In the following sections, we
describe the variety of observed phenomena. In the following sections
linear offsets are given with the $\gtrsim$ symbol to indicate that
the measured values are lower limits since only the projected
separation is available (a distance of 1.7 kpc is assumed).

\subsection{SMA 1.3 mm Continuum}

The naturally weighted 1.3 mm SMA continuum image of I(N) is presented in
Figure~\ref{contwithmasers} (center). This image is qualitatively
similar to that presented in \citet{Hunter06}, but the calibration has
been refined for improved signal-to-noise. These data resolve the
presence of at least seven 1.3~mm ``cores'' in I(N). As pointed out in
\citet{Hunter06}, this region has the hallmarks of a ``protocluster'',
i.e. a region that will form a cluster of intermediate to massive
stars. The integrated flux density within the $3\sigma$ contour (30
\mjb\/, see Fig.~\ref{contwithmasers}) of the naturally weighted
1.3~mm continuum image is $7.4\pm 0.8$ Jy \citep[this is 1.6 times
  higher than reported in][]{Hunter06}; this estimate includes both
statistical and $10\%$ calibration uncertainties. \citet{Sandell00}
find a 1.3~mm single dish flux density of $14.2\pm 0.2$ Jy, implying
that we have recovered about $50\%$ of the total flux with the SMA.

We have used the uniformly weighted image corrected for primary beam
attenuation to fit the seven peaks identified in
Figure~\ref{contwithmasers} with a single 2-D Gaussian (using JMFIT in
AIPS). The fitted positions and flux densities of the 1.3~mm cores are
listed in Table~\ref{mmpos}.  SMA1, SMA4, and SMA6 in particular, are
not well-fit by a single Gaussian component, and instead appear to
have a compact (unresolved, FWHM $< 1\arcsec$) region of emission,
embedded in a more extended though still compact (FWHM $< 3\arcsec$)
region. However, attempts to uniquely disentangle these two (or more)
components with the current resolution and sensitivity were not
successful. The JMFIT sizes listed in Table~\ref{mmpos} are
representative of the more extended (though still compact on the scale
of the I(N) protocluster) component, and the integrated intensities
and other quantities calculated from it ($T_b$ and mass) are likely to
be underestimates. These sizes while quite significant from a
signal-to-noise perspective should be viewed as upper limits.

\subsection{ATCA 3.4~mm Continuum}

As first discussed by \citet{Beuther08}, the 3.4~mm continuum emission
toward I(N) (not shown) is qualitatively similar to that at
1.3~mm. With our new imaging of the original \citet{Beuther08} data,
we have been able to increase the signal-to-noise and reliably detect
SMA4 at this wavelength.  The fitted (JMFIT)
parameters of the compact 3~mm continuum emission are given in
Table~\ref{mmpos}.

\subsection{VLA Water Masers and Continuum Emission}

The association of water masers with molecular outflows make them a
good tracer of protostellar activity \citep{Tofani95,Hofner96}.  We
find strong 22~GHz \water\ maser emission toward 11 distinct regions
in \ngcin , listed in Table~\ref{watertable} in order of increasing
declination.  Masers that lie in close proximity to each other ($<
1\arcsec$) are collectively called by ``group'' names C1,
C2...C11. The intensity weighted positions of these groups are listed
in Table~\ref{watertable} and are plotted on the 1.3~mm continuum
image in Figure~\ref{contwithmasers}, along with average spectra for
each group (the position of each distinct \water\ maser and its
velocity are presented in Table~\ref{bigwatertable}, available in its
entirety online).  Eight of the groups lie within 2\arcsec\ of the
1.3~mm sources SMA1, SMA2, SMA4, and SMA6, whose positions are listed
in Table~\ref{mmpos}.  Two of the maser groups (C1 and C2) lie near
the SW edge of the SMA primary beam. These masers are coincident with
the single-dish millimeter continuum source SM2, identified in JCMT
submillimeter continuum images by \citet{Sandell00}.  Maser group C4
appears in a region with no compact millimeter continuum emission at
the present sensitivity. Instead it lies at the southern edge of the
$4.5~\mu$m nebula that extends to the southwest from SMA4
\citep[][also see \S3.6]{Hunter06}. More detailed phenomenology for
the SMA 1.3~mm cores (except SMA5 and SMA7 which are weak and fairly
diffuse) are given below.

\subsubsection{SMA1, SMA2, and SMA3 Regions}

\begin{figure*}
\includegraphics[scale=0.65,angle=-90]{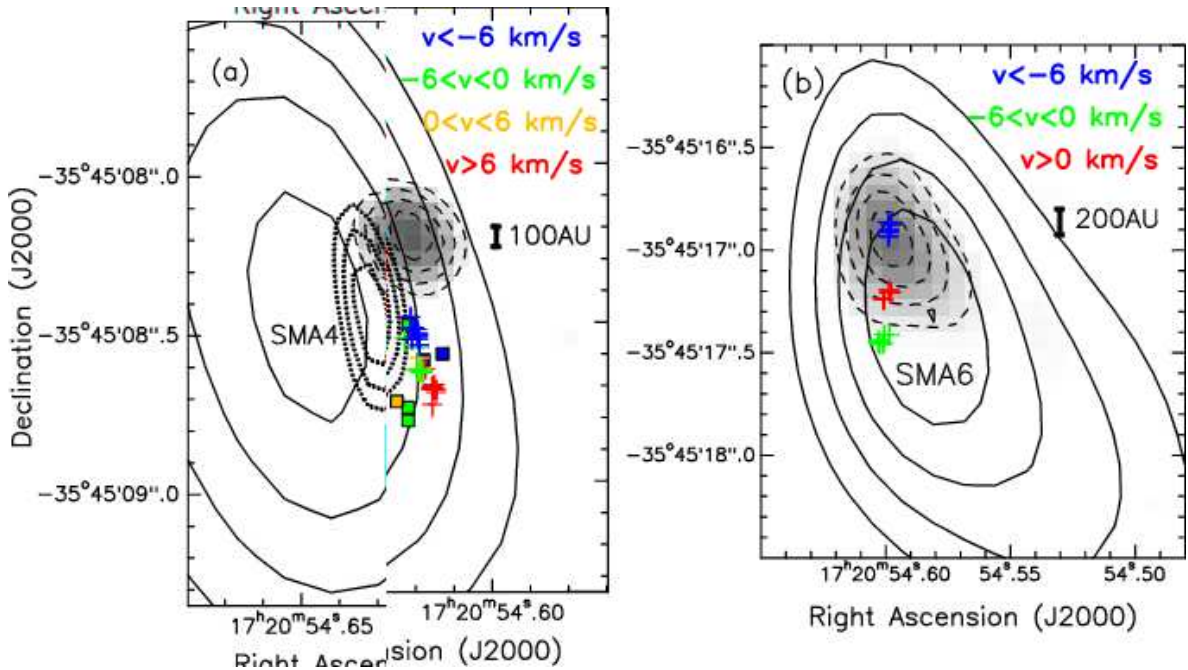}
 \caption{(a) Zoom of the SMA4 region with the grayscale and dashed
   contours showing the 1.3~cm continuum emission, dotted contours
   showing 3.6~cm, and the solid contours showing the uniformly
   weighted 1.3~mm continuum. The 3.6~cm contours are 0.12, 0.16, and
   0.20 \mjb\/, the 1.3~cm contour levels are 0.19, 0.25, 0.31, and
   0.37~\mjb\/, and the 1.3~mm contour levels are 100, 120, 140, and
   160~\mjb.  (b) Zoom of the SMA6 region with the greyscale and
   dashed contours showing 7~mm continuum emission and the solid
   contours showing the 1.3~mm continuum. The 7~mm contour levels are
   0.6, 0.8, 1.0, and 1.2 \mjb\/, and the uniformly weighted 1.3~mm
   contour levels are 100, 140, 180, 220, and 260 \mjb\/. The cross
   symbols show the C5 (a) and C3 (b) water maser positions with S/N
   ratio $> 7$, color-coded by velocity. The filled square symbols in
   (a) show the location of 6.7~GHz \methanol\ masers from
   \citet{Walsh98}, with same velocity coding as \water\/ masers.
\label{c5c3color}}
\end{figure*}

Figure~\ref{qbandabsolute}a shows a close-up view of the region from
SMA1 through SMA3 in both 1.3 and 7~mm continuum. This region also
contains \water\ maser groups C6 through C11. Following
\citet{Rodriguez07} we call lower frequency continuum emission that
appears to be associated with a particular SMA source by the ``SMA
name''. If more than one source is associated with a 1.3~mm source,
the SMA name is appended by a,b,c etc.  Lower frequency continuum
emission that is not clearly associated with any particular SMA source
is called ``VLA'' followed by the number of the closest SMA source.
From the improvement in the 7~mm image quality over that shown in
\citet{Rodriguez07}, several new features are present: (1) The 7~mm
source SMA1b is resolved into two distinct sources separated by $\sim
0\farcs45$ ($\gtrsim 800$ AU, also see Figure~\ref{qbandabsolute}a),
we shall refer to the eastern component as SMA1d; (2) SMA1a is
elongated NE/SW and appears to point toward SMA1b; (3) Two new 7~mm
features are detected to the north of SMA2 and SMA3, the stronger of
the two is closest to SMA3 ($\sim 1\arcsec$ north), and is thus called
VLA3, while the other is barely above the $3\sigma$ level and is not
discussed further. The flux densities and positions of all of the
continuum sources from 1.3~mm to 3.6~cm are provided in
Table~\ref{mmpos}.

Figure~\ref{qbandabsolute}b further zooms in on the continuum emission
from SMA1b and SMA1d and the maser groups C8, C9, and C10.  From this
figure it is clear that SMA1d is also distinct from SMA1b at
1.3~cm. From our reanalysis of the 3.6~cm data originally reported by
\citet{Carral02} we also detect SMA1b at this frequency.  As mentioned
in \S2.2 there appears to be a systematic $0\farcs2$ position
offset of the 3.6~cm data which if removed would cause it to match the
1.3~cm and 7~mm position for SMA1b. Maser group C8, located $\sim
0\farcs3$ south of SMA1b, contains the brightest \water\ maser in the
whole region and is resolved into a linear north/south structure
$\gtrsim 800$~AU in length.  The velocity structure of this linear
group of masers is broad and complex, although the northern part is
mostly blueshifted while the southern part is mostly redshifted with
respect to the $\sim -3$ \kms\/ systemic velocity of SMA1 (see \S3.5
for discussion of systemic velocities).  These characteristics are
similar to the brightest \water\ maser component (C6) in the
protocluster AFGL~5142 \citep{Hunter99} but on a somewhat larger
physical scale.  A 44~GHz maser with a velocity of $\sim -0.9$ \kms\/
is also coincident with this linear \water\/ maser feature.

The redshifted and kinetically narrow \water\/ maser group C9 is
coincident with SMA1d and contains the second brightest maser feature
in I(N).  A weak blueshifted group of masers (C10) lies $\gtrsim 1200$
AU southeast of SMA1b.  Back on the larger scale of
Fig.~\ref{qbandabsolute}a, group C11 is another blueshifted narrow
feature that lies $\gtrsim 1300$~AU northwest of SMA1c. Group C6 is
$\gtrsim 400$ AU northwest of SMA2 and is blueshifted with respect to
the systemic velocity of SMA2 ($\sim -5$ \kms\/; see \S3.5). Group C7
is a narrow but fairly strong component $\gtrsim 1500$ AU northwest of
SMA2, and is near the systemic velocity.

\subsubsection{SMA4 Region}

Looking further south toward SMA4, the spatial/velocity structure of
water maser group C5, as well as the 1.3~mm and 1.3 and 3.6~cm
continuum emission in this region is shown in Figure~\ref{c5c3color}a.
There are three distinct spatial clusters of \water\/ masers within
the C5 group, corresponding to the three different velocity
components.  These three clusters of masers are oriented roughly
north/south (P.A.=$17\arcdeg$) with blueshifted emission to the north,
and redshifted emission to the south. The C5 group lies within
$0\farcs4$ ($\gtrsim 700$ AU) of the 1.3~mm to 3.6~cm continuum
emission peaks, just within the combined astronomical uncertainties of
the data.  The C5 group is also coincident with a cluster of Class~II
6.7~GHz \methanol\ masers \citep{Walsh98,Caswell2009} which have a similar
velocity gradient. The 1.3 and 3.6~cm emission has been reported
previously by \citet{Rodriguez07} and \citet{Carral02}, respectively.
We report the detection of SMA4 at 7~mm (not shown,
see Table~\ref{mmpos}). As mentioned in \S2.2 there appears to be a
systematic offset of $0\farcs2$ of the 3.6~cm data which if removed
would cause it to match the 1.3~cm and 7~mm position. It is currently
unclear if the maser and 1.3~mm offsets are significant (or simply due
to absolute position uncertainties).

\subsubsection{SMA6 Region}

A zoomed in view of the SMA6 region is shown in
Figure~\ref{c5c3color}b.  The C3 group of water masers is coincident
with SMA6 and also has three distinct velocity clusters (see
Fig.~\ref{c5c3color}b). The kinematics of C3 are curious, with the
three clusters lying along a roughly north-south line with systemic
velocity to the south, then redshifted in the middle, and then
blueshifted emission to the north. The 7~mm continuum peak toward SMA6 lies
$\gtrsim 700$ AU north of the 1.3~mm peak, coincident with the northern
blueshifted cluster of C3 masers. SMA6 is not detected at wavelengths
longer than 7~mm.  The 1.3~mm morphology of SMA6 is suggestive of
unresolved substructure in the north-south direction, and as mentioned
in \S3.1 is not well fit by a single Gaussian. Interestingly, although
the 7~mm data has significantly better spatial resolution than the
1.3~mm data, both images show a similar morphology.

\subsection{VLA 44 GHz Methanol masers}

\begin{figure}
\plotone{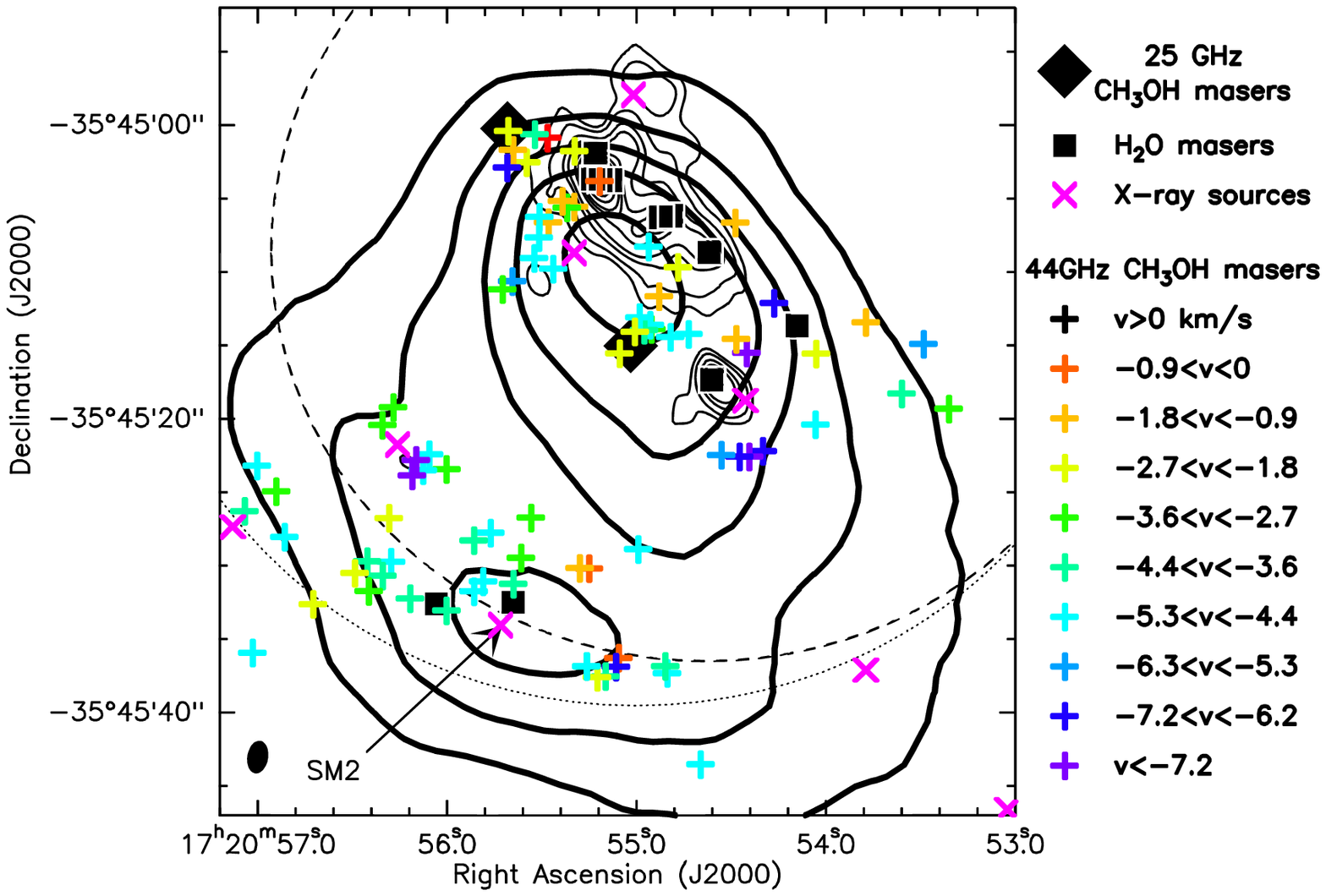}
\caption{SMA 1.3~mm continuum contours of I(N) (thin black contours)
  with thick black contours showing SCUBA 450~\mum\ emission
  (G. Sandell, private communication). The 1.3~mm continuum image and
  contour levels are the same as Fig.~\ref{contwithmasers}. The SCUBA
  450~\mum\ resolution is $8\arcsec$ and the contour levels are 60,
  80, 100, 120, 160 \jyb.  The tip of the arrow for SM2 is the
  position reported by \citet{Sandell00}. The colored crosses
  show the locations of 44~GHz Class I methanol masers from this
  work. The locations of the water maser groups (also see
  Fig.~\ref{contwithmasers}) are shown by $\blacksquare$ symbols and
  the 25~GHz \methanol\/ masers reported by \citet{Beuther05} are shown
  by $\blacklozenge$ symbols. The X-ray sources from \citet{Feigelson09}
  are marked by magenta X's.  The dashed arc indicates the SMA
  primary beam and the dotted arc indicates the VLA 44 GHz primary
  beam. The 1.3~mm SMA continuum resolution is shown in the lower left.
  \label{methanolmasers}}
\end{figure}

Strong 44~GHz \methanol\ maser emission from \ngcin\ was first
detected in the single dish survey of \citet[46\arcsec\/
beam;][]{Haschick90}.  \citet{Kogan98} resolved this emission into 23
spots using ten antennas of the VLA, but they were unable to determine
accurate absolute positions.  Our new observations with the full VLA
provide a deeper more accurate view of the maser emission.  By
examining and comparing the details of the emission in each channel
and from channel to channel, we have identified 83 individual maser
spots. These are listed in Table~\ref{methanoltable}, available in its
entirety online.  The values of $v_{\rm peak}$ refer to the center
velocity of the channel in which the peak emission occurs. The
tabulated positions are the fitted position in the peak channel for
each component. The peak brightness temperatures of these masers (see
Table~\ref{methanoltable}) lie far above the energy level of this
transition (65~K, 43.7~\percm), with values as high as $T_b=3.6\times
10^5$ K with the current spatial resolution. While we have detected
far more spots of emission, the spatial and velocity distribution of
the \citet{Kogan98} spots are quite consistent with our data if one
shifts the \citet{Kogan98} positions by 7.3\arcsec\ to the
south-southwest. At the current level of sensitivity, no thermal
emission was detected.

The maser positions are shown as velocity color-coded crosses in
Figure~\ref{methanolmasers}.  The peak maser velocities are in the
range $-8.49$ to $-0.35$ \kms\/, close to the systemic velocity of the
region.  Figure~\ref{methanolmasers} also shows a 450~\mum\/ JCMT
continuum image with $8\arcsec$ resolution from \citet{Sandell00}. The
masers are distributed over much of the area of dust emission, as
delineated by the lowest $450~\mu$m contour.  The masers can be
divided into two main concentrations: the general area of the compact
SMA continuum sources, and the area around the single-dish source SM2
\citep{Sandell00}.  The former concentration contains a more diverse
range of velocities, though no obvious trends in kinematics versus
position are seen. It is notable that while these masers are clustered
around I(N) 1.3~mm continuum sources, only one is positionally
coincident with a compact SMA source (SMA1, see
Fig.~\ref{qbandabsolute}). The latter concentration towards the single
dish continuum source SM2 is not associated with any 1.3 mm SMA
continuum emission but this region lies outside the half-power point
of the SMA data, significantly reducing the sensitivity in this
region. Two 25~GHz Class I \methanol\/ masers were also detected
toward I(N) by \citet{Beuther05}, and both of these are coincident
with 44~GHz maser emission and have comparable velocities as shown in
Fig.~\ref{methanolmasers}.  

The position of the X-ray sources recently reported for this region
\citep{Feigelson09} are also shown in Figure~\ref{methanolmasers}.
The median photon energy of many of these X-ray sources is relatively
hard (E > 4 keV), suggesting deeply embedded sources. These authors
have already noted the possible association of X-ray source 1454 with
SMA7, and 1447 with SMA6 (although the latter is less likely given
that its median photon energy is only 1 keV).  With our new data, we
note four additional possible associations: X-ray source 1470 is
within $1\farcs8$ of the water maser component C2; and X-ray sources
1463, 1477 and 1487 each reside within a few arcseconds of a cluster
of methanol masers.  Source 1463 lies SSE of SMA1, just beyond the
lowest millimeter contour.  Source 1477 aligns with the shoulder of
$450~\mu$m emission northeast of SM2, while source 1487 lies just east
of the lowest $450~\mu$m contour

\subsection{SMA 1.3 Millimeter Molecular Line Data}

A total of 79 lines from 19 molecular species are detected toward SMA1
(Table~\ref{sma1lines}) along with 14 unidentified features within the
4 GHz of total bandwidth observed by the SMA. The spectra toward the
SMA1 1.3~mm continuum peak are shown in Figure~\ref{dsb}.  The lines
were indentified using the CDMS \citep{Muller01}, JPL
\citep{Pickett98}, and
Splatalogue\footnote{http://www.cv.nrao.edu/php/aremijan/splat\_beta/}
spectral line databases. The large number of organic transitions
detected, coupled with modest line brightness temperatures of 30~K
suggests that this source harbors a moderate ``hot core''
{\citep{vandishoeck98}.  Many of the stronger lines detected toward
SMA1 are also detected toward SMA2, though most are weaker, suggesting
that this source also harbors a weak hot core.  Other sources observed
with similar linear resolution with this SMA setup have yielded both
less diverse \citep[S255N,][]{Cyganowski07}, as well as much richer
hot core spectra (NGC6334I, Brogan et al., in prep). The broadband
SEST single-dish molecular line survey toward NGC~6334I and I(N) also
shows that while source I(N) exhibits significant hot core emission,
source I shows a much richer array of molecular line species along
with stronger emission \citep{Thorwirth03}.

\begin{figure*}
\includegraphics[scale=0.8,angle=0]{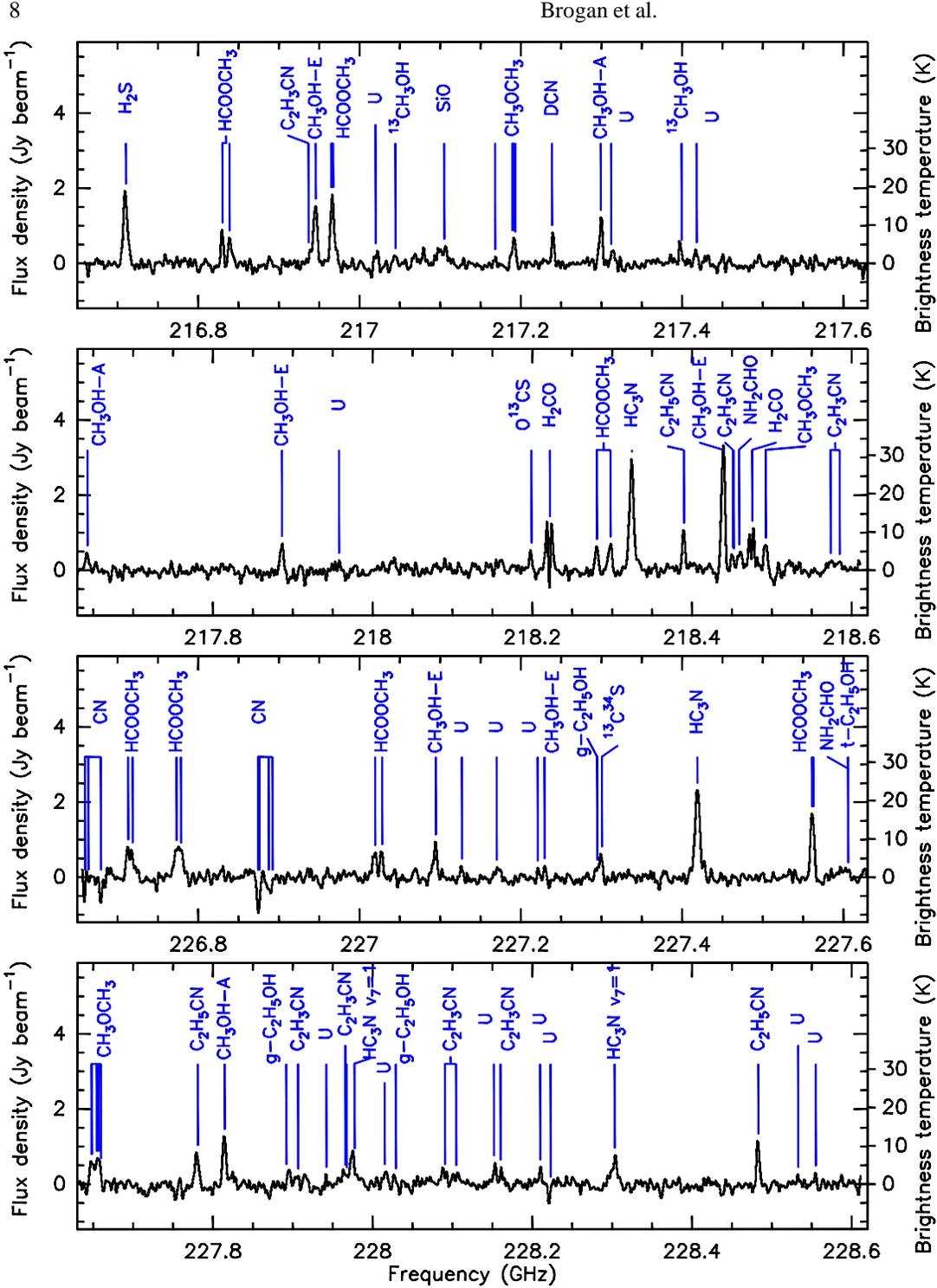}
\caption{SMA spectra of the upper and lower sidebands from the 1.3 mm
  continuum peak of NGC6334I(N)-SMA1. The spectra have been Hanning
  smoothed.
\label{dsb}}
\end{figure*}

\subsubsection{Extended Molecular Line Emission}

\begin{figure}
\plotone{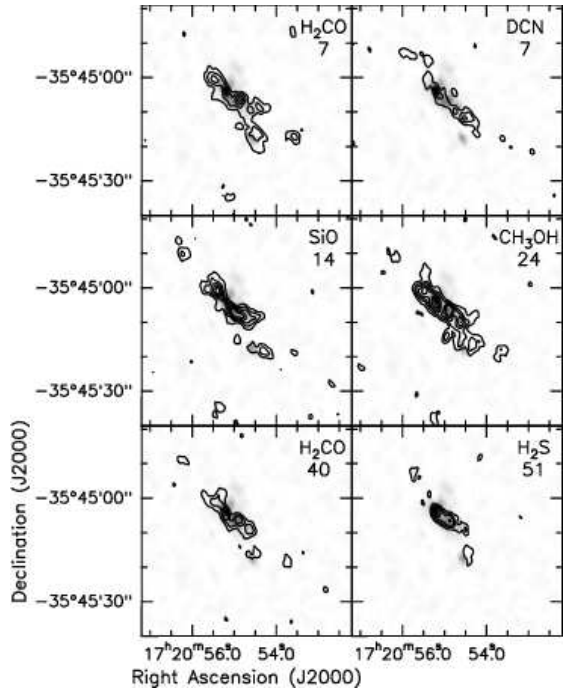}
\caption{In each panel, the colorscale is the 1.3~mm continuum and the
  contours show the integrated intensity of the indicated spectral
  lines.  The dashed circle is the FWHP of the SMA primary beam; these 
  images have not been corrected for primary beam attenuation.
  The contour levels are \formaldehyde\ ($3_{0,3}-2_{0,2}$): 4.35, 7.25,
  10.15, 13.05 \jykms ; DCN: 1.77, 2.95, 4.13 \jykms ; SiO: 6, 10, 14,
  18, 22 \jykms ; \methanol-E ($4_{+2,2}-3_{+1,2}$): 4.8, 8, 11.2,
  14.4, 17.6 \jykms; \formaldehyde\ ($3_{2,2}-2_{2,1}$): 3.24, 5.4,
  7.56 \jykms; \htwos: 2.55, 4.25, 5.95, 9.35 \jykms. The number below each
  molecular name is the $E_{\rm lower}$ of the transition in units of
  \percm.
  \label{molportext}}
\end{figure}

\begin{figure*}
\includegraphics[scale=0.7,angle=0]{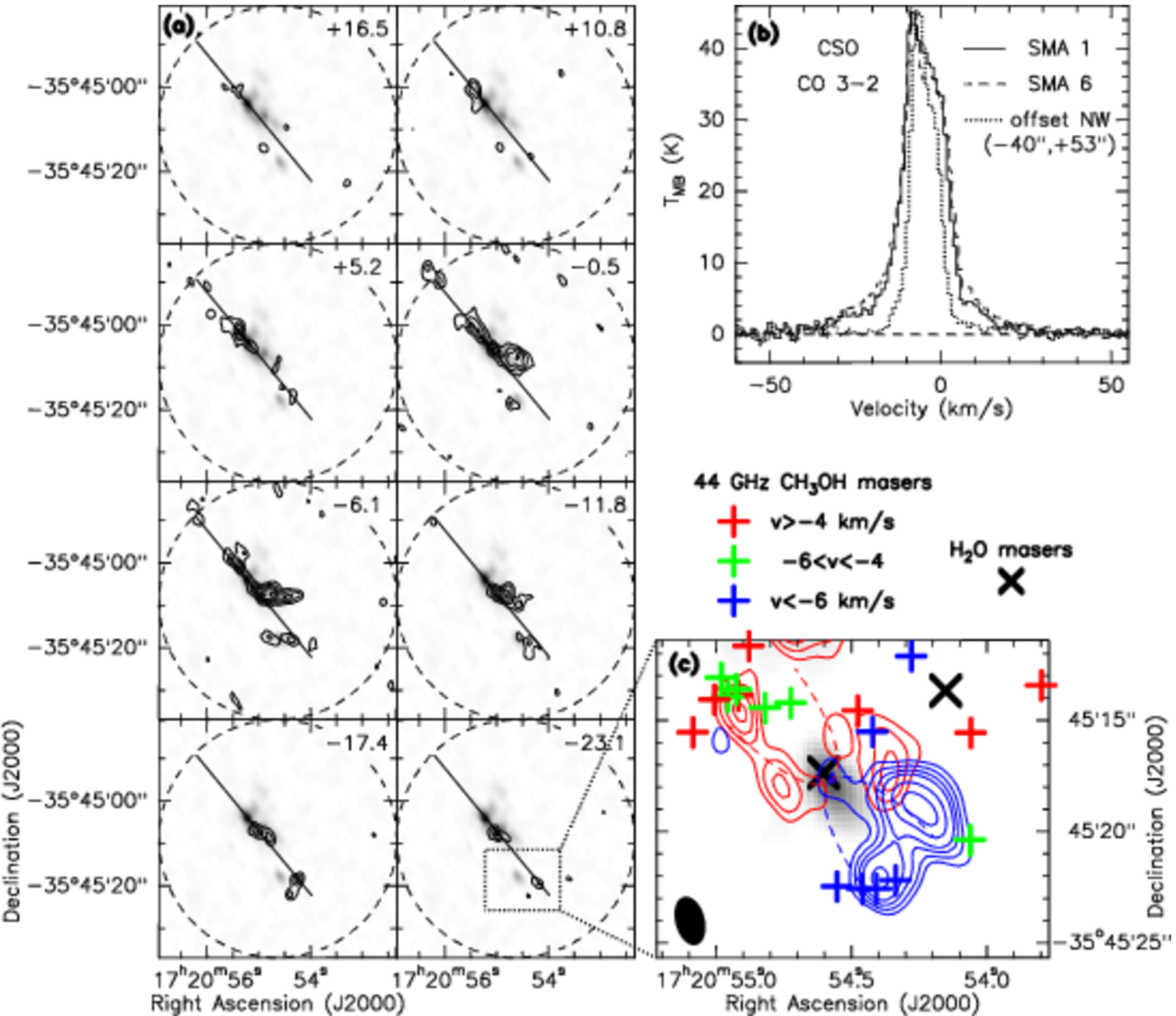}
\caption{(a) Channel maps of SiO 5--4.  In each panel, the greyscale
  is the 1.3~mm continuum emission and the contours are SiO emission
  (levels = 0.29, 0.48, 0.67, 0.86, 1.05 Jy).  The outflow axis is drawn at position
  angle $+39\arcdeg$.  The dashed circle is the SMA primary beam.  (b)
  CSO CO (3--2) spectra taken near SMA1, SMA6 and a position offset
  $-40\arcsec,+53\arcsec$ to the northwest (away from any obvious
  outflow emission). The SMA1 spectrum was taken 2.9\arcsec\ from
  SMA1, and the SMA6 spectrum is the average of the three spectra
  taken closest to SMA6 (5\arcsec, 7\arcsec, and 8\arcsec\ away).  (c)
  Close up of SiO (5--4) outflow around SMA6.  The contours show SiO
  (5--4) integrated emission, red contours show integrated intensity
  from +17 to -6 \kms\ and blue contours show integrated intensity
  from from -8 to -31 \kms\/.
The \methanol\/ and \hto\/ maser
positions are also shown.  The solid black ellipse is the SMA 
synthesized beam.
\label{sio8chanmaps}}
\end{figure*}

Many of the lines from low excitation states ($E_l <$ 70 ~\percm) show
emission extending to the NE and SW of the SMA continuum emission (see
Figure~\ref{molportext}). This emission is clearly tracing one or more
outflows emanating from the protocluster. Interpretation of the
extended emission is complicated by the fact that the SMA data are not
sensitive to smooth emission on sizescales $\gtrsim 20\arcsec$ (due to
spatial filtering by the interferomter). Generally, the emission to
the northeast is red-shifted, while that to the southwest is
blueshifted, but the kinematics become very confused in the southern
part of I(N) with both red and blueshifted emission present. SiO in
particular shows a striking bipolar appearance centered on SMA1.
Channel maps of SiO (Figure~\ref{sio8chanmaps}a) reveal the complex
velocity field of the extended emission.  A lobe of molecular material
also extends westward from SMA4 at low, but blueshifted velocities in
SiO and several other species shown in Fig.~\ref{molportext} (H$_2$CO,
\methanol, and H$_2$S for example).

A more compact SiO velocity gradient also appears to be centered on
SMA6. CSO CO(3--2) spectra with 20\arcsec\ resolution show that the
broad line wings toward SMA6 are of similar magnitude to those seen
toward SMA1 (see Fig.~\ref{sio8chanmaps}b), suggesting both objects
harbor outflows. The integrated red and blue-shifted emission around
SMA6 is shown in detail in Figure~\ref{sio8chanmaps}c, and is more or
less centered on the CN and H$_2$CO ($E_{l}=7$ \percm) absorption
velocity of $\sim -4$ \kms\/ (see \S3.5.2).  This figure also shows
that the sense of the SiO gradient is reflected in the position and
velocity of the 44~GHz methanol masers surrounding SMA6. However, on
smaller scales the C3 group of water masers is oriented in a more
north-south direction (Fig.~\ref{c5c3color}b). It is unclear how the
curious kinematics of the C3 water masers (from south to north: systemic,
then red, then blueshifted features) fits in with the compact NE-SW
outflow.

\subsubsection{Compact Molecular Line Emission and Absorption}

\begin{figure}
\plotone{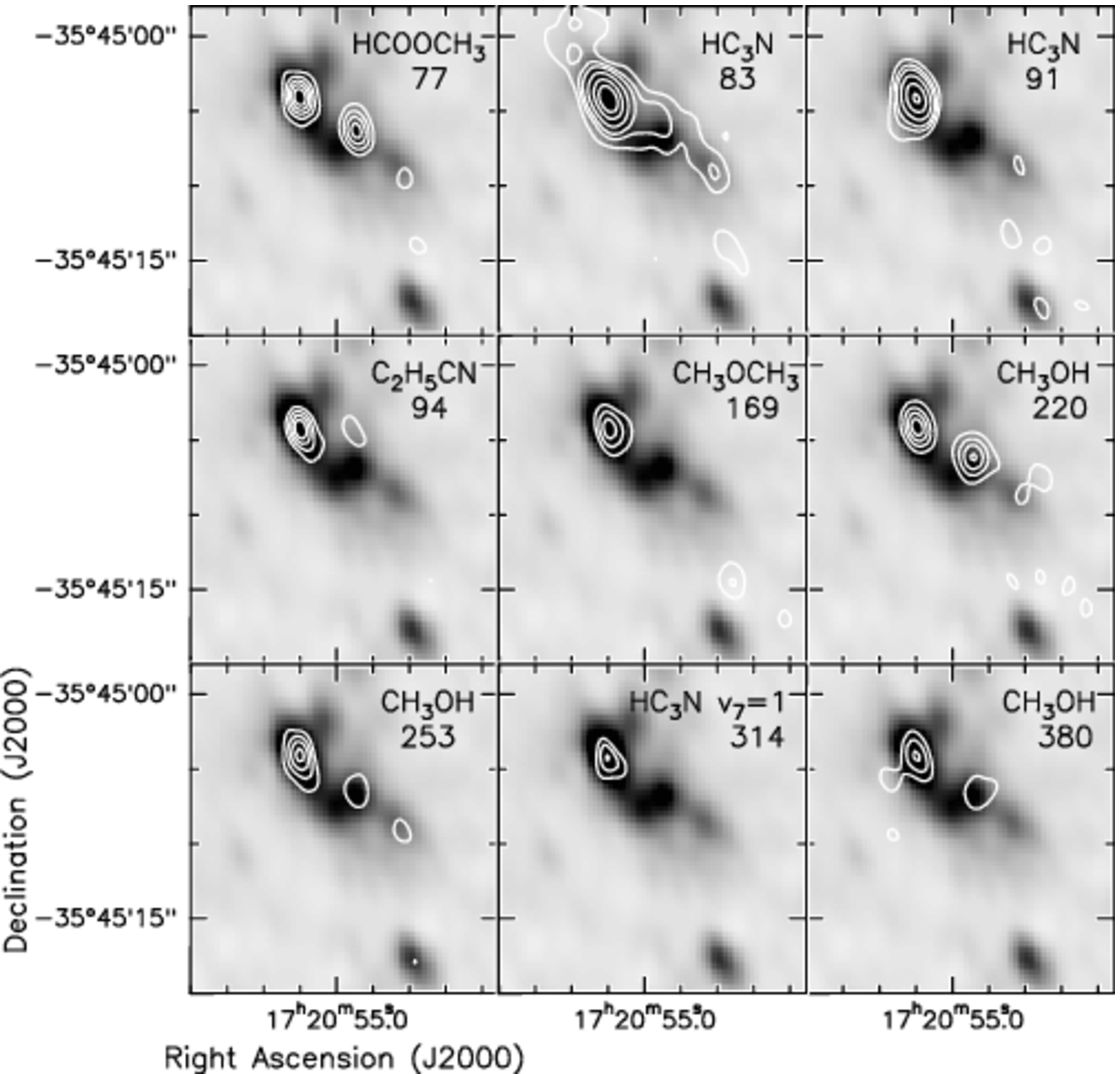}
\caption{In each panel, the greyscale is the 1.3~mm continuum and the
  contours are drawn from moment-zero images of spectral lines.
  The contour levels are: \methylformate: 2.1, 4.2, 6.3, 8.4, 10.5
  \jykms; \hcccn: 2.55, 4.25, 5.95, 9.35, 14.45, 19.55 \jykms;
  \ethylcyanide: 1.8, 3.0, 4.2, 5.4 \jykms; \dimethylether: 1.35,
  2.25, 3.15 \jykms; \methanol(220): 1.89, 3.78, 5.67, 7.56 \jykms;
  2.31, 3.85, 5.39, 6.93 \jykms; \hcccn(v$_7$=1): 1.5, 2.5, 3.5
  \jykms; \methanol(380): 2.1, 3.5, 4.9 \jykms.  The numbers below each
  molecular name is the $E_{\rm lower}$ of the transition in units of
  \percm.
\label{molport}}
\end{figure}

As seen in Figure~\ref{molport}, emission from the higher excitation
lines ($E_l >$ 70~\percm) is concentrated toward SMA1.  In several
species, a secondary peak of emission appears toward SMA2. Indeed, the
SMA2 peak in methyl formate and the lower lying transitions of
methanol is almost equal to that of SMA1. A few vibrational states of
\methanol\/ and HC$_3$N are detected exclusively toward SMA1,
suggesting that it is warmer and/or denser than SMA2. Several species
also show weak emission toward SMA4 and to the north and west of SMA6.
Figure~\ref{mom1} shows the first moment maps for several
representative species detected in emission, along with the CN
molecule which is detected in absorption. Because of the possibility
of confusion with outflowing or infalling gas, it is difficult to
conclusively determine the systemic velocities of all of the 1.3~mm
cores.  Both SMA1 and SMA2 show enough consistency in a range of
species with compact emission to estimate that $V_{lsr}({\rm
  SMA1})=-2.8\pm 0.5$ \kms\/ and $V_{lsr}({\rm SMA2})=-4.5\pm 0.5$
\kms\/.  Compact emission from SMA4 is weak and shows a peak emission
velocity of $V_{lsr}({\rm SMA4})=-5\pm 1$ \kms\/. No {\em emission}
centered on SMA6 was detected.

The CN(2--1) lines are seen in absorption against most of the stronger
compact continuum sources, while most of its extended emission is
resolved out by the interferometer. None of the 1.3~mm continuum
brightness temperatures ($T_b$) are high enough at the current angular
resolution to detect the $E_l=40$~\percm\/ H$_2$CO transition in
absorption (negative) against it, though self-absorption is seen
toward SMA1 and SMA2.  Only SMA1 has a continuum $T_b$ ($\sim 15$ K at
current resolution) high enough to detect the H$_2$CO $E_l=7$ \percm\/
transition in (negative) absorption.  Figure~\ref{spectra}a and b show
spectra of \meth\/ ($E_l=77$~\percm\/), \form\/ at $E_l=7$ and 40
\percm\/, and CN ($E_l=3.8$~\percm\/) toward the SMA1 and SMA2 1.3~mm
continuum peaks.  The absorption velocity of CN toward SMA1 is in good
agreement with $V_{lsr}({\rm SMA1})=-2.8\pm 0.5$ \kms\/. The 
absorption seen in the H$_2$CO ($E_l=7$~\percm) line is slightly
redshifted by $\sim 0.5$ \kms\/ compared to CN and the
$V_{lsr}$. Interestingly, the $E_l=40$~\percm\/ H$_2$CO line shows a
self-absorption dip redshifted by $\sim 1$~\kms\/ relative to CN and
the $V_{lsr}$ (see Fig.~\ref{spectra}a).

SMA2 shows CN absorption redshifted by $\sim 4$ \kms\/ from the
$V_{lsr}({\rm SMA2})=-4.5\pm 0.5$ \kms\/, suggesting strong infall is
present towards this source (see Fig.~\ref{spectra}b).  In contrast,
the $E_{l}=40$~\percm\/ H$_2$CO emission peak is slightly blueshifted
($\sim 0.5$ \kms\/) compared to other emission lines (see for example
\meth\/ in Fig.\ref{spectra}b), while the $E_{l}=7$~\percm\/ H$_2$CO
transition's peak is blueshifted by $2$ \kms\/. Neither \form\/ line is
Gaussian in shape, with the redshifted side of both transitions
cutting off very sharply as might be expected from very strong,
red-shifted self-absorption. Weak blueshifted ($-5$ to $-15$ \kms\/)
H$_2$O masers were also detected toward SMA2 so these masers and
H$_2$CO could be tracing a pole-on outflow, with the $V_{lsr}$
emission missing due to self-absorption and the redshifted H$_2$CO
outflow emission obscured by the continuum.

No compact molecular line emission is detected toward SMA3 so it is
unclear if the CN absorption toward this core with a velocity of $\sim
-2$ \kms\/ is also tracing infall or the $V_{lsr}$.  Toward SMA4, CN
shows absorption at $\sim -5$ \kms\/ in good agreement with the weak
emission lines detected toward this source including $E_{l}=40$~\percm\
H$_2$CO ($V_{lsr}({\rm SMA4})=-5\pm 1$ \kms\/).  The $E_{l}=7$
\percm\/ H$_2$CO transition toward SMA4 is difficult to interpret as
it is non-Gaussian in shape with a peak at $\sim -2.5$ \kms\/ and a
significant blue-shifted shoulder.  The CN absorption toward SMA6 has
a velocity of $-4.0\pm 0.5$ \kms\/, in good agreement with the
velocity of a self-absorption dip seen in $E_{l}=7$~\percm\/
H$_2$CO. Since no compact emission is detected towards this source it
is unclear how this absorption velocity compares to the $V_{lsr}$ of
SMA6.

Also of interest is the barely resolved SE-NW (or sometimes more E-W)
velocity gradient across SMA1 evident in a number of compact species
(see for example HCOOCH$_3$ and CH$_3$OH in Fig.~\ref{mom1}). The full
width of the velocity gradient is about $\sim 5$ \kms\/ (from
inspecting the line cubes), in reasonable agreement with the gradient
inferred by \citet{Beuther07} for NH$_3$(6,6) based on a double peaked
profile. This gradient is discussed further in \S4.5.

\begin{figure}
\plotone{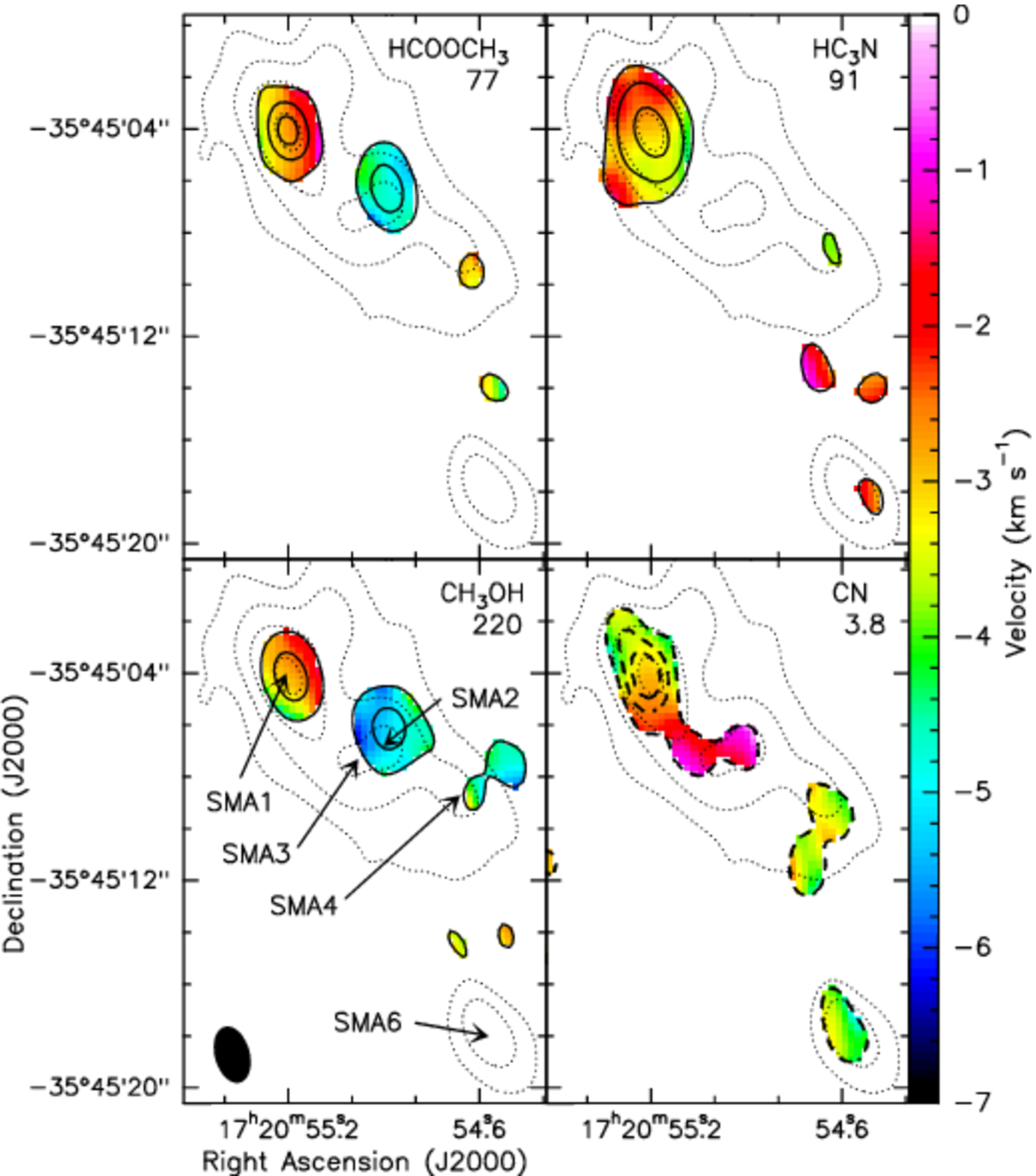}
\caption{In each panel, the color scale shows the first moment map of the
indicated molecular line, while the solid black contours depict the
integrated intensity image of the line emission.  In the case of CN,
the dashed contours indicate that the line appears in absorption.  The
levels are: \methylformate = 2.1, 6.3, 10.5 \jykms; \hcccn = 2.55,
5.95, 14.45 \jykms; \methanol = 1.89, 5.67 \jykms; CN = -1.2, -2.4,
-3.6 \jykms.  In each panel, the thin dotted contours show the 1.3~mm
continuum emission (levels = 60, 180, 390, and 900 \mjb).  The numbers
below each molecular name is the $E_{\rm lower}$ of the transition in
units of \percm. \label{mom1}}
\end{figure}

\begin{figure}
\plotone{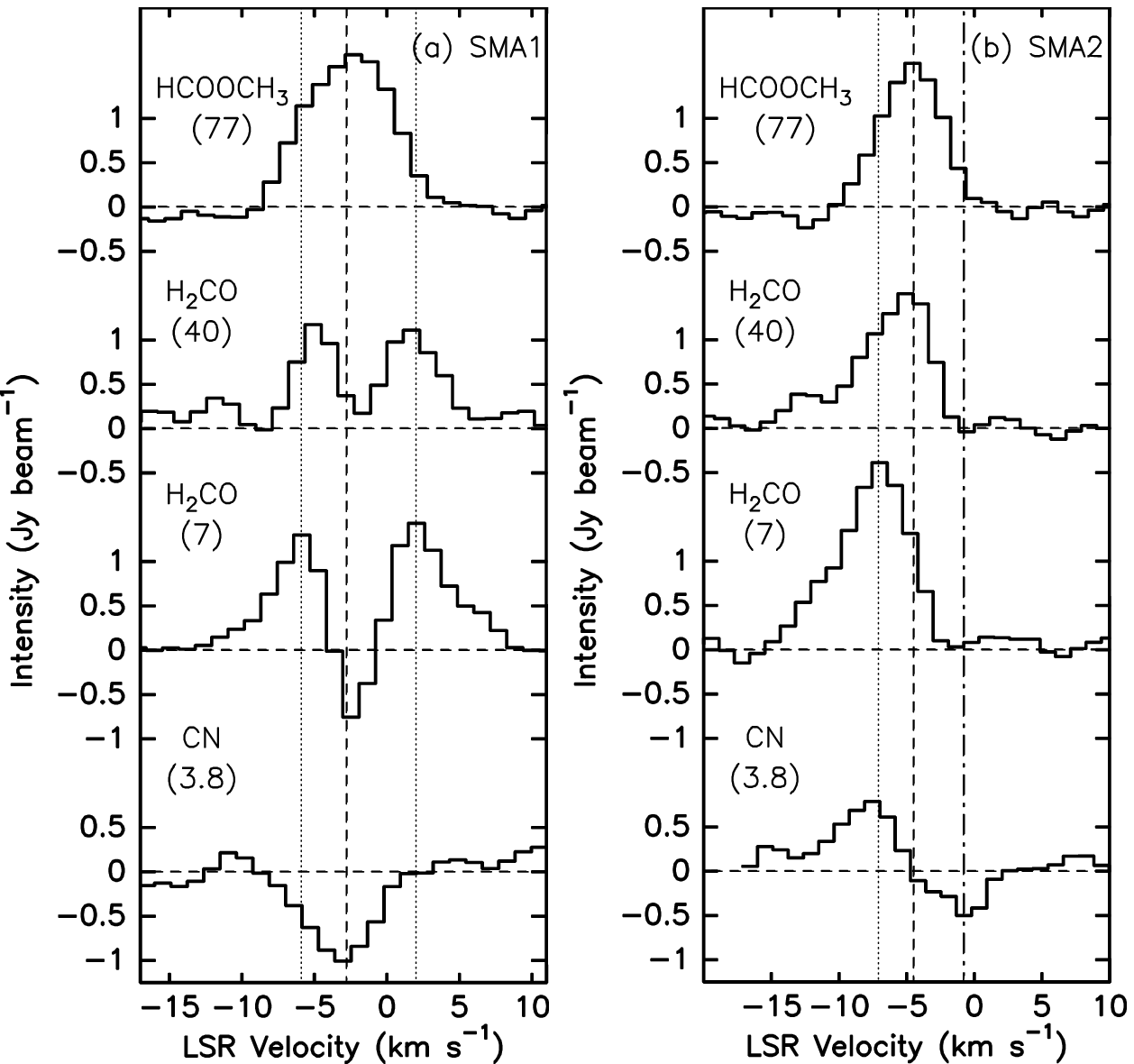}
\caption{(a) Sample spectra toward the 1.3~mm continuum peak of
  SMA1. (b) Sample spectra toward the 1.3~mm continuum peak of SMA2.
  In both plots the dashed line indicates the $V_{lsr}$ and the numbers
  in parenthesis under the molecule names are the lower state energy in
  cm$^{-1}$. In (a) the two dotted lines indicate the velocity of the
  peak red and blueshifted H$_2$CO ($E_l=7$ cm$^{-1}$) emission. In
  (b) the dotted line indicates the velocity of the peak
  blueshifted H$_2$CO ($E_l=7$ cm$^{-1}$) emission and the dot-dashed
  line indicates the velocity of maximum CN absorption.
\label{spectra}}
\end{figure}

\subsection{Mid-infrared Emission}

Figure~\ref{irac} shows a three color mid-IR image of the I(N) region
with methanol maser positions and integrated SiO and 1.3 mm continuum
contours superposed.  The extended 4.5 \mum\/ emission to the SW of
the I(N) continuum has been reported previously by
\citet{Hunter06}. In addition, a weak region of predominantly 4.5
\mum\/ emission is also visible between the SMA6 continuum peak and
the arc of 44 GHz \methanol\/ masers $5\arcsec$ to the SW; this region
is coincident with the blueshifted side of the SMA6 outflow
(Fig.~\ref{sio8chanmaps}c). Clumpy knots of extended 4.5 \mum\/
emission are also present to the NE of SMA1, and in the vicinity of
SMA7. We also present for the first time, an unresolved source of 24
\mum\/ emission located $\sim 2\arcsec$ SW of SMA4.  This figure
demonstrates that with the exception of extended 4.5 \mum\/ emission,
and a single source of 24 \mum\/ emission, \ngcin\/ is relatively dim
in the mid-IR. Indeed, SMA4 appears to be the only millimeter
continuum source directly associated with a mid-IR source. In
contrast, the bright haze of red emission in the SW corner of
Fig.~\ref{irac} emanates from the periphery of the saturated mid-IR
bright protocluster \ngci\/. The nature of the mid-IR emission is
discussed in detail in \S4.1 and \S4.2.

\begin{figure*}
\includegraphics[scale=1.0,angle=0]{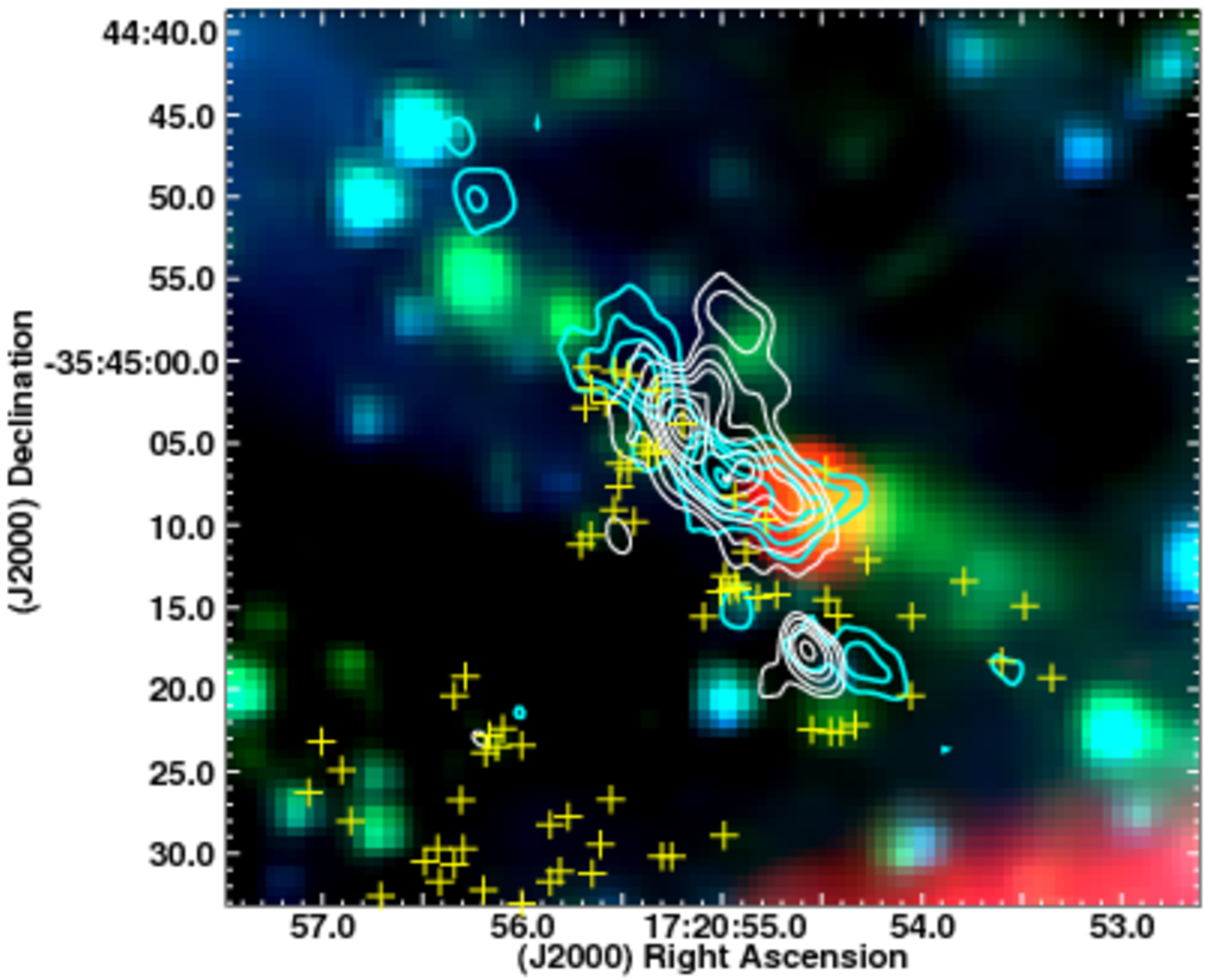}
\caption{{\em Spitzer} 3-color image with RGB mapped to 24, 4.5, and
  3.6 \mum\/. White 1.3~mm continuum and cyan SiO integrated intensity
  contours are overlaid, along with yellow $+$ symbols showing the
  locations of the 44~GHz \methanol\/ masers. The 1.3~mm continuum
  contour levels are the same as Fig.~1, and the SiO (5--4) integrated
  emission contour levels are 6, 10, 14, 20, 26 \mjb\/*\kms\/.
\label{irac}}
\end{figure*}

\section{DISCUSSION} 

\subsection{Presence of Multiple Outflows}

Due to the ubiquity of bipolar outflows from protostars, one would
expect to find multiple outflows in a protostellar cluster.  As
discussed in \citet{Cyganowski08} (and references therein), the 4.5
\mum\/ IRAC passband can be dominated by molecular line emission from
H$_2$ and vibrationally-excited CO in regions of strong shocks such as
those found in massive molecular outflows. To the NE of SMA1, the
extended 4.5 \mum\/ emission appears to be an extension of the
large-scale NE-SW SMA1 outflow traced by SiO and other tracers of the
extended emission (see
Figs.~\ref{molportext},\ref{sio8chanmaps}a,\ref{irac}). To the south
where the outflows from SMA1 and SMA4 (westward flow) overlap, it is
unclear which outflow the 4.5 \mum\/ emission is arising
from. Interestingly, the 4.5~\mum\/ emission extends further than the
SiO does to the west and SW. As described in \S3.5, SMA2 may harbor a
more or less pole-on outflow as traced by blueshifted H$_2$O masers
and possibly H$_2$CO. As described in \S3.6 and \S3.7, a smaller scale
NE-SW outflow also appears to emanate from SMA6, with blueshifted SiO
and 44~GHz \methanol\ maser emission located to the south coincident
with 4.5 \mum\/ emission (Figs.~\ref{sio8chanmaps}c,\ref{irac}).

The ATCA molecular line data reported by \citet{Beuther07,Beuther08}
(HCN(1--0) and NH$_3$ for example) show good overall morphological
agreement with the largescale emission emanating from SMA1 observed
with the SMA (see for example
Figs.~\ref{molportext},\ref{sio8chanmaps},\ref{irac}). In particular,
the complex kinematics of the extended emission were also noted by
Beuther et al.\ and ascribed to possible precession. This idea is in
good agreement with the discovery of a possible tight binary in SMA1b
and SMA1d (separated by $\gtrsim 800$ AU) if these two sources do in
fact represent two different protostars \citep[see for
example][]{Chandler05,Anglada07,Cunningham09}. Although
Figs.~\ref{molportext},\ref{sio8chanmaps}a,b,c,\ref{irac} show SiO and
other molecules in the vicinity of SMA6, \citet{Beuther07,Beuther08}
did not detect any significant molecular emission in this
region. However, SMA6 was near the edge of the ATCA primary beam.

It is notable that the orientation of the NE-SW outflow inferred to
originate from SMA1 is different from the largescale SE-NW outflow
direction inferred from single-dish SiO(5--4) data with $23\arcsec$
resolution presented by \citet{Megeath99}. In their single dish data,
blueshifted emission is located to the SE of I(N) in the vicinity of
SM2 (Fig.~\ref{methanolmasers}), while blueshifted and redshifted
emission overlap in the vicinity of I(N). Toward I(N), the P-V diagram
presented by \citet{Megeath99} shows good agreement with the velocity
range observed in the SMA SiO(5--4) data ($-30$ to $+17$ \kms\/,
Fig.~\ref{sio8chanmaps}a,b), and indeed the SEST data did not have
sufficient spatial resolution to resolve the SMA sizescale flow from
SMA1. Moreover, \citet{Megeath99} also observe a clump of shocked
2.2~\mum\/ H$_2$ emission coincident with the northeastern lobe of the
SMA1 outflow and extended 4.5~\mum\/ emission. Thus, it seems likely
that the single outflow reported by \citet{Megeath99}, is in fact at
least two outflows -- one consistent with the NE-SW flow detected by
the SMA and a second in the vicinity of SM2 and the C1 and C2 water
masers. This second outflow would also explain the widespread 44~GHz
\methanol\ maser emission observed in this region.  While this seems
the most probable interpretation of the data, it is also possible that
spatial filtering of the SMA data on sizescales $\gtrsim 20\arcsec$
have played a role in the apparent discrepancy.  Unfortunately, the
SM2 area is beyond the primary beam of the SMA observations, so
additional data will be required to explore this region of activity
further.

\subsection{Nature of Mid-infrared Emission}

To assess the nature of the mid-IR emission in more detail we analyzed
the 1-8~\mum\ flux densities of several knots in the vicinity of I(N)
that morphologically appear to be dominated by shock line emission. We 
included data from the {\em Spitzer} IRAC bands \citep[see for
example][]{ybarra}, but also near-infrared $J$, $H$, and $K_s$ fluxes
\citep[see for example][]{smith95}.  To derive the physical
conditions, we have implemented a set of shock models similar to that
of \citet{ybarra} \citep[see also][]{neufeld,smith05,smith06}.  We
calculate the equilibrium excitation of molecular hydrogen (H$_2$) in
a grid of temperatures (1000-5000~K) and densities (10$^2$-10$^6$~\cc)
using the escape probability code Radex \citep[][since the optical
depths are low, the escape probability formalism is not critical, but
the code was still very useful to read the extensive tables of
molecular data]{radex}.  We use excitation rates from collisions with
H$_2$ and He from \citet{lebourlot,lebourlot2} including reactive
collisions as prescribed therein, and from collisions with H from
\citet{wrathmall}.  Quadrupole transition probabilities are from
\citet{wolk}.  We convolve the line intensities with {\em Spitzer}
\citep{reach} and
2MASS\footnote{\url{http://www.ipac.caltech.edu/2mass/releases/allsky/doc/sec3\_1b1.html}}
filter transmission profiles to calculate band-average fluxes as
specified in appendix A of \citet{robitaille}.

The analysis of the shocked emission is complicated by diffuse PAH
emission, especially in the 5.8 and 8.0~\mum\ images.  We performed
several different background subtractions to attempt to isolate the
H$_2$ emission, with the assumption that most of the emission at
4.5~\mum\ is H$_2$ (there are no PAH emission features in that band).
Naturally this decomposition is subject to uncertainty, so the
precision of shock physical parameters is limted by this observational
fact.  We find that the near and mid-IR broadband flux ratios are most
consistent with the hottest post-shock temperatures ($\sim$4000K),
indicating that these are strong shocks as would be expected from a
powerful massive stellar outflow.  The density in the shock cooling
zone is harder to constrain because density changes the broadband flux
ratios less than does temperature \citep[see Fig.~2 in ][]{ybarra},
and because foreground extinction can mimic a decrease in emitting
material density.  The fluxes here are broadly consistent with
densities of 10$^3$--$10^5$~cm$^{-3}$.  Once we have constrained the
physical parameters in the H$_2$ emitting zone, we can translate
mid-infrared fluxes into total shock luminosity.  For shock
recombination zones (T$\simeq$4000K) our models indicate that 6\% of
the total H$_2$ line emission comes out in the IRAC 4.5~\mum\ band.
The NGC6334I(N) outflow emits 15$\pm$3~Jy over 100$\pm$20 square
arcseconds (sum of both lobes).  At a distance of 1.7~kpc, that
corresponds to 2300$\pm$300~L$_\sun$ total luminosity in shocked
molecular hydrogen, with an average column density of
3.5$\pm$0.5$\times$10$^{21}$cm$^{-2}$, for a total of 0.1~M$_\sun$ of
hot shocked material (that which is emitting at the highest
temperatures -- a much larger mass of cooler and entrained material is
present in the outflow as well, simply not emitting in the near
infrared).

SMA4 is the only source in I(N) associated with 6.7~GHz \methanol\/
masers (see \S3.3.2), and in \S3.6 we report the detection of a
24~\mum\/ source $2\arcsec$ to the SW of the 1.3~mm continuum
peak. Class~II 6.7~GHz \methanol\/ masers are thought to be
radiatively pumped by warm dust, and are found almost exclusively in
regions of massive star formation \citep{Cragg02,Cragg05,Minier03},
providing strong evidence that SMA4 is a massive protostar. Two
scenarios exist for the nature of the 24~\mum\/ emission: (1) it may
trace emission from hot dust near a central protostar and is the
24~\mum\ counterpart to SMA4 or (2) it traces emission from hot dust
in the walls of the cavity formed by the outflow traced by extended
4.5 \mum\/ and molecular line emission, as well as 44 GHz methanol
masers W/SW of SMA4. Recent work by \citet{deBuizer05,deBuizer07}
favor the latter interpretation. Higher-resolution mid-infrared data
would help to distinguish between these possibilities for SMA4.
The 24~\mum\ flux is 0.45$\pm$0.04~Jy, or 2~$L_\sun$ (just in the
24~\mum\ band).  Protostars emit roughly 5\% of their luminosity in
the 24~\mum\ band over a relatively large range of evolutionary state,
so if this source represents a significant fraction of the luminosity
of the driving protostar, the latter would have an approximate
luminosity of only about 40~$L_\sun$.  This low luminosity lends
support to the interpretation that the 24~\mum\/ emission is coming
from hot dust away from the central source, such as on an outflow
cavity, and that the driving source itself is still highly embedded
and extinguished even at 24~\mum.

\subsection{What are the 44 GHz Methanol Masers Tracing?}

The 44~GHz Class~I \methanol\ maser line is often found in regions of
high-mass star formation
\citep{Haschick90,Slysh94,Kurtz04,Valtts07,Pratap08}. Like other Class
I masers \citep{Plambeck90}, it is often found separated from compact
\HII\/ regions and in many cases appears to be associated with shocked
gas created by bipolar outflows \citep[e.g.][]{Kurtz04,Sandell05}.
While this maser has been found within other millimeter protoclusters
such as S255N and G31.41+0.31 \citep{Cyganowski07,Araya08}, the number
and spatial distribution of maser spots is particularly large in
\ngcin.  In all of these studies including the current one, the 44~GHz
\methanol\ masers tend to lie within a few \kms\/ of the systemic
velocity even though they typically trace powerful outflows. This
dichotomy is likely due to the fact that an enhanced column of
\methanol\ gas with sufficient velocity coherence along the
line-of-sight is most likely to be found where a shock is moving
perpendicular to the line-of-sight (i.e. a shock viewed edge-on).  The
same effect is believed to give rise to the narrow \ammonia\ (3,3)
maser emission located near the ends of the protostellar jet from
IRAS~20126+4104 \citep{Zhang99}.  

As described in \S4.1, it is possible that the 44~GHz masers located
in the SE portion of the observed field of view are associated with an
outflow originating from the relatively unexplored region of continuum
emission SM2.  The discovery of an X-ray source near SM2 indicates
that additional cluster members may be present in this area.  A more
general conclusion can be drawn from the relatively widespread extent
($45''\sim 1.2 \times 10^{18}$~cm) of 44~GHz maser emission, an extent
that is shared by ammonia (1,1) and (2,2) line and 450~\mum\ continuum
emission (see Fig.~\label{methanolmasers}).  For a single dominant YSO
to maintain this region at a conservative minimum temperature of 30~K
(the ammonia temperature from \citet{Kuiper95} and average dust
temperature from \citet{Sandell00}) would require a central luminosity
of $5 \times 10^5$~\lsun\ \citep[e.g.][]{Scoville76} which is 30 times
higher than the value ($1.7\times 10^4$ \lsun\/) inferred from the
far-infrared spectral energy distribution for this region
\citep{Sandell00}.  Thus, these phenomena must be driven by the
collective activity of a number of cluster members.

\subsection{Temperatures of the Hot Core Sources SMA1 and SMA2}

\begin{figure}
\plotone{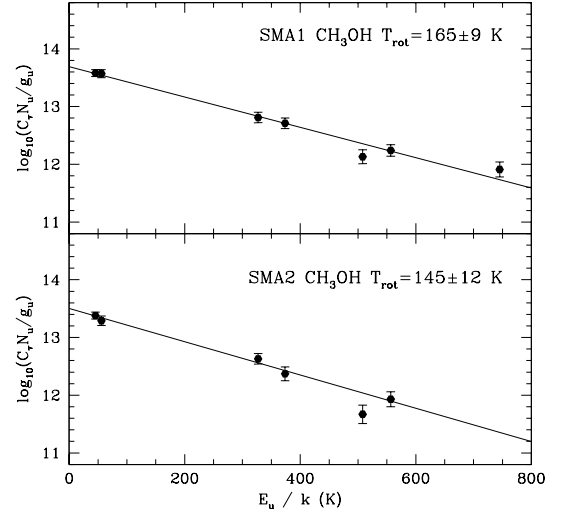}
\caption{Rotation diagrams for the \methanol\/ transitions observed
  for SMA1 ({\em top}) and SMA2 ({\em bottom}); the fitted
  temperatures are indicated. The column densities have been corrected
  for optical depth effects as described in \S 4.3.
\label{meth_rot}}
\end{figure}

When several transitions of the same molecule spanning a range of
energies are observed, temperatures can be inferred from the rotation
diagram method \citep[see for example][]{Goldsmith99}, using the
relations
\begin{equation}
\frac{N_u}{g_u} = \frac{3k}{8\pi^3\nu g_Ig_K}\frac{1}{\mu^2S}\int S_{\nu}dv,
\end{equation}
and
\begin{equation}
{\rm log}(N_u/g_u) = {\rm log}(N_{tot}/Q(T_{rot})) - 0.4343E_u/kT_{rot},
\end{equation}
where k is Boltzmann's constant, $\nu$ is the frequency, $g_I$ and
$g_K$ are the degeneracies associated with the nuclear spin and k
quantum number respectively, $\mu^2$ is the square of the dipole matrix
element, $S$ is the statistical line strength, $\int
S_{\nu}dv$ is the observed integrated intensity of the transition,
$N_{tot}$ is the total column density, $Q(T_{rot}$) is the partition
function evaluated at the rotation temperature $T_{rot}$, and $E_u$ is
the upper state energy of the transition.  There are seven transitions
of \methanol\/ detected toward SMA1 and six toward SMA2 (including
both A and E type transitions) that can be used for this purpose. The
$\int S_{\nu}dv$ for each transition was calculated from Gaussian
fits to the line emission at the 1.3~mm continuum peaks. The rotation
temperature resulting from a weighted least squares fit to the data
is $189\pm 19$~K for SMA1 and $154\pm 14$~K for SMA2.

However, as demonstrated by \citet{Goldsmith99} and others, line
optical depth can artificially increase the derived rotation
temperature if not corrected for. In general the line optical depth
in LTE can be calculated from
\begin{equation}
\tau =
\frac{8\pi^3}{3h}\frac{N_o}{g_o}\mu^2Se^{-E_u/kT_{ex}}(e^{h\nu/kT_{rot}}-1)
\frac{1}{\Delta
  v},
\end{equation}
where $h$ is Planck's constant, $N_o/g_o$ is the column density of the
ground state divided by the ground state degeneracy, $T_{ex}$ is the
excitation temperature, and $\Delta v$ is the FWHM line width.  Then
the left hand side of Eq. 2 is modified to ${\rm
log}(C_{\tau}N_u/g_u)$ where $C_{\tau}= \tau/(1-e^{-\tau})$. Since we
do not know the ground state column density, this equation cannot be
used directly to calculate $\tau$ for each transition. Unfortunately,
in the I(N) SMA dataset we do not have a suitable isotopologue that
can be used to infer the \methanol\/ optical depth either \citep[see
for example][]{Brogan07}. Instead, we have iteratively solved for the
optical depth and \Tr\/ that produces the best $\chi^2$ fit to the
data (i.e. minimizes the scatter). Using this technique we find the
best fit occurs for a \Tr\/=$165\pm 9$~K for SMA1 and \Tr\/=$145\pm
12$~K for SMA2 (see Figure~\ref{meth_rot}a), and modest optical depths
of 1.75 for SMA1 and 0.5 for SMA2 for the observed transition that
would have the highest optical depth (\methanol\/-E
($4_{+2,2}-3_{+1,2}$)) at 218.4400~GHz). While not dramatically
different from the \Tr\/ derived without optical depth correction,
these corrected temperatures are clearly more accurate.  We do not
have enough transitions to do a more sophisticated non-LTE analysis or
to compare the results for $A$ and $E$ transitions independently
\citep[see for example][]{Sutton04}. However, we note that for the
\methanol\/-E ($5_{+1,4}-4_{+2,2}$) transition (at 216.9456 GHz) which
we have in common with the \citet{Sutton04} methanol survey of the hot
core source W3(OH) with comparable spatial resolution, the derived
opacity is similar for the two sources (0.45 for SMA1 in I(N) compared
to 0.65 for W3(OH)). Indeed the derived temperatures for SMA1 and SMA2
are also very similar to that of W3(OH) \citep[\Tr\/ = $140\pm 10$
K,][]{Sutton04}.

The optical depth implied by the brightness temperature of the
strongest methanol transition ($4_{+2,2}-3_{+1,2}$, 24 K) toward SMA1
is only 0.16 assuming that $T_{kin}$=\Tr\/=165 K and the emitting
region is the size of the LSB synthesized beam ($2\farcs39\times
1\farcs55$). An optical depth of 1.75 for this transition implies that
the emitting region is actually only $0\farcs8$ in size (1400
AU). This size is in good agreement with the fitted size
($0\farcs9\times 0\farcs7$, PA=$162\arcdeg$) of the integrated
intensity measured from the compact, but still strong
\methanol\/-A$^+$($16_{1,16}-15_{2,13}$) transition ($E_l=219.8$
\percm\/)at 227.8147~GHz. This size implies that the hot core region
encompasses both SMA1b and SMA1d (see Fig.~\ref{qbandabsolute}b).

The column densities of \methanol\/ for SMA1 and SMA2 are $(1.0\pm
0.4)\times 10^{17}$ \ct\/ and $(5.6\pm 0.6)\times 10^{16}$ \ct\/,
respectively, an order of magnitude less than \citet{Sutton04} find
for W3(OH). \citet{Sutton04} find a \methanol\/ abundance of $\sim
2\times 10^{-6}$ for W3(OH). Using single dish observations of
infrared dark clouds (IRDCs), \citet{Leurini07} find abundances an
order of magnitude smaller for the IRDC cores. Using this range of
abundance and assuming that the hot core size is 1400 AU, we estimate
that the H$_2$ column density is $(0.5 - 5)\times 10^{23}$ \ct\/ and
the H$_2$ density is $(0.2 - 2)\times 10^{7}$ \cc\/ for SMA1 and about
a factor of two smaller for SMA2.

Using the ATCA, \citet{Beuther08} observed CH$_3$CN(5$_K$--4$_K$) up
to K=4 toward SMA1 and SMA2 and derived a rotation temperature of
\Tr\/$=170\pm 50$~K for SMA2 in good agreement with our CH$+3$OH
measurement ($145\pm 12$ K). The optical depth in SMA1 was too high to
obtain a CH$_3$CN \Tr\/ for SMA1 in good agreement with our finding
that this source has a higher column density than SMA2. Also using the
ATCA, \citet{Beuther07} detected NH$_3$ (5,5), and (6,6) emission
toward both SMA1 and SMA2 confirming that $T_k$ is greater than 100~K,
though it was not possible to derive more accurate estimates from
those data.

\subsection{The Nature of the 1.3 mm Sources}

\subsubsection{Spectral Energy Distributions}

\begin{figure*}
\includegraphics[scale=0.9,angle=0]{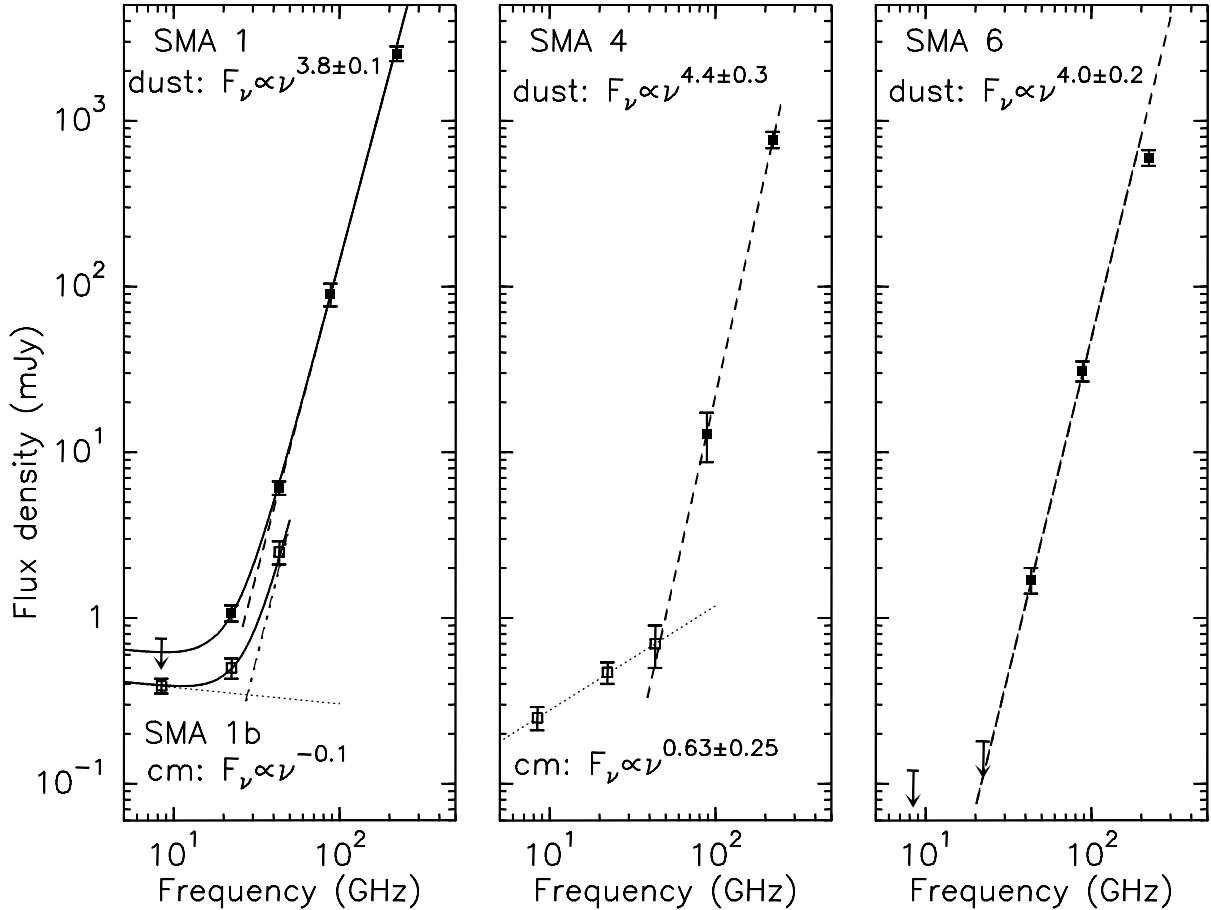}
\caption{Spectral energy distributions for sources SMA1, SMA4, and
  SMA6. For SMA1, the sum of the emission of SMA1a,b,c, and d are
  shown as solid squares, while the flux density as a function of
  frequency for SMA1b is shown as open squares for $\nu\leq 43$
  GHz. Open squares are also used for the $\nu\leq 43$ GHz emission
  for SMA4 to indicate that they may not be from the same source as
  the higher frequency data, the fits in this case assume case 3
  described in S4.5.1.  Emission models from dust are shown as dashed
  (or dot-dash in the case of SMA1b) lines, while emission models for
  free-free emission are shown as dotted lines. For SMA1 the sum of
  the dust and free-free emission models, described in detail in \S
  4.4.1, is shown as a solid line.
  \label{sed}}
\end{figure*}

Because no higher frequency arcsecond-resolution data exist on this
region, the determination of the nature of the 1.3~mm sources hinges
entirely upon accurate measurements of their spectral index toward
longer wavelengths (thus motivating our careful re-reduction of the
available data).  The comparatively poor angular resolution of the
current 1.3 and 3.4~mm data compared to the 7~mm, 1.3~cm, and 3.6~cm
data makes it challenging to disentangle the free-free versus dust
contributions to these sources. Figure~\ref{sed} shows the spectral
energy distributions (SEDs) of SMA1, SMA4, and SMA6 based on the flux
densities provided in Table~\ref{mmpos}.  The different resolutions of
the different datasets are definitely a concern. However, in many cases
the emission does not appear to be resolved even in the highest
resolution observations of SMA4, and to some degree SMA6. In that
case, as long as no significant emission is resolved out, the data
will still be comparable despite the differing resolutions. Caveats
for a few sources are described in detail below. 

For SMA1, which contains at least four components at the longer
wavelengths, we show both the integrated emission over the whole SMA1
region, and SMA1b by itself at wavelengths longer than 7~mm for
comparison.  The integrated emission from SMA1 shows evidence for both
a free-free and an optically thin dust component. The integrated dust
spectral index of $\alpha_{dust}=3.8\pm 0.1$ ($S_{\nu}\propto
\nu^{\alpha}$) was determined by a linear least-squares fit to the
three flux densities from 7~mm to 1.3~mm (for SMA1 at 7~mm this is the
sum of emission from SMA1a,b,c, and d; the free-free contribution is
an order of magnitude smaller and hence negligible in the dust fit).
On smaller scales, we also find that the 7~mm emission from SMA1b is
dominated by dust, while the 3.6~cm emission is consistent with
optically-thin free-free emission with a spectral index of
$\nu^{-0.1}$, such as that from an \HII\/ region with an electron
density of $3 \times 10^4$ \cc, temperature of 10000~K and a diameter
of $0.3\arcsec$ (510 AU).  At the intermediate wavelength of 1.3~cm,
these two emission mechanisms are comparable for SMA1b, with $\approx
30$\% of the emission arising from dust. This is a notable result
since 1.3~cm data are typically assumed to be free of dust.

The emission from SMA1a may all be due to dust since it is not
detected longward of 7~mm; the $3\sigma$ 1.3~cm upper limit of $\sim
0.18$ mJy implies that the spectral index must be $>2.4$.  In
contrast, both SMA1c and SMA1d are detected at 1.3~cm and have a
1.3~cm to 7~mm spectral index of $sim 2$. This spectral index could be
due to optically thick dust emission which seems very unlikely at
these wavelengths, optically thick free-free emission such as one
might observe around a very dense hypercompact \HII\/ region, or a
combination of optically thin free-free and dust emission. If we take
the total observed flux density of SMA1 at 1.3~cm (1.07~mJy, i.e. the
sum of SMA1a upper limit and SMA1b,c,and d detections) and subtract
the total dust model (0.50~mJy) and the SMA1b free-free emission model
(0.35~mJy), we find that 0.22~mJy remains, suggesting that the
integrated dust model cannot account for all of the SMA1c+SMA1d
emission. Thus, one or both of SMA1c and SMA1d must have a weak
free-free component, and since these objects have 1.3~cm flux
densities of 0.27 and 0.30~mJy, respectively (greater than the
residual), both must also have a dust component. If {\em all} of the
free-free residual (0.22 mJy) belonged to either SMA1c or SMA1d, the
$3\sigma$ 3.6~cm upper limit of $\sim 0.12$ mJy implies a free-free
spectral index $>0.6$, consistent with a thermal jet, stellar wind, or
hypercompact \HII\/ region. If instead the residual free-free emission
is about equally split between SMA1c and SMA1d both could harbor small
optically thin \HII\/ regions in addition to a dust component.

Understanding the nature of SMA4 is complicated by the lack of
positional agreement between the 1.3 - 3.4~mm data compared to the
longer wavelength data (3.6~cm, 1.3~cm, and 7~mm offset $\sim
0\farcs4$ to the NW; Fig.~\ref{c5c3color}).  From the spectral break
in the SMA4 SED around 7~mm shown in Fig.~\ref{sed} it is clear that
both dust and free-free emission is present in this region. The
spectral index of the SMA4 emission between 1.3 and 3.4~mm is
$\alpha_{dust}=4.4\pm 0.3$, while the spectral index between 3.6~cm
and 7~mm is $\alpha_{ff}=0.63\pm 0.25$ suggesting either a
hypercompact \HII\/ region with a turnover around 7~mm \citep[see for
  example the review by ][]{Lizano08}, or a thermal jet/stellar wind
\citep{Reynolds86}.  Overall, the data are consistent with the
following three possibilities: (1) the emission all arises from the
{\em same} source which contains both dust and a hypercompact \HII\/
region (i.e. the position offsets are just within the combined
position uncertainties); (2) Two sources are present: one is a compact
dust source with no detectable free-free emission (1.3 - 3.4~mm
emission) and the other (7~mm - 3.6~cm emission) is a hypercompact
\HII\/ region ($\gtrsim 700$ AU away); or (3) the free-free emission
arises from a one-sided thermal jet driven by the dust source of case
(2).  If we assume case (1) and fit the centimeter wavelength emission
from SMA4 after extrapolating and subtracting the dust spectrum, we
find that a free-free component with a constant density, temperature,
and size of $1.2\times 10^6$ \cc\/, 8000~K, and 53~AU best
approximates the 3.6 and 1.3 cm data points \citep[see][for the
  details of the free-free emission model]{Hunter08}. However, this
model overpredicts the 7~mm flux density by a significant margin.
More sophisticated free-free emission models, such as those with a
density gradient and/or a shell geometry, may provide a better match
to the 7~mm data \citep[see for example][]{Avalos06}.  However, the
present data are insufficient in angular resolution and number of
spectral measurements to meaningfully constrain a more detailed
model. For case (2), it is even more difficult to fit the 7~mm data
with a hypercompact \HII\/ region model because the spectrum from
3.6~cm to 7~mm is so linear, and yet not steep enough to still be
optically thick (i.e. spectral index of 2).  Thus, case (3) seems to
best fit the current data (see Fig.~\ref{sed}, with the 3.6~cm to 7~mm
data having a typical jet or stellar wind spectral index
\citep[$\alpha_{ff}=0.6$][]{Reynolds86}, and no dust component. We
note that in the jet interpretation for the cm-$\lambda$ emission
(case 3) the $\sim 0\farcs4$ offset of the cm-$\lambda$ to the west is
in good agreement with the fact that a westward outflow appears to
emanate from SMA4.

Although there is an offset $\sim 0\farcs3$ to the north between the
7~mm and 1.3~mm emission toward SMA6, the fact that the shape of the
continuum contours at these two wavelengths is similar, and most of
the apparent offset is along the long axis of the 1.3~mm beam makes it
quite plausible that they are in fact coincident (also note that the
offset is not in the same direction as that of SMA4). Indeed, the
emission from SMA6 can be explained completely by optically-thin dust
emission (at the current level of sensitivity).  The dust spectral
index from a linear fit to the three flux densities from 7~mm to
1.3~mm is $\alpha=4.0\pm 0.2$. If we assume all the 1.3~mm emission
arises from a region with the size measured from the higher resolution
7~mm image, ($0\farcs9\times 0\farcs7$), the brightness temperature is
$T_b=25$ K setting a lower limit on the dust temperature.  There is a
hint of a break towards a flatter spectral index at 1.3~mm for SMA6,
as would be expected if the dust emission were becoming optically
thick, but higher frequency data are required to confirm this trend.

SMA2 and SMA3 are the only strong, compact millimeter sources not to
be detected at wavelengths longward of 3.4~mm. Between 1.3 and 3.4~mm
the spectral index of SMA2 is $\alpha=4.0\pm 0.3$, and that of SMA3 is
$\alpha=4.3\pm 0.3$, consistent with optically thin dust
emission. Extrapolating these spectral indices to 7~mm, we would have
expected to detect SMA2 and SMA3 at the 1.4 and 0.9 mJy level,
respectively (amounting to $7\sigma$ and $4.5\sigma$) {\em if the
  emission were unresolved at 7~mm}. The sizes derived for SMA2 and
SMA3 at 1.3~mm are $2\farcs1\times 1\farcs3$ and $2\farcs5\times
1\farcs9$. If these are the true sizes of the emitting regions, the
peak 7~mm intensity would be diminished by a factor of $\sim 10$ over
that predicted here, plausibly explaining the non-detections. The
emission denoted VLA3 is only detected at 7~mm, and thus its nature is
very uncertain. The non-detection of VLA3 at 1.3~cm implies that its
spectral index is $> 2$, and we estimate a 1.3~mm upper limit of $\sim
150$ mJy suggesting that its spectral index must also be less than
$\sim 3$. These limits suggest that VLA3 is predominately due to dust
emission, but contribution from a hypercompact \HII\/ region cannot be
excluded.

In summary, of the nine compact sources detected in I(N), all appear
to have a dust component.  Four of these sources (SMA1b, SMA1c, SMA1d,
and SMA4) also show evidence for free-free emission at our current
level of sensitivity.  Three sources (SMA1b, SMA1d, and SMA2) and
possibly a fourth (SMA4) show modest hot core molecular line emission,
also suggesting the presence of a central powering source. One source
with neither free-free nor hot core emission does appear to be
powering an outflow (SMA6). From this circumstantial evidence it
appears that at least six of the compact sources may harbor a central
source. However, as described by \citet{Zinnecker2007}, in deeply
embedded regions it is extraordinarily difficult to determine whether
a massive protostar or zero age main sequence star (ZAMS) has formed,
as this designation is dependent upon detecting the source in the
near-IR (a wavelength regime in which I(N) is still obscured). In
principle, the detection of free-free emission is often taken as
indicative of at least an early B-type star. However, even the
presence of free-free emission is ambiguous since the accretion
luminosity itself can be sufficient to ionize the gas around a
protostar. Thus, while it is clear that a number of massive to
intermediate stars are forming in I(N), the exact number and
evolutionary state is uncertain.

A few comments on the previously published flux measurements for these
sources are in order. When the more accurate absolute flux calibration
presented here for the original \citet{Carral02} 3.6~cm data is taken
into account, the agreement for SMA4 is quite good. However, it is
unclear why the detection of SMA1b was not reported by these authors,
since it is 1.6 times brighter than SMA4. We obtain 7~mm flux
densities between 2.2 (SMA6) to 8 (SMA1a) times {\em smaller} than
those reported by \citet{Rodriguez07}, though our 1.3~cm flux
densities are comparable. This discrepancy comes in spite of the fact
that the flux densities derived for the phase calibrators at both 7
and 1.3~cm are within $2\%$ for our reduction compared to that
reported in \citet{Rodriguez07}.  As described in \S2.2.3, continuum
emission from the strong ultracompact \HII\/ region in source I causes
significant contamination of the I(N) region if not included in the
cleaning; however our detailed tests suggest that at most it could
erroneously increase the integrated fluxes by a factor of $\sim 2$,
and thus cannot account for the large discrepancy. In addition, the
integrated flux densities reported in \citet{Rodriguez07} do not
appear consistent with the contour levels plotted in their Figures 2
and 3, indeed the peak values are reasonably consistent with those
reported here so the problem may have occurred in their integrated
Gaussian fits.  Finally, our re-imaging of the 3.4~mm data from
\citet{Beuther08} has increased the peak flux densities by a factor of
about 1.5 due to our self-calibration and correction for the primary
beam attenuation.

\subsubsection{Mass estimates from the dust emission}

In Table~\ref{mass} we give estimates for the gas mass $M_{gas}$,
molecular hydrogen column density $N_{H_2}$, and molecular hydrogen
density $n_{H_2}$ for all seven 1.3~mm dust sources. It should be
noted that if a protostar or zero age main sequence star has already
formed, the millimeter data is only sensitive to the {\em
  circum-(proto)stellar} material. The gas masses were calculated from
\begin{equation}
M_{gas} = \frac{3.24\times 10^{-3}S_{\nu}({\rm Jy})D^2({\rm
    kpc})R C_{\tau_{dust}}}{J(\nu,T_{dust})\nu^3({\rm THz})\kappa_{\nu}}  ~{\rm M_{\odot}},
\end{equation}
where $S_{\nu}$ is the integrated flux density from Table~\ref{mmpos},
$D$ is the distance (assumed to be 1.7 kpc), $R$ is the gas to dust
ratio (assumed to be 100), $J(\nu,T)=1/({\rm exp}(h\nu/T_{dust}k)-1)$, and
$C_{dust}$ is the correction factor for the dust opacity
$C_{\tau_{dust}}= \tau_{dust}/(1-e^{-\tau_{dust}})$.  We have used
$\kappa_{\rm 1.3~mm}= 1$ cm$^{2}$~g$^{-1}$ which is the average of the dust
opacities derived by \citet{Ossenkopf94} for the case of thin ice
mantles and densities between $10^{6}$ to $10^{8}$ \cc\/ at
1.3~mm. We have estimated a range of dust temperatures based on the
observed spectral line properties of each source: we use the derived
\Tr\/ values calculated in \S4.4 for SMA1 and SMA2, and a lower
temperature range for SMA3 to SMA7 based on the non-detection of hot
core spectral lines towards these sources. The column densities were
estimated from
\begin{equation}
N_{H_2} = \frac{2.4\times 10^{30}S_{\nu}({\rm Jy})R
  C_{\tau_{dust}}}{J(\nu,T_{dust})\nu^3({\rm
    THz})\theta_1(\arcsec)\theta_2(\arcsec)\kappa_{\nu}}
~{\rm cm}^{-2},
\end{equation}
where $\theta_1$ and $\theta_2$ are the fitted sizes from
Table~\ref{mmpos}. The $n_{H_2}$ was calculated by dividing $N_{H_2}$
by the geometric mean of the fitted size, assuming a distance of 1.7
kpc.  The estimated dust masses are similar to those reported in
\citet{Hunter06} when the differences in absolute flux calibration and
assumed dust temperatures are taken into account. As with all such
estimates, the $M_{gas}$, $N_{H_2}$, $n_{H_2}$ reported in
Table~\ref{mass} are uncertain to at least a factor of 2. The total
mass estimated for the {\em compact} emission of I(N) is between 89
and 232 \msun\/. These dust emission based masses do not include the
mass of any protostars or ZAMSs that may already have formed within
the compact cores. It is also notable that \citet{Matthews08} estimate
that the mass of the whole I(N) core is $\sim 2780$ \msun\/, so there
is still a considerable reservoir of material that has not yet
collapsed.

The values for SMA1 are particularly uncertain because the current
1.3~mm resolution is insufficient to distinguish between the four dust
sources detected at 7~mm (SMA1a, SMA1b, SMA1c, and SMA1d). This
unresolved structure makes the apparent size of this source quite
large -- significantly affecting all derived quantities. For example,
if we assume that the majority of dust emission comes from a region
half the size (in both directions) of the current fit ($2\farcs8\times
1\farcs9$), the mass increases only a little (due to increase in
$C_{\tau_{dust}}$), but $N_{H_2}$ and $n_{H_2}$ both increase by about
an order of magnitude.  An additional source of uncertainty is the
$T_{dust}$. As described in \S4.3, the hot core line emission appears
to arise from a $0\farcs8$ region encompassing only SMA1b and
SMA1d. Thus, the $T_{dust}$ range derived from the \methanol\/
rotation diagram may not be appropriate for SMA1a or SMA1c. Using
$\kappa_{\rm 7~mm}= 0.05$ cm$^{2}$~g$^{-1}$ (obtained by scaling
$\kappa_{\rm 1.3~mm}= 1$ by (7~mm/1.3~mm)$^{-\beta}$ with
$\beta=\alpha_{dust}-2$ where $\alpha_{dust}=3.8$ from
Fig.~\ref{sed}), and $T_{dust}=155-175$~K we find that the sum of the
masses derived for SMA1a,b,c, and d at 7~mm is 20 - 18 \msun\/ in good
agreement with the 1.3~mm estimate for SMA1 (16 - 14 \msun\/). Indeed,
the agreement is nearly perfect if we assume that the dust emission at
1.3~mm arises from an area $5\times$ smaller than the current size
estimate (i.e. Table~\ref{mmpos}), which is certainly
plausible. Higher resolution 1.3~mm observations are required to
better constrain the temperatures of SMA1a and SMA1c, as well as the
sizes of all four SMA1 sources.

\subsection{Evidence for Accretion and Infall Around SMA1?}

As described in \S3.4, a SE-NW velocity gradient of $\sim 5$ \kms\/ is
observed toward SMA1 in some (but not all) of the species with compact
emission \citep[a similar gradient was inferred from a double peaked
NH$_3$(6,6) profile by][]{Beuther07}. Since the gradient is barely
spatially resolved with the current SMA resolution, the nature of this
gradient is unclear. Its direction is more or less perpendicular to
the large scale NE-SW outflow suggestive of a possible disk
interpretation, but it also has the same orientation as a line joining
the SMA1b and SMA1d sources, and thus could simply represent
unresolved emission from these two sources if they have different
systemic velocities \citep[as was found for CephA-East in a similar
situation by][]{Brogan07}. We note that this velocity gradient was not
observed in the CH$_3$CN(5$_K$--4$_K$) transitions observed by
\citet{Beuther08} with the ATCA.  Higher spatial resolution data will
be needed to further explore this velocity gradient.

As described in \S3.5.2, there is a trend of increasingly redshifted
absorption with increasing line excitation temperature toward SMA1,
using CN ($E_l$=3.8 \percm), \form ($E_l$=7 \percm), and \form\/
($E_l$=40 \percm) as tracers (see Fig.~\ref{spectra}a). In \S4.3 we
find that the hot core of SMA1 has a temperature of $165\pm 9$~K,
demonstrating that this source is definitely centrally heated,
suggesting higher excitation lines will be found closer to the
center. We suggest that the observed redshift/temperature trend is
tracing infall that is accelerating with depth into the core. It would
be very interesting to follow-up this tentative result with higher
angular and spectral resolution. In particular if the continuum
emission were resolved, one could be assured that all
emission/absorption originates from only the front side of the source,
removing any ambiguities \citep[see for example][]{Chandler05}.

\subsection{The Location of the Protocluster}

Finally, it is interesting to note that the massive protocluster
appears to be forming near the northwestern edge of the parent cloud
rather than toward the center.  This offset is apparent in both the
JCMT 450 \mum\/ continuum image (Fig.~\ref{methanolmasers}) and in the
ammonia (1,1) and (2,2) images \citep{Beuther05}.  It is notable that
the archival SCUBA 450 \mum\/ images recently produced by
\citet{difrancesco08} show very good positional agreement (to within
$3\arcsec$) with the earlier images produced by \citet{Sandell00} (and
shown in Fig.~\ref{methanolmasers}).  The ridge of dust emission
between the protocluster and the masers associated with the SM2
continuum source coincides with a dark area in the {\em Spitzer}
3-color image (Fig.~\ref{irac}), i.e. an infrared dark cloud
\citep[IRDC; see for example][]{Egan98}.  Clearly this region is
optically-thick in the mid-infrared, and contains a significant amount
of potential star-forming material.  Future observations with broader
bandwidth and greater sensitivity with the EVLA and ALMA may reveal
whether this area is truly quiescent or contains embedded,
pre-protostellar cores.

\section{CONCLUSIONS}

Our multiwavelength radio through near-infrared investigation of the
massive protocluster NGC6334I(N) provides new morphological and
quantitative detail on the star formation activity of the individual
massive cores.  We show that I(N) is undergoing copious massive
star formation with many of the dominant sources detected only at
millimeter wavelengths.  The principal source SMA1 is resolved into
four components of dust and/or free-free emission (SMA1a, b, c, and
d).  Delineated by centimeter continuum and water masers, SMA1b and
SMA1d form a close central pair ($\gtrsim 800$ AU) which powers a hot
core (\Tr\/=$165\pm 9$~K) of molecular line emission and likely drives
the dominant large-scale bipolar outflow which exhibits evidence of
precession. We find a spatially unresolved $\sim 5$ \kms\/ velocity
gradient across SMA1 (encompassing all 4 subarcsecond components). The
origin of this velocity gradient (rotation, or cluster kinematics)
requires higher resolution millimeter wavelength follow-up. We also
detect possible infall that is accelerating with depth into the SMA1
core; higher angular resolution is also required to confirm this
finding. The mass of gas around SMA1 as traced by {\em compact}
emission is estimated to be $\sim 15$ \msun\/, but if this core is
still actively accreting, this may grow significantly larger in the
future.

A weaker hot core (\Tr\/=$145\pm 12$~K) is found toward SMA2 which
also exhibits redshifted CN absorption along with with highly
blueshifted water masers, suggesting simultaneous infall and outflow
in this source.  SMA4 consists of a dust core with a probable
centimeter jet, water maser emission, and blueshifted SiO emanating to
the west suggesting an outflow which terminates in a reflection nebula
traced by the 24 \mum\/ source.  Mostly lacking in molecular lines,
SMA6 exhibits a centimeter-millimeter SED consistent with pure dust
emission and drives an outflow traced by SiO, CO and 44~GHz methanol
masers.  Overall, the narrow velocity range and spatial arrangement of
the 44~GHz methanol masers with respect to the protocluster members
are consistent with their origin in low-velocity gas impacted by
outflows from these objects.

 The presence of multiple outflows in the region suggests that many
 protostars are forming simultaneously.  At the same time, the variety
 of phenomena associated with the individual cluster members suggests
 a spread in either age or total luminosity.  Also of interest is the
 spread in systemic velocities observed between the cluster members,
 this spread is at least 2.5 \kms\/ and may provide an important clue
 about the dynamics of cluster formation.  A more quantitative
 assessment of these key properties will require broader wavelength
 coverage to better constrain the SEDs, and more sensitive continuum
 and line observations enabled by the EVLA and ALMA in the near
 future.  At the same time, a deeper exploration of the mostly
 featureless, infrared-dark region between the protocluster and SM2
 may prove fruitful in the study of pre-protostellar cores.

\acknowledgments

We thank G\"oran Sandell for providing us with the calibrated JCMT 450
\mum\/ image.  This research has made use of the NASA's Astrophysics
Data System Bibliographic Services, the Cologne Database for Molecular
Spectroscopy, Splatalogue, and the JPL line catalog.



\begin{thebibliography}{75}
\expandafter\ifx\csname natexlab\endcsname\relax\def\natexlab#1{#1}\fi

\bibitem[{{Anglada} {et~al.}(2007){Anglada}, {L{\'o}pez}, {Estalella},
  {Masegosa}, {Riera}, \& {Raga}}]{Anglada07}
{Anglada}, G., {L{\'o}pez}, R., {Estalella}, R., {Masegosa}, J., {Riera}, A.,
  \& {Raga}, A.~C. 2007, \aj, 133, 2799

\bibitem[{{Araya} {et~al.}(2008){Araya}, {Hofner}, {Kurtz}, {Olmi}, \&
  {Linz}}]{Araya08}
{Araya}, E., {Hofner}, P., {Kurtz}, S., {Olmi}, L., \& {Linz}, H. 2008, \apj,
  675, 420

\bibitem[{{Avalos} {et~al.}(2006){Avalos}, {Lizano}, {Rodr{\'{\i}}guez},
  {Franco-Hern{\'a}ndez}, \& {Moran}}]{Avalos06}
{Avalos}, M., {Lizano}, S., {Rodr{\'{\i}}guez}, L.~F., {Franco-Hern{\'a}ndez},
  R., \& {Moran}, J.~M. 2006, \apj, 641, 406

\bibitem[{{Beuther} {et~al.}(2005){Beuther}, {Thorwirth}, {Zhang}, {Hunter},
  {Megeath}, {Walsh}, \& {Menten}}]{Beuther05}
{Beuther}, H., {Thorwirth}, S., {Zhang}, Q., {Hunter}, T.~R., {Megeath}, S.~T.,
  {Walsh}, A.~J., \& {Menten}, K.~M. 2005, \apj, 627, 834

\bibitem[{{Beuther} {et~al.}(2007){Beuther}, {Walsh}, {Thorwirth}, {Zhang},
  {Hunter}, {Megeath}, \& {Menten}}]{Beuther07}
{Beuther}, H., {Walsh}, A.~J., {Thorwirth}, S., {Zhang}, Q., {Hunter}, T.~R.,
  {Megeath}, S.~T., \& {Menten}, K.~M. 2007, \aap, 466, 989

\bibitem[{{Beuther} {et~al.}(2008){Beuther}, {Walsh}, {Thorwirth}, {Zhang},
  {Hunter}, {Megeath}, \& {Menten}}]{Beuther08}
---. 2008, ArXiv e-prints, 801

\bibitem[{{Blundell} {et~al.}(1998)}]{Blundell98}
{Blundell}, R. {et~al.} 1998, in Institute of Electrical and Electronics
  Engineers, Inc. Conference, 246--247

\bibitem[{{Brogan} {et~al.}(2007){Brogan}, {Chandler}, {Hunter}, {Shirley}, \&
  {Sarma}}]{Brogan07}
{Brogan}, C.~L., {Chandler}, C.~J., {Hunter}, T.~R., {Shirley}, Y.~L., \&
  {Sarma}, A.~P. 2007, \apjl, 660, L133

\bibitem[{{Carey} {et~al.}(2009){Carey}, {Noriega-Crespo}, {Mizuno}, {Shenoy},
  {Paladini}, {Kraemer}, {Price}, {Flagey}, {Ryan}, {Ingalls}, {Kuchar},
  {Pinheiro Gon{\c c}alves}, {Indebetouw}, {Billot}, {Marleau}, {Padgett},
  {Rebull}, {Bressert}, {Ali}, {Molinari}, {Martin}, {Berriman}, {Boulanger},
  {Latter}, {Miville-Deschenes}, {Shipman}, \& {Testi}}]{Carey09}
{Carey}, S.~J., {Noriega-Crespo}, A., {Mizuno}, D.~R., {Shenoy}, S.,
  {Paladini}, R., {Kraemer}, K.~E., {Price}, S.~D., {Flagey}, N., {Ryan}, E.,
  {Ingalls}, J.~G., {Kuchar}, T.~A., {Pinheiro Gon{\c c}alves}, D.,
  {Indebetouw}, R., {Billot}, N., {Marleau}, F.~R., {Padgett}, D.~L., {Rebull},
  L.~M., {Bressert}, E., {Ali}, B., {Molinari}, S., {Martin}, P.~G.,
  {Berriman}, G.~B., {Boulanger}, F., {Latter}, W.~B., {Miville-Deschenes},
  M.~A., {Shipman}, R., \& {Testi}, L. 2009, \pasp, 121, 76

\bibitem[{{Carral} {et~al.}(2002){Carral}, {Kurtz}, {Rodr{\'{\i}}guez},
  {Menten}, {Cant{\'o}}, \& {Arceo}}]{Carral02}
{Carral}, P., {Kurtz}, S.~E., {Rodr{\'{\i}}guez}, L.~F., {Menten}, K.,
  {Cant{\'o}}, J., \& {Arceo}, R. 2002, \aj, 123, 2574

\bibitem[Caswell(2009)]{Caswell2009} 
Caswell, J.~L.\ 2009, arXiv:0907.5255 

\bibitem[{{Chandler} {et~al.}(2005){Chandler}, {Brogan}, {Shirley}, \&
  {Loinard}}]{Chandler05}
{Chandler}, C.~J., {Brogan}, C.~L., {Shirley}, Y.~L., \& {Loinard}, L. 2005,
  \apj, 632, 371

\bibitem[{{Cheung} {et~al.}(1978){Cheung}, {Frogel}, {Hauser}, \&
  {Gezari}}]{Cheung78}
{Cheung}, L., {Frogel}, J.~A., {Hauser}, M.~G., \& {Gezari}, D.~Y. 1978, \apjl,
  226, L149

\bibitem[{{Cragg} {et~al.}(2002){Cragg}, {Sobolev}, \& {Godfrey}}]{Cragg02}
{Cragg}, D.~M., {Sobolev}, A.~M., \& {Godfrey}, P.~D. 2002, \mnras, 331, 521

\bibitem[{{Cragg} {et~al.}(2005){Cragg}, {Sobolev}, \& {Godfrey}}]{Cragg05}
---. 2005, \mnras, 360, 533

\bibitem[{{Cunningham} {et~al.}(2009){Cunningham}, {Moeckel}, \&
  {Bally}}]{Cunningham09}
{Cunningham}, N.~J., {Moeckel}, N., \& {Bally}, J. 2009, \apj, 692, 943

\bibitem[{{Cyganowski} {et~al.}(2007){Cyganowski}, {Brogan}, \&
  {Hunter}}]{Cyganowski07}
{Cyganowski}, C.~J., {Brogan}, C.~L., \& {Hunter}, T.~R. 2007, \aj,
134, 346

\bibitem[{{Cyganowski} {et~al.}(2009){Cyganowski}, {Brogan}, {Hunter},
    \& {Churchwell}}]{Cyganowski09} {Cyganowski}, C.~J., {Brogan},
  C.~L., {Hunter}, T.~R., \& Churchwell, E.\ 2007, \apj, 702, 1615

\bibitem[{{Cyganowski} {et~al.}(2008){Cyganowski}, {Whitney}, {Holden},
  {Braden}, {Brogan}, {Churchwell}, {Indebetouw}, {Watson}, {Babler},
  {Benjamin}, {Gomez}, {Meade}, {Povich}, {Robitaille}, \&
  {Watson}}]{Cyganowski08}
{Cyganowski}, C.~J., {Whitney}, B.~A., {Holden}, E., {Braden}, E., {Brogan},
  C.~L., {Churchwell}, E., {Indebetouw}, R., {Watson}, D.~F., {Babler}, B.~L.,
  {Benjamin}, R., {Gomez}, M., {Meade}, M.~R., {Povich}, M.~S., {Robitaille},
  T.~P., \& {Watson}, C. 2008, \aj, 136, 2391


\bibitem[{{De Buizer}(2007)}]{deBuizer07}
{De Buizer}, J. 2007, in IAU Symposium, Vol. 242, IAU Symposium, ed. J.~M.
  {Chapman} \& W.~A. {Baan}, 102--109

\bibitem[{{De Buizer} {et~al.}(2005){De Buizer}, {Osorio}, \&
  {Calvet}}]{deBuizer05}
{De Buizer}, J.~M., {Osorio}, M., \& {Calvet}, N. 2005, \apj, 635, 452

\bibitem[{{Di Francesco} {et~al.}(2008){Di Francesco}, {Johnstone}, {Kirk},
  {MacKenzie}, \& {Ledwosinska}}]{difrancesco08}
{Di Francesco}, J., {Johnstone}, D., {Kirk}, H., {MacKenzie}, T., \&
  {Ledwosinska}, E. 2008, \apjs, 175, 277

\bibitem[{{Egan} {et~al.}(1998){Egan}, {Shipman}, {Price}, {Carey}, {Clark}, \&
  {Cohen}}]{Egan98}
{Egan}, M.~P., {Shipman}, R.~F., {Price}, S.~D., {Carey}, S.~J., {Clark},
  F.~O., \& {Cohen}, M. 1998, \apjl, 494, L199+

\bibitem[{{Ellison} {et~al.}(1989){Ellison}, {Schaffer}, {Schaal}, {Miller}, \&
  {Vail}}]{Ellison89}
{Ellison}, B.~N., {Schaffer}, P.~L., {Schaal}, W., {Miller}, R.~E., \& {Vail},
  D. 1989, International Journal of Infrared and Millimeter Waves, 10, 937

\bibitem[{{Fazio} {et~al.}(2004)}]{Fazio04}
{Fazio}, G.~G. {et~al.} 2004, \apjs, 154, 10

\bibitem[{{Feigelson} {et~al.}(2009){Feigelson}, {Martin}, {McNeill}, {Broos},
  \& {Garmire}}]{Feigelson09}
{Feigelson}, E.~D., {Martin}, A.~L., {McNeill}, C.~J., {Broos}, P.~S., \&
  {Garmire}, G.~P. 2009, ArXiv e-prints

\bibitem[{{Gezari}(1982)}]{Gezari82}
{Gezari}, D.~Y. 1982, \apjl, 259, L29

\bibitem[{{Goldsmith} \& {Langer}(1999)}]{Goldsmith99}
{Goldsmith}, P.~F. \& {Langer}, W.~D. 1999, \apj, 517, 209

\bibitem[{{Haschick} {et~al.}(1990){Haschick}, {Menten}, \&
  {Baan}}]{Haschick90}
{Haschick}, A.~D., {Menten}, K.~M., \& {Baan}, W.~A. 1990, \apj, 354, 556

\bibitem[{{Hofner} \& {Churchwell}(1996)}]{Hofner96}
{Hofner}, P. \& {Churchwell}, E. 1996, \aaps, 120, 283

\bibitem[{{Hunter} {et~al.}(2008){Hunter}, {Brogan}, {Indebetouw}, \&
  {Cyganowski}}]{Hunter08}
{Hunter}, T.~R., {Brogan}, C.~L., {Indebetouw}, R., \& {Cyganowski}, C.~J.
  2008, \apj, 680, 1271

\bibitem[{{Hunter} {et~al.}(2006){Hunter}, {Brogan}, {Megeath}, {Menten},
  {Beuther}, \& {Thorwirth}}]{Hunter06}
{Hunter}, T.~R., {Brogan}, C.~L., {Megeath}, S.~T., {Menten}, K.~M., {Beuther},
  H., \& {Thorwirth}, S. 2006, \apj, 649, 888

\bibitem[{{Hunter} {et~al.}(1999){Hunter}, {Testi}, {Zhang}, \&
  {Sridharan}}]{Hunter99}
{Hunter}, T.~R., {Testi}, L., {Zhang}, Q., \& {Sridharan}, T.~K. 1999, \aj,
  118, 477

\bibitem[{{Kogan} \& {Slysh}(1998)}]{Kogan98}
{Kogan}, L. \& {Slysh}, V. 1998, \apj, 497, 800

\bibitem[{{Kuiper} {et~al.}(1995){Kuiper}, {Peters}, {Forster}, {Gardner}, \&
  {Whiteoak}}]{Kuiper95}
{Kuiper}, T.~B.~H., {Peters}, W.~L., {Forster}, J.~R., {Gardner}, F.~F., \&
  {Whiteoak}, J.~B. 1995, \apj, 446, 692

\bibitem[{{Kurtz} {et~al.}(2004){Kurtz}, {Hofner}, \& {{\'A}lvarez}}]{Kurtz04}
{Kurtz}, S., {Hofner}, P., \& {{\'A}lvarez}, C.~V. 2004, \apjs, 155, 149

\bibitem[{{Le Bourlot} {et~al.}(1999){Le Bourlot}, {Pineau des For{\^e}ts}, \&
  {Flower}}]{lebourlot}
{Le Bourlot}, J., {Pineau des For{\^e}ts}, G., \& {Flower}, D.~R. 1999, \mnras,
  305, 802

\bibitem[{{Le Bourlot} {et~al.}(2002){Le Bourlot}, {Pineau des For{\^e}ts},
  {Flower}, \& {Cabrit}}]{lebourlot2}
{Le Bourlot}, J., {Pineau des For{\^e}ts}, G., {Flower}, D.~R., \& {Cabrit}, S.
  2002, \mnras, 332, 985

\bibitem[{{Leurini} {et~al.}(2007){Leurini}, {Schilke}, {Wyrowski}, \&
  {Menten}}]{Leurini07}
{Leurini}, S., {Schilke}, P., {Wyrowski}, F., \& {Menten}, K.~M. 2007, \aap,
  466, 215

\bibitem[{{Lizano}(2008)}]{Lizano08}
{Lizano}, S. 2008, in Astronomical Society of the Pacific Conference Series,
  Vol. 387, Massive Star Formation: Observations Confront Theory, ed.
  H.~{Beuther}, H.~{Linz}, \& T.~{Henning}, 232--+

\bibitem[{{Matthews} {et~al.}(2008){Matthews}, {McCutcheon}, {Kirk}, {White},
  \& {Cohen}}]{Matthews08}
{Matthews}, H.~E., {McCutcheon}, W.~H., {Kirk}, H., {White}, G.~J., \& {Cohen},
  M. 2008, \aj, 136, 2083

\bibitem[{{Megeath} \& {Tieftrunk}(1999)}]{Megeath99}
{Megeath}, S.~T. \& {Tieftrunk}, A.~R. 1999, \apjl, 526, L113

\bibitem[{{Minier} {et~al.}(2003){Minier}, {Ellingsen}, {Norris}, \&
  {Booth}}]{Minier03}
{Minier}, V., {Ellingsen}, S.~P., {Norris}, R.~P., \& {Booth}, R.~S. 2003,
  \aap, 403, 1095

\bibitem[{{Moran} \& {Rodriguez}(1980)}]{Moran80}
{Moran}, J.~M. \& {Rodriguez}, L.~F. 1980, \apjl, 236, L159

\bibitem[{{M{\"u}ller} {et~al.}(2001){M{\"u}ller}, {Thorwirth}, {Roth}, \&
  {Winnewisser}}]{Muller01}
{M{\"u}ller}, H.~S.~P., {Thorwirth}, S., {Roth}, D.~A., \& {Winnewisser}, G.
  2001, \aap, 370, L49

\bibitem[{{Neckel}(1978)}]{Neckel78}
{Neckel}, T. 1978, \aap, 69, 51

\bibitem[{{Neufeld} \& {Yuan}(2008)}]{neufeld}
{Neufeld}, D.~A. \& {Yuan}, Y. 2008, \apj, 678, 974

\bibitem[{{Ossenkopf} \& {Henning}(1994)}]{Ossenkopf94}
{Ossenkopf}, V. \& {Henning}, T. 1994, \aap, 291, 943

\bibitem[{{Persi} \& {Tapia}(2008)}]{Persi08}
{Persi}, P. \& {Tapia}, M. 2008, {Star Formation in NGC 6334}, ed.
  B.~{Reipurth}, 456--+

\bibitem[{{Pickett} {et~al.}(1998){Pickett}, {Poynter}, {Cohen}, {Delitsky},
  {Pearson}, \& {Muller}}]{Pickett98}
{Pickett}, H.~M., {Poynter}, I.~R.~L., {Cohen}, E.~A., {Delitsky}, M.~L.,
  {Pearson}, J.~C., \& {Muller}, H.~S.~P. 1998, Journal of Quantitative
  Spectroscopy and Radiative Transfer, 60, 883

\bibitem[{{Pirogov} {et~al.}(2003){Pirogov}, {Zinchenko}, {Caselli},
  {Johansson}, \& {Myers}}]{Pirogov03}
{Pirogov}, L., {Zinchenko}, I., {Caselli}, P., {Johansson}, L.~E.~B., \&
  {Myers}, P.~C. 2003, \aap, 405, 639

\bibitem[{{Plambeck} \& {Menten}(1990)}]{Plambeck90}
{Plambeck}, R.~L. \& {Menten}, K.~M. 1990, \apj, 364, 555

\bibitem[{{Pratap} {et~al.}(2008){Pratap}, {Shute}, {Keane}, {Battersby}, \&
  {Sterling}}]{Pratap08}
{Pratap}, P., {Shute}, P.~A., {Keane}, T.~C., {Battersby}, C., \& {Sterling},
  S. 2008, \aj, 135, 1718

\bibitem[{{Reach} {et~al.}(2005){Reach}, {Megeath}, {Cohen}, {Hora}, {Carey},
  {Surace}, {Willner}, {Barmby}, {Wilson}, {Glaccum}, {Lowrance}, {Marengo}, \&
  {Fazio}}]{reach}
{Reach}, W.~T., {Megeath}, S.~T., {Cohen}, M., {Hora}, J., {Carey}, S.,
  {Surace}, J., {Willner}, S.~P., {Barmby}, P., {Wilson}, G., {Glaccum}, W.,
  {Lowrance}, P., {Marengo}, M., \& {Fazio}, G.~G. 2005, \pasp, 117, 978

\bibitem[{{Reynolds}(1986)}]{Reynolds86}
{Reynolds}, S.~P. 1986, \apj, 304, 713
\bibitem[{{Robitaille} {et~al.}(2007){Robitaille}, {Whitney}, {Indebetouw}, \&
  {Wood}}]{robitaille}
{Robitaille}, T.~P., {Whitney}, B.~A., {Indebetouw}, R., \& {Wood}, K. 2007,
  \apjs, 169, 328

\bibitem[{{Rodr{\'{\i}}guez} {et~al.}(2007){Rodr{\'{\i}}guez}, {Zapata}, \&
  {Ho}}]{Rodriguez07}
{Rodr{\'{\i}}guez}, L.~F., {Zapata}, L.~A., \& {Ho}, P.~T.~P. 2007, \apjl, 654,
  L143

\bibitem[{{Sandell}(2000)}]{Sandell00}
{Sandell}, G. 2000, \aap, 358, 242

\bibitem[{{Sandell} {et~al.}(2005){Sandell}, {Goss}, \& {Wright}}]{Sandell05}
{Sandell}, G., {Goss}, W.~M., \& {Wright}, M. 2005, \apj, 621, 839

\bibitem[{{Scoville} \& {Kwan}(1976)}]{Scoville76}
{Scoville}, N.~Z. \& {Kwan}, J. 1976, \apj, 206, 718

\bibitem[{{Skrutskie} {et~al.}(2006){Skrutskie}, {Cutri}, {Stiening},
  {Weinberg}, {Schneider}, {Carpenter}, {Beichman}, {Capps}, {Chester},
  {Elias}, {Huchra}, {Liebert}, {Lonsdale}, {Monet}, {Price}, {Seitzer},
  {Jarrett}, {Kirkpatrick}, {Gizis}, {Howard}, {Evans}, {Fowler}, {Fullmer},
  {Hurt}, {Light}, {Kopan}, {Marsh}, {McCallon}, {Tam}, {Van Dyk}, \&
  {Wheelock}}]{skrutskie}
{Skrutskie}, M.~F., {Cutri}, R.~M., {Stiening}, R., {Weinberg}, M.~D.,
  {Schneider}, S., {Carpenter}, J.~M., {Beichman}, C., {Capps}, R., {Chester},
  T., {Elias}, J., {Huchra}, J., {Liebert}, J., {Lonsdale}, C., {Monet}, D.~G.,
  {Price}, S., {Seitzer}, P., {Jarrett}, T., {Kirkpatrick}, J.~D., {Gizis},
  J.~E., {Howard}, E., {Evans}, T., {Fowler}, J., {Fullmer}, L., {Hurt}, R.,
  {Light}, R., {Kopan}, E.~L., {Marsh}, K.~A., {McCallon}, H.~L., {Tam}, R.,
  {Van Dyk}, S., \& {Wheelock}, S. 2006, \aj, 131, 1163

\bibitem[{{Slysh} {et~al.}(1994){Slysh}, {Kalenskii}, {Valtts}, \&
  {Otrupcek}}]{Slysh94}
{Slysh}, V.~I., {Kalenskii}, S.~V., {Valtts}, I.~E., \& {Otrupcek}, R. 1994,
  \mnras, 268, 464

\bibitem[{{Smith} {et~al.}(2006){Smith}, {Hora}, {Marengo}, \&
  {Pipher}}]{smith06}
{Smith}, H.~A., {Hora}, J.~L., {Marengo}, M., \& {Pipher}, J.~L. 2006, \apj,
  645, 1264

\bibitem[{{Smith}(1995)}]{smith95}
{Smith}, M.~D. 1995, \aap, 296, 789

\bibitem[{{Smith} \& {Rosen}(2005)}]{smith05}
{Smith}, M.~D. \& {Rosen}, A. 2005, \mnras, 357, 1370
\bibitem[{{Straw} \& {Hyland}(1989)}]{Straw89a}
{Straw}, S.~M. \& {Hyland}, A.~R. 1989, \apj, 340, 318

\bibitem[{{Sutton} {et~al.}(2004){Sutton}, {Sobolev}, {Salii}, {Malyshev},
  {Ostrovskii}, \& {Zinchenko}}]{Sutton04}
{Sutton}, E.~C., {Sobolev}, A.~M., {Salii}, S.~V., {Malyshev}, A.~V.,
  {Ostrovskii}, A.~B., \& {Zinchenko}, I.~I. 2004, \apj, 609, 231

\bibitem[{{Thorwirth} {et~al.}(2003){Thorwirth}, {Winnewisser}, {Megeath}, \&
  {Tieftrunk}}]{Thorwirth03}
{Thorwirth}, S., {Winnewisser}, G., {Megeath}, S.~T., \& {Tieftrunk}, A.~R.
  2003, in ASP Conf. Ser. 287: Galactic Star Formation Across the Stellar Mass
  Spectrum, 257--260

\bibitem[{{Tofani} {et~al.}(1995){Tofani}, {Felli}, {Taylor}, \&
  {Hunter}}]{Tofani95}
{Tofani}, G., {Felli}, M., {Taylor}, G.~B., \& {Hunter}, T.~R. 1995, \aaps,
  112, 299

\bibitem[{{Val'Tts} \& {Larionov}(2007)}]{Valtts07}
{Val'Tts}, I.~E. \& {Larionov}, G.~M. 2007, Astronomy Reports, 51, 519

\bibitem[{{van der Tak} {et~al.}(2007){van der Tak}, {Black}, {Sch{\"o}ier},
  {Jansen}, \& {van Dishoeck}}]{radex}
{van der Tak}, F.~F.~S., {Black}, J.~H., {Sch{\"o}ier}, F.~L., {Jansen}, D.~J.,
  \& {van Dishoeck}, E.~F. 2007, \aap, 468, 627

\bibitem[{{van Dishoeck} \& {Blake}(1998)}]{vandishoeck98}
{van Dishoeck}, E.~F. \& {Blake}, G.~A. 1998, \araa, 36, 317

\bibitem[{{Walsh} {et~al.}(1998){Walsh}, {Burton}, {Hyland}, \&
  {Robinson}}]{Walsh98}
{Walsh}, A.~J., {Burton}, M.~G., {Hyland}, A.~R., \& {Robinson}, G. 1998,
  \mnras, 301, 640

\bibitem[{{Wolniewicz} {et~al.}(1998){Wolniewicz}, {Simbotin}, \&
  {Dalgarno}}]{wolk}
{Wolniewicz}, L., {Simbotin}, I., \& {Dalgarno}, A. 1998, \apjs, 115, 293

\bibitem[{{Wrathmall} {et~al.}(2007){Wrathmall}, {Gusdorf}, \&
  {Flower}}]{wrathmall}
{Wrathmall}, S.~A., {Gusdorf}, A., \& {Flower}, D.~R. 2007, \mnras, 382, 133

\bibitem[{{Ybarra} \& {Lada}(2009)}]{ybarra}
{Ybarra}, J.~E. \& {Lada}, E.~A. 2009, \apjl, 695, L120

\bibitem[{{Zhang} {et~al.}(1999){Zhang}, {Hunter}, {Sridharan}, \&
  {Cesaroni}}]{Zhang99}
{Zhang}, Q., {Hunter}, T.~R., {Sridharan}, T.~K., \& {Cesaroni}, R. 1999,
  \apjl, 527, L117

\bibitem[Zinnecker \& Yorke(2007)]{Zinnecker2007} 
Zinnecker, H., \& Yorke, H.~W.\ 2007, \araa, 45, 481 



\end{thebibliography}

\begin{deluxetable}{lllccc}
\tablecaption{Observed Properties of Compact Continuum Sources \label{mmpos}}  
\tablecolumns{6}
\tablewidth{0pc}
\tabletypesize{\footnotesize}
\tablehead{
  & \multicolumn{2}{c}{Centroid position$^a$} & \colhead{Peak$^b$}
      & \colhead{Integrated$^b$} &  \colhead{Estimated} \\ 
\colhead{Source} & \colhead{RA (J2000)} & \colhead{Dec (J2000)}
      & \colhead{Intensity} & \colhead{Flux Density} & \colhead{Size} \\ 
  &  \colhead{$\alpha$ ($^{\rm h}~~^{\rm m}~~^{\rm s}$)}   
      & \colhead{$\delta$ ($^{\circ}~~{'}~~{''}$)} & \colhead{(\mjb\/)} & \colhead{(mJy)} 
      & \colhead{($\arcsec\times \arcsec$ [P.A. $\arcdeg$])}
}
\startdata
\multicolumn{6}{c}{1.3 Millimeter$^c$}\\ 
SMA1 & 17 20 55.188 & -35 45 03.92 & 826 (6) & 2550(50) & $ 2.8\times 1.9~ [23]$ \\ 
SMA2 & 17 20 54.890 & -35 45 06.71 & 324 (6) &  980 (25) & $ 2.1\times 1.3~ [137]$ \\ 
SMA3 & 17 20 54.990 & -35 45 07.29 & 258 (6) & 1060 (31) & $ 2.5\times 1.9~[9]$ \\  
SMA4 & 17 20 54.667 & -35 45 08.38 & 141 (6) &  770 (40) & $ 3.0\times 1.5~ [41] $ \\ 
SMA5 & 17 20 55.054 & -35 45 01.49 & 140 (6) &  440 (100)& $ \leq 2.0 $  \\ 
SMA6 & 17 20 54.563 & -35 45 17.71 & 282 (9) &  600 (29) & $ 2.2\times 0.9~ [38]$ \\  
SMA7 & 17 20 54.931 & -35 44 57.79 &  60 (6) &  360 (40) & $ 3.2\times 2.3~ [47]$ \\  
\hline
\multicolumn{6}{c}{3.4 Millimeter$^{e,f}$}\\  
SMA1 & 17 20 55.194 & -35 45 03.79 &  55 (5) & 90 (10) & $\leq 3.0$ \\
SMA2 & 17 20 54.86  & -35 45 06.5  &  24 (5) & \nodata & $<1.6$ \\
SMA3 & 17 20 54.99  & -35 45 07.0  &  20 (5) & \nodata & $<1.6$ \\ 
SMA4$^d$ & 17 20 54.69  & -35 45 07.5  &  13 (3) & \nodata & $<1.6$ \\ 
SMA5$^d$ & 17 20 55.04  & -35 45 01.3  &  15 (3) & \nodata & $<1.6$ \\ 
SMA6 & 17 20 54.588 & -35 45 17.86  & 31 (3) & \nodata & $<1.6$ \\ 
\hline
\multicolumn{6}{c}{7 Millimeter}\\  
SMA1a & 17 20 55.150  & -35 45 05.28  & 0.9 (0.2)  & 1.7 (0.3)     & $\leq 1$   \\
SMA1b & 17 20 55.194  & -35 45 03.50  & 1.3 (0.2)  & 2.5 (0.4)     & $0.6\times 0.4~[130]$\\
SMA1c & 17 20 55.275  & -35 45 02.46  & 0.9 (0.2)  & \nodata  & $<0.4$ \\
SMA1d$^d$ & 17 20 55.229  & -35 45 03.68  & 1.0 (0.2)  & \nodata  & $<0.4$ \\
VLA3$^d$  & 17 20 54.999  & -35 45 06.48  & 1.0 (0.2)  & 1.8 (0.3)     &  $\leq 1$  \\
SMA4$^d$  & 17 20 54.624  & -35 45 08.24   & 0.7 (0.2)  & \nodata  & $<0.4$  \\
SMA6  & 17 20 54.585  & -35 45 16.97  & 1.3 (0.2)  & 1.7 (0.3)     & $0.9\times 0.7~ [15]$ \\
\hline
\multicolumn{6}{c}{1.3 Centimeter}\\ 
SMA1a &  \nodata  &  \nodata          & $<0.18$ (0.06)  & \nodata    & \nodata \\
SMA1b & 17 20 55.191  & -35 45 03.53  & 0.50 (0.07)  & \nodata  & $<0.3$  \\
SMA1c & 17 20 55.270  & -35 45 02.45  & 0.27 (0.06)  & \nodata  & $<0.4$  \\
SMA1d$^d$ & 17 20 55.229  & -35 45 03.68  & 0.30 (0.06)  & \nodata  & $<0.3$  \\
SMA4  & 17 20 54.623  & -35 45 08.25  & 0.47 (0.07)  & \nodata  & $<0.3$  \\
SMA6 &  \nodata  &  \nodata          & $<0.18$ (0.06)  & \nodata    & \nodata \\ 
\hline
\multicolumn{6}{c}{3.6 Centimeter$^e$}\\ 
SMA1a &  \nodata  &  \nodata          & $<0.12$ (0.04)  & \nodata    & \nodata \\
SMA1b$^d$ & 17 20 55.202  & -35 45 03.74  & 0.39 (0.04) & \nodata & $<0.3$  \\
SMA1c &  \nodata  &  \nodata          & $<0.12$ (0.04)  & \nodata & \nodata \\
SMA1d &  \nodata  &  \nodata          & $<0.12$ (0.04)  & \nodata    & \nodata \\
SMA4  & 17 20 54.623  & -35 45 08.25  & 0.25 (0.04) & \nodata  & $<0.3$  \\
SMA6 &  \nodata  &  \nodata          & $<0.12$ (0.04)  & \nodata    & \nodata 
\enddata
\tablenotetext{a}{Relative uncertainties are indicated by the
  number of significant figures shown. Absolute uncertainties are at
  least an order of magnitude larger.}
\tablenotetext{b}{Because the data have different resolutions, the
  peak intensity values should only be compared for cases where the
  source is unresolved (indicated by \nodata symbols in Integrated
  Flux Density column).Estimated statistical uncertainties indicated in parenthesis.}
\tablenotetext{c}{Fits based on uniformly weighted image. SMA1, SMA4,
  and SMA6 appear to also have a more compact core so the fitted
  peaks are underestimated.}
\tablenotetext{d}{New detection based on archival data.}
\tablenotetext{e}{Primary beam correction makes a significant
  difference for these images; the original papers do not appear to
  have performed this correction based on reported flux densities.}
\tablenotetext{f}{Extra uncertainty for the peak and integrated fluxes
  for SMA1 and SMA5, as well as SMA2 and SMA3 were added because these 
  source pairs were barely resolved from each other.}
\end{deluxetable}


\begin{deluxetable}{ccccccc} 
\tablecolumns{7}
\tablewidth{0pc} 
\tablecaption{22.235~GHz Water Maser Groups in \ngcin\ \label{watertable}}  
\tabletypesize{\footnotesize} 
\tablehead{
       & \multicolumn{2}{c}{Centroid position$^a$} & \colhead{Associated} & \colhead{Peak} & \colhead{Peak} & \colhead{Velocity} \\ 
\#  & \colhead{RA (J2000}) & \colhead{Dec (J2000)} & \colhead{mm source}  & \colhead{Intensity} & \colhead{Velocity} & \colhead{Range} \\ 
        &  \colhead{$\alpha$ ($^{\rm h}~~^{\rm m}~~^{\rm s}$)}   & \colhead{$\delta$ ($^{\circ}~~{'}~~{''}$}) & \colhead{(offset)} 
       & \colhead{(\jyb\/)} & \colhead{(\kms\/)}  & \colhead{(\kms\/)}
}
 1 &  17 20 56.058 & -35 45 32.59 & SM2\tablenotemark{b} (4.50\arcsec) & $38.3$  & $-6.0$ & $+5.2$ to $-16.2$\\
 2 &  17 20 55.650 & -35 45 32.48 & SM2\tablenotemark{b} (1.78\arcsec) & $9.2$ & $-9.0$ & $+7.2$ to $-15.6$ \\
 3 &  17 20 54.600 & -35 45 17.27 & SMA6 (0.67\arcsec) & $8.8$ & $-2.2$  & $+10.4$ to $-12.6$ \\ 
 4 &  17 20 54.152 & -35 45 13.66 & -         & $29.0$  & $-2.7$  & $-0.7$ to $-4.4$\\
 5 &  17 20 54.618 & -35 45 08.65 & SMA4 (2.11\arcsec) & $17.7$  & $+0.2$  & $+12.4$ to $-12.9$\tablenotemark{c}\\
 6 &  17 20 54.870 & -35 45 06.32 & SMA2 (0.55\arcsec) & $2.7$ & $-13.3$ & $-4.7$ to $-15.6$\\
 7 &  17 20 54.819 & -35 45 06.18 & SMA2 (1.11\arcsec) & $5.3$ & $-0.7$  & $+0.2$ to $-1.7$ \\
 8 &  17 20 55.192 & -35 45 03.77 & SMA1b (0.24\arcsec) & $78.1$ & $-6.3$ & $+13.1$ to $-26.8$\tablenotemark{d}\\
 9 &  17 20 55.228 & -35 45 03.66 & SMA1d (0.02\arcsec) & $75.7$ & $+3.9$ & $+5.2$ to $+2.2$\\
10 &  17 20 55.147 & -35 45 03.78 & SMA1 (0.53\arcsec) & $3.4$ & $-8.6$ & $-7.7$ to $-9.6$\\ 
11 &  17 20 55.215 & -35 45 02.02 & SMA1 (1.96\arcsec) & $5.5$ & $-4.4$ & $-3.7$ to $-6.3$\\
\enddata
\tablenotetext{a}{Intensity weighted positions for masers listed under each group name in Table~\ref{bigwatertable}.}
\tablenotetext{b}{See \citet{Sandell00}.}
\tablenotetext{c}{Emission extends beyond the positive velocity extent of the bandpass.}
\tablenotetext{d}{Emission extends beyond the negative velocity extent of the bandpass.}
\end{deluxetable}

\begin{deluxetable}{cccc}
\tablecolumns{4}
\tablewidth{0pc} 
\tablecaption{22.235~GHz Water Masers Detected in \ngcin\  \label{bigwatertable}} 
\tabletypesize{\footnotesize} 
\tablehead{ 
\colhead{Velocity} & \multicolumn{2}{c}{Centroid position} & \colhead{Peak} \\ 
\colhead{(\kms\/)} & \colhead{R.A. (J2000)} & \colhead{Dec (J2000)}              & \colhead{Intensity} \\  
         &  \colhead{$\alpha$ ($^{\rm h}~~^{\rm m}~~^{\rm s}$)}   & \colhead{$\delta$ ($^{\circ}~~{'}~~{''}$)} 
         & \colhead{(\jyb\/)}
}
\startdata
\multicolumn{4}{c}{Group C1}\\ 
+5.2 & 17 20 56.0583 & -35 45 32.4934 & 0.384\\ 
+4.9 & 17 20 56.0599 & -35 45 32.5678 & 0.548\\ 
+4.5 & 17 20 56.0585 & -35 45 32.5920 & 0.586\\ 
+4.2 & 17 20 56.0578 & -35 45 32.5805 & 0.893\\ 
+3.9 & 17 20 56.0581 & -35 45 32.5797 & 1.373\\ 
+3.5 & 17 20 56.0586 & -35 45 32.5776 & 1.764\\ 
+3.2 & 17 20 56.0579 & -35 45 32.6134 & 1.675\\ 
+2.9 & 17 20 56.0580 & -35 45 32.5594 & 1.534\\ 
+2.5 & 17 20 56.0581 & -35 45 32.5946 & 1.185\\ 
+2.2 & 17 20 56.0584 & -35 45 32.5566 & 0.934\\ 
+1.9 & 17 20 56.0584 & -35 45 32.5619 & 0.639\\ 
+1.6 & 17 20 56.0614 & -35 45 32.4557 & 0.508\\ 
+1.2 & 17 20 56.0617 & -35 45 32.4066 & 0.381\\ 
+0.9 & 17 20 56.0555 & -35 45 32.5404 & 0.535\\ 
+0.6 & 17 20 56.0579 & -35 45 32.5826 & 0.504\\ 
+0.2 & 17 20 56.0590 & -35 45 32.5242 & 0.497\\ 
-0.1 & 17 20 56.0607 & -35 45 32.4663 & 0.422\\ 
-0.4 & 17 20 56.0584 & -35 45 32.5610 & 0.400\\ 
-0.7 & 17 20 56.0597 & -35 45 32.5347 & 0.558\\ 
-1.1 & 17 20 56.0566 & -35 45 32.6202 & 0.652\\ 
-1.4 & 17 20 56.0588 & -35 45 32.5413 & 0.578\\ 
-1.7 & 17 20 56.0581 & -35 45 32.6271 & 0.749\\ 
-2.1 & 17 20 56.0587 & -35 45 32.5695 & 1.507\\ 
-2.4 & 17 20 56.0582 & -35 45 32.5774 & 3.357\\ 
-2.7 & 17 20 56.0580 & -35 45 32.5891 & 6.453\\ 
-3.0 & 17 20 56.0581 & -35 45 32.5907 & 11.608\\ 
-3.4 & 17 20 56.0579 & -35 45 32.5977 & 16.950\\ 
-3.7 & 17 20 56.0579 & -35 45 32.5953 & 13.072\\ 
-4.0 & 17 20 56.0579 & -35 45 32.5929 & 10.177\\ 
-4.4 & 17 20 56.0579 & -35 45 32.5925 & 13.265\\ 
-4.7 & 17 20 56.0577 & -35 45 32.5900 & 16.513\\ 
-5.0 & 17 20 56.0578 & -35 45 32.5893 & 20.479\\ 
-5.4 & 17 20 56.0578 & -35 45 32.5910 & 28.570\\ 
-5.7 & 17 20 56.0579 & -35 45 32.5903 & 38.006\\ 
-6.0 & 17 20 56.0578 & -35 45 32.5935 & 38.297\\ 
-6.3 & 17 20 56.0578 & -35 45 32.5900 & 28.243\\ 
-6.7 & 17 20 56.0577 & -35 45 32.5910 & 19.718\\ 
-7.0 & 17 20 56.0578 & -35 45 32.5912 & 22.008\\ 
-7.3 & 17 20 56.0578 & -35 45 32.5952 & 31.096\\ 
-7.7 & 17 20 56.0577 & -35 45 32.5945 & 35.008\\ 
-8.0 & 17 20 56.0579 & -35 45 32.5918 & 28.781\\ 
-8.3 & 17 20 56.0579 & -35 45 32.5943 & 18.181\\ 
-8.6 & 17 20 56.0580 & -35 45 32.5908 & 10.865\\ 
-9.0 & 17 20 56.0582 & -35 45 32.5926 & 7.395\\ 
-9.3 & 17 20 56.0582 & -35 45 32.5850 & 4.048\\ 
-9.6 & 17 20 56.0582 & -35 45 32.5758 & 1.460\\ 
-10.0 & 17 20 56.0600 & -35 45 32.6461 & 0.521\\ 
-15.6 & 17 20 56.0558 & -35 45 32.4942 & 0.536\\ 
-15.9 & 17 20 56.0566 & -35 45 32.5314 & 0.608\\ 
-16.2 & 17 20 56.0558 & -35 45 32.5633 & 0.515\\ 
-21.5 & 17 20 56.0600 & -35 45 32.3669 & 0.375\\ 
+5.2 & 17 20 56.0585 & -35 45 32.4918 & 0.432\\ 
+4.9 & 17 20 56.0597 & -35 45 32.5667 & 0.562\\ 
+4.5 & 17 20 56.0587 & -35 45 32.5882 & 0.681\\ 
+4.2 & 17 20 56.0579 & -35 45 32.5851 & 1.028\\ 
+3.9 & 17 20 56.0582 & -35 45 32.5809 & 1.426\\ 
+3.5 & 17 20 56.0586 & -35 45 32.5788 & 1.805\\ 
+3.2 & 17 20 56.0579 & -35 45 32.6131 & 1.706\\ 
+2.9 & 17 20 56.0580 & -35 45 32.5580 & 1.577\\ 
+2.5 & 17 20 56.0581 & -35 45 32.5956 & 1.228\\ 
+2.2 & 17 20 56.0584 & -35 45 32.5573 & 0.963\\ 
+1.9 & 17 20 56.0583 & -35 45 32.5609 & 0.652\\ 
+1.6 & 17 20 56.0613 & -35 45 32.4586 & 0.509\\ 
+0.9 & 17 20 56.0556 & -35 45 32.5405 & 0.566\\ 
+0.6 & 17 20 56.0580 & -35 45 32.5863 & 0.547\\ 
+0.2 & 17 20 56.0595 & -35 45 32.5222 & 0.460\\ 
-0.1 & 17 20 56.0603 & -35 45 32.4752 & 0.459\\ 
-0.4 & 17 20 56.0583 & -35 45 32.5707 & 0.450\\ 
-0.7 & 17 20 56.0600 & -35 45 32.5378 & 0.579\\ 
-1.1 & 17 20 56.0565 & -35 45 32.6301 & 0.636\\ 
-1.4 & 17 20 56.0588 & -35 45 32.5398 & 0.573\\ 
-1.7 & 17 20 56.0581 & -35 45 32.6371 & 0.745\\ 
-2.1 & 17 20 56.0587 & -35 45 32.5701 & 1.514\\ 
-2.4 & 17 20 56.0582 & -35 45 32.5774 & 3.370\\ 
-2.7 & 17 20 56.0580 & -35 45 32.5890 & 6.416\\ 
-3.0 & 17 20 56.0581 & -35 45 32.5906 & 11.555\\ 
-3.4 & 17 20 56.0579 & -35 45 32.5976 & 16.873\\ 
-3.7 & 17 20 56.0579 & -35 45 32.5952 & 13.029\\ 
-4.0 & 17 20 56.0579 & -35 45 32.5929 & 10.164\\ 
-4.4 & 17 20 56.0579 & -35 45 32.5924 & 13.184\\ 
-4.7 & 17 20 56.0577 & -35 45 32.5899 & 16.471\\ 
-5.0 & 17 20 56.0578 & -35 45 32.5893 & 20.463\\ 
-5.4 & 17 20 56.0578 & -35 45 32.5909 & 28.475\\ 
-5.7 & 17 20 56.0579 & -35 45 32.5903 & 37.967\\ 
-6.0 & 17 20 56.0578 & -35 45 32.5935 & 38.098\\ 
-6.3 & 17 20 56.0578 & -35 45 32.5899 & 28.084\\ 
-6.7 & 17 20 56.0577 & -35 45 32.5910 & 19.678\\ 
-7.0 & 17 20 56.0578 & -35 45 32.5912 & 22.025\\ 
-7.3 & 17 20 56.0578 & -35 45 32.5952 & 31.077\\ 
-7.7 & 17 20 56.0577 & -35 45 32.5945 & 34.860\\ 
-8.0 & 17 20 56.0579 & -35 45 32.5918 & 28.709\\ 
-8.3 & 17 20 56.0579 & -35 45 32.5942 & 18.050\\ 
-8.6 & 17 20 56.0580 & -35 45 32.5905 & 10.632\\ 
-9.0 & 17 20 56.0582 & -35 45 32.5924 & 7.253\\ 
-9.3 & 17 20 56.0582 & -35 45 32.5845 & 3.913\\ 
-9.6 & 17 20 56.0582 & -35 45 32.5756 & 1.456\\ 
-10.0 & 17 20 56.0601 & -35 45 32.6682 & 0.427\\ 
-15.6 & 17 20 56.0561 & -35 45 32.4888 & 0.532\\ 
-15.9 & 17 20 56.0568 & -35 45 32.5344 & 0.641\\ 
-16.2 & 17 20 56.0559 & -35 45 32.5610 & 0.525\\ 
\multicolumn{4}{c}{Group C2}\\
+7.2 & 17 20 55.6493 & -35 45 32.3938 & 0.492\\ 
+6.8 & 17 20 55.6472 & -35 45 32.4863 & 0.457\\ 
+3.5 & 17 20 55.6427 & -35 45 32.5464 & 1.094\\ 
+3.2 & 17 20 55.6433 & -35 45 32.5315 & 4.374\\ 
+2.9 & 17 20 55.6431 & -35 45 32.5330 & 6.952\\ 
+2.5 & 17 20 55.6431 & -35 45 32.5417 & 4.822\\ 
+2.2 & 17 20 55.6431 & -35 45 32.5434 & 2.022\\ 
+1.9 & 17 20 55.6432 & -35 45 32.5221 & 0.675\\ 
-4.4 & 17 20 55.6459 & -35 45 32.4432 & 0.545\\ 
-4.7 & 17 20 55.6441 & -35 45 32.5102 & 0.683\\ 
-8.3 & 17 20 55.6546 & -35 45 32.4731 & 3.048\\ 
-8.6 & 17 20 55.6550 & -35 45 32.4425 & 7.213\\ 
-9.0 & 17 20 55.6551 & -35 45 32.4383 & 9.175\\ 
-9.3 & 17 20 55.6551 & -35 45 32.4329 & 7.673\\ 
-9.6 & 17 20 55.6546 & -35 45 32.4394 & 5.907\\ 
-10.0 & 17 20 55.6545 & -35 45 32.4422 & 4.096\\ 
-10.3 & 17 20 55.6523 & -35 45 32.4669 & 1.958\\ 
-10.6 & 17 20 55.6482 & -35 45 32.5098 & 0.999\\ 
-11.0 & 17 20 55.6459 & -35 45 32.5250 & 0.701\\ 
-11.3 & 17 20 55.6437 & -35 45 32.5237 & 0.784\\ 
-11.6 & 17 20 55.6458 & -35 45 32.4930 & 0.754\\ 
-11.9 & 17 20 55.6450 & -35 45 32.5199 & 0.638\\ 
-14.9 & 17 20 55.6440 & -35 45 32.5229 & 0.790\\ 
-15.2 & 17 20 55.6451 & -35 45 32.4784 & 0.854\\ 
-15.6 & 17 20 55.6436 & -35 45 32.5214 & 0.476\\ 
\multicolumn{4}{c}{Group C3}\\
+10.4 & 17 20 54.6005 & -35 45 17.2406 & 0.399\\ 
+10.1 & 17 20 54.5980 & -35 45 17.2058 & 3.031\\ 
+9.8 & 17 20 54.5976 & -35 45 17.2063 & 5.251\\ 
+9.5 & 17 20 54.5983 & -35 45 17.1966 & 1.719\\ 
-1.7 & 17 20 54.6008 & -35 45 17.4507 & 0.855\\ 
-2.1 & 17 20 54.6024 & -35 45 17.4397 & 3.840\\ 
-2.4 & 17 20 54.6021 & -35 45 17.4564 & 8.732\\ 
-2.7 & 17 20 54.6025 & -35 45 17.4485 & 4.544\\ 
-3.0 & 17 20 54.6003 & -35 45 17.4098 & 0.653\\ 
-9.3 & 17 20 54.5985 & -35 45 16.9099 & 0.865\\ 
-9.6 & 17 20 54.5987 & -35 45 16.8621 & 0.834\\ 
-10.0 & 17 20 54.5982 & -35 45 16.9016 & 0.765\\ 
-10.3 & 17 20 54.5986 & -35 45 16.9311 & 0.881\\ 
-10.6 & 17 20 54.5976 & -35 45 16.8737 & 0.774\\ 
-11.0 & 17 20 54.5980 & -35 45 16.8495 & 0.493\\ 
+10.4 & 17 20 54.6008 & -35 45 17.2373 & 0.450\\ 
+10.1 & 17 20 54.5980 & -35 45 17.2058 & 3.101\\ 
+9.8 & 17 20 54.5976 & -35 45 17.2064 & 5.347\\ 
+9.5 & 17 20 54.5982 & -35 45 17.1969 & 1.771\\ 
-1.7 & 17 20 54.6007 & -35 45 17.4528 & 0.841\\ 
-2.1 & 17 20 54.6024 & -35 45 17.4397 & 3.854\\ 
-2.4 & 17 20 54.6021 & -35 45 17.4564 & 8.787\\ 
-2.7 & 17 20 54.6025 & -35 45 17.4485 & 4.480\\ 
-3.0 & 17 20 54.5982 & -35 45 17.4122 & 0.540\\ 
-9.3 & 17 20 54.5985 & -35 45 16.9085 & 0.885\\ 
-9.6 & 17 20 54.5987 & -35 45 16.8620 & 0.833\\ 
-10.0 & 17 20 54.5982 & -35 45 16.9021 & 0.757\\ 
-10.3 & 17 20 54.5986 & -35 45 16.9327 & 0.906\\ 
-10.6 & 17 20 54.5976 & -35 45 16.8730 & 0.751\\ 
-11.0 & 17 20 54.5978 & -35 45 16.8534 & 0.499\\ 
-12.6 & 17 20 54.5957 & -35 45 16.9127 & 0.486\\ 
\multicolumn{4}{c}{Group 4}\\
-0.7 & 17 20 54.1519 & -35 45 13.6356 & 0.443\\ 
-1.1 & 17 20 54.1526 & -35 45 13.6565 & 1.487\\ 
-1.4 & 17 20 54.1522 & -35 45 13.6502 & 2.202\\ 
-1.7 & 17 20 54.1519 & -35 45 13.6713 & 2.687\\ 
-2.1 & 17 20 54.1523 & -35 45 13.6613 & 6.767\\ 
-2.4 & 17 20 54.1520 & -35 45 13.6625 & 18.031\\ 
-2.7 & 17 20 54.1521 & -35 45 13.6614 & 29.025\\ 
-3.0 & 17 20 54.1522 & -35 45 13.6595 & 28.199\\ 
-3.4 & 17 20 54.1522 & -35 45 13.6622 & 17.975\\ 
-3.7 & 17 20 54.1524 & -35 45 13.6671 & 9.020\\ 
-4.0 & 17 20 54.1525 & -35 45 13.6757 & 3.567\\ 
-4.4 & 17 20 54.1534 & -35 45 13.6654 & 1.005\\ 
\multicolumn{4}{c}{Group C5}\\
+12.4 & 17 20 54.6167 & -35 45 08.6918 & 0.338\\ 
+12.1 & 17 20 54.6147 & -35 45 08.7214 & 1.340\\ 
+11.8 & 17 20 54.6156 & -35 45 08.7082 & 4.033\\ 
+11.4 & 17 20 54.6157 & -35 45 08.7105 & 8.967\\ 
+11.1 & 17 20 54.6158 & -35 45 08.7065 & 13.269\\ 
+10.8 & 17 20 54.6157 & -35 45 08.7107 & 10.566\\ 
+10.4 & 17 20 54.6157 & -35 45 08.7023 & 4.252\\ 
+10.1 & 17 20 54.6153 & -35 45 08.7110 & 1.658\\ 
+9.8 & 17 20 54.6151 & -35 45 08.6967 & 1.425\\ 
+9.5 & 17 20 54.6156 & -35 45 08.7093 & 3.465\\ 
+9.1 & 17 20 54.6157 & -35 45 08.7130 & 4.332\\ 
+8.8 & 17 20 54.6155 & -35 45 08.6989 & 1.695\\ 
+8.5 & 17 20 54.6147 & -35 45 08.7116 & 0.531\\ 
+8.1 & 17 20 54.6157 & -35 45 08.7595 & 0.395\\ 
+7.8 & 17 20 54.6175 & -35 45 08.6477 & 0.487\\ 
+7.5 & 17 20 54.6146 & -35 45 08.6738 & 0.351\\ 
+3.2 & 17 20 54.6190 & -35 45 08.6127 & 1.603\\ 
+2.9 & 17 20 54.6185 & -35 45 08.6493 & 4.546\\ 
+2.5 & 17 20 54.6185 & -35 45 08.6554 & 7.838\\ 
+2.2 & 17 20 54.6184 & -35 45 08.6586 & 12.628\\ 
+1.9 & 17 20 54.6184 & -35 45 08.6540 & 16.060\\ 
+1.6 & 17 20 54.6188 & -35 45 08.6520 & 10.792\\ 
+1.2 & 17 20 54.6192 & -35 45 08.6552 & 8.127\\ 
+0.9 & 17 20 54.6193 & -35 45 08.6480 & 12.375\\ 
+0.6 & 17 20 54.6187 & -35 45 08.6574 & 15.588\\ 
+0.2 & 17 20 54.6185 & -35 45 08.6585 & 17.735\\ 
-0.1 & 17 20 54.6185 & -35 45 08.6545 & 15.340\\ 
-0.4 & 17 20 54.6184 & -35 45 08.6601 & 8.512\\ 
-0.7 & 17 20 54.6184 & -35 45 08.6571 & 4.007\\ 
-1.1 & 17 20 54.6183 & -35 45 08.6541 & 2.817\\ 
-1.4 & 17 20 54.6185 & -35 45 08.6568 & 1.573\\ 
-1.7 & 17 20 54.6191 & -35 45 08.5802 & 0.444\\ 
-4.4 & 17 20 54.6272 & -35 45 08.3544 & 0.444\\ 
-4.7 & 17 20 54.6230 & -35 45 08.5517 & 0.481\\ 
-5.4 & 17 20 54.6159 & -35 45 08.7635 & 0.472\\ 
-5.7 & 17 20 54.6198 & -35 45 08.6503 & 0.718\\ 
-6.0 & 17 20 54.6201 & -35 45 08.6755 & 0.629\\ 
-9.0 & 17 20 54.6214 & -35 45 08.4841 & 1.125\\ 
-9.3 & 17 20 54.6204 & -35 45 08.5182 & 2.909\\ 
-9.6 & 17 20 54.6202 & -35 45 08.5223 & 4.261\\ 
-10.0 & 17 20 54.6206 & -35 45 08.5225 & 2.696\\ 
-10.3 & 17 20 54.6210 & -35 45 08.5548 & 0.889\\ 
-10.6 & 17 20 54.6190 & -35 45 08.5728 & 0.897\\ 
-11.0 & 17 20 54.6193 & -35 45 08.5426 & 3.239\\ 
-11.3 & 17 20 54.6195 & -35 45 08.5426 & 7.580\\ 
-11.6 & 17 20 54.6196 & -35 45 08.5391 & 7.289\\ 
-11.9 & 17 20 54.6195 & -35 45 08.5532 & 3.543\\ 
-12.3 & 17 20 54.6193 & -35 45 08.5529 & 2.313\\ 
-12.6 & 17 20 54.6194 & -35 45 08.5359 & 1.537\\ 
-12.9 & 17 20 54.6197 & -35 45 08.5228 & 0.657\\ 
\multicolumn{4}{c}{Group C6}\\
-4.7 & 17 20 54.8707 & -35 45 06.2958 & 0.422\\ 
-5.0 & 17 20 54.8711 & -35 45 06.3301 & 1.437\\ 
-5.4 & 17 20 54.8708 & -35 45 06.3243 & 2.054\\ 
-5.7 & 17 20 54.8701 & -35 45 06.3103 & 1.200\\ 
-6.0 & 17 20 54.8710 & -35 45 06.3411 & 1.289\\ 
-6.3 & 17 20 54.8711 & -35 45 06.3146 & 1.332\\ 
-6.7 & 17 20 54.8719 & -35 45 06.3697 & 0.873\\ 
-9.3 & 17 20 54.8706 & -35 45 06.4191 & 0.588\\ 
-9.6 & 17 20 54.8699 & -35 45 06.3908 & 0.826\\ 
-10.0 & 17 20 54.8704 & -35 45 06.3552 & 0.745\\ 
-10.3 & 17 20 54.8721 & -35 45 06.3812 & 0.921\\ 
-10.6 & 17 20 54.8717 & -35 45 06.3420 & 0.905\\ 
-11.0 & 17 20 54.8709 & -35 45 06.3195 & 0.870\\ 
-11.3 & 17 20 54.8728 & -35 45 06.2861 & 0.994\\ 
-11.6 & 17 20 54.8712 & -35 45 06.3566 & 0.638\\ 
-12.6 & 17 20 54.8695 & -35 45 06.3256 & 0.612\\ 
-12.9 & 17 20 54.8702 & -35 45 06.3279 & 1.445\\ 
-13.3 & 17 20 54.8706 & -35 45 06.3372 & 2.679\\ 
-13.6 & 17 20 54.8700 & -35 45 06.3435 & 2.571\\ 
-13.9 & 17 20 54.8711 & -35 45 06.3392 & 1.449\\ 
-14.2 & 17 20 54.8713 & -35 45 06.3218 & 1.047\\ 
-14.6 & 17 20 54.8713 & -35 45 06.3419 & 0.527\\ 
-14.9 & 17 20 54.8726 & -35 45 06.3400 & 0.620\\ 
-15.2 & 17 20 54.8715 & -35 45 06.3181 & 0.747\\ 
-15.6 & 17 20 54.8691 & -35 45 06.3873 & 0.504\\ 
\multicolumn{4}{c}{Group C7}\\
+0.2 & 17 20 54.8194 & -35 45 06.1839 & 1.050\\ 
-0.1 & 17 20 54.8185 & -35 45 06.1763 & 3.217\\ 
-0.4 & 17 20 54.8186 & -35 45 06.1898 & 5.107\\ 
-0.7 & 17 20 54.8186 & -35 45 06.1865 & 5.317\\ 
-1.1 & 17 20 54.8185 & -35 45 06.1769 & 3.830\\ 
-1.4 & 17 20 54.8194 & -35 45 06.1844 & 1.766\\ 
-1.7 & 17 20 54.8197 & -35 45 06.2083 & 0.639\\ 
\multicolumn{4}{c}{Group C8}\\
+13.1 & 17 20 55.1917 & -35 45 03.9750 & 1.794\\ 
+12.8 & 17 20 55.1924 & -35 45 03.9824 & 1.054\\ 
+12.4 & 17 20 55.1915 & -35 45 04.0071 & 1.229\\ 
+12.1 & 17 20 55.1910 & -35 45 04.0080 & 2.023\\ 
+11.8 & 17 20 55.1902 & -35 45 04.0175 & 2.637\\ 
+11.4 & 17 20 55.1906 & -35 45 03.9988 & 2.325\\ 
+11.1 & 17 20 55.1904 & -35 45 03.9124 & 2.139\\ 
+10.8 & 17 20 55.1909 & -35 45 03.8332 & 3.507\\ 
+10.4 & 17 20 55.1916 & -35 45 03.8332 & 4.087\\ 
+10.1 & 17 20 55.1919 & -35 45 03.8792 & 3.133\\ 
+9.8 & 17 20 55.1921 & -35 45 03.9153 & 2.291\\ 
+9.5 & 17 20 55.1919 & -35 45 03.8692 & 2.191\\ 
+9.1 & 17 20 55.1919 & -35 45 03.8214 & 2.536\\ 
+8.8 & 17 20 55.1914 & -35 45 03.8482 & 2.113\\ 
+8.5 & 17 20 55.1917 & -35 45 03.8784 & 1.497\\ 
+8.1 & 17 20 55.1933 & -35 45 03.9278 & 1.653\\ 
+7.8 & 17 20 55.1921 & -35 45 03.9726 & 2.156\\ 
+7.5 & 17 20 55.1925 & -35 45 03.9627 & 1.673\\ 
+7.2 & 17 20 55.1925 & -35 45 03.9406 & 1.079\\ 
+6.8 & 17 20 55.1925 & -35 45 03.8447 & 0.933\\ 
+6.5 & 17 20 55.1911 & -35 45 03.7066 & 1.481\\ 
+6.2 & 17 20 55.1923 & -35 45 03.7210 & 1.139\\ 
+5.8 & 17 20 55.1929 & -35 45 03.8972 & 0.628\\ 
+13.1 & 17 20 55.1917 & -35 45 03.9750 & 1.794\\ 
+12.8 & 17 20 55.1924 & -35 45 03.9824 & 1.054\\ 
+12.4 & 17 20 55.1915 & -35 45 04.0071 & 1.229\\ 
+12.1 & 17 20 55.1910 & -35 45 04.0080 & 2.023\\ 
+11.8 & 17 20 55.1902 & -35 45 04.0175 & 2.637\\ 
+11.4 & 17 20 55.1906 & -35 45 03.9988 & 2.325\\ 
+11.1 & 17 20 55.1904 & -35 45 03.9124 & 2.139\\ 
+10.8 & 17 20 55.1909 & -35 45 03.8332 & 3.507\\ 
+10.4 & 17 20 55.1916 & -35 45 03.8332 & 4.087\\ 
+10.1 & 17 20 55.1919 & -35 45 03.8792 & 3.133\\ 
+9.8 & 17 20 55.1921 & -35 45 03.9153 & 2.291\\ 
+9.5 & 17 20 55.1919 & -35 45 03.8692 & 2.191\\ 
+9.1 & 17 20 55.1919 & -35 45 03.8214 & 2.536\\ 
+8.8 & 17 20 55.1914 & -35 45 03.8482 & 2.113\\ 
+8.5 & 17 20 55.1917 & -35 45 03.8784 & 1.497\\ 
+8.1 & 17 20 55.1933 & -35 45 03.9278 & 1.653\\ 
+7.8 & 17 20 55.1921 & -35 45 03.9726 & 2.156\\ 
+7.5 & 17 20 55.1925 & -35 45 03.9627 & 1.673\\ 
+7.2 & 17 20 55.1925 & -35 45 03.9406 & 1.079\\ 
+6.8 & 17 20 55.1925 & -35 45 03.8447 & 0.933\\ 
+6.5 & 17 20 55.1911 & -35 45 03.7066 & 1.481\\ 
+6.2 & 17 20 55.1923 & -35 45 03.7210 & 1.139\\ 
+5.8 & 17 20 55.1929 & -35 45 03.8972 & 0.628\\ 
+2.9 & 17 20 55.1879 & -35 45 03.9111 & 7.802\\ 
+2.5 & 17 20 55.1868 & -35 45 03.8148 & 9.693\\ 
+2.2 & 17 20 55.1902 & -35 45 03.8415 & 12.735\\ 
+1.9 & 17 20 55.1906 & -35 45 03.8025 & 18.090\\ 
+1.6 & 17 20 55.1895 & -35 45 03.7526 & 22.575\\ 
+1.2 & 17 20 55.1895 & -35 45 03.7069 & 24.907\\ 
+0.9 & 17 20 55.1898 & -35 45 03.7138 & 16.233\\ 
+0.6 & 17 20 55.1902 & -35 45 03.8094 & 7.740\\ 
+0.2 & 17 20 55.1907 & -35 45 03.7977 & 6.720\\ 
-0.1 & 17 20 55.1899 & -35 45 03.7710 & 7.543\\ 
-0.4 & 17 20 55.1896 & -35 45 03.7864 & 9.031\\ 
-0.7 & 17 20 55.1904 & -35 45 03.8088 & 11.988\\ 
-1.1 & 17 20 55.1917 & -35 45 03.8732 & 13.519\\ 
-1.4 & 17 20 55.1929 & -35 45 03.9320 & 17.916\\ 
-1.7 & 17 20 55.1933 & -35 45 03.9473 & 16.399\\ 
-2.1 & 17 20 55.1926 & -35 45 03.9406 & 7.498\\ 
-2.4 & 17 20 55.1901 & -35 45 03.9437 & 5.003\\ 
-2.7 & 17 20 55.1899 & -35 45 03.9648 & 4.951\\ 
-3.0 & 17 20 55.1904 & -35 45 03.9530 & 3.646\\ 
-3.4 & 17 20 55.1897 & -35 45 03.9070 & 2.848\\ 
-3.7 & 17 20 55.1906 & -35 45 03.8096 & 2.187\\ 
-4.0 & 17 20 55.1912 & -35 45 03.6872 & 1.939\\ 
-4.4 & 17 20 55.1909 & -35 45 03.7404 & 2.241\\ 
-4.7 & 17 20 55.1914 & -35 45 03.8146 & 4.136\\ 
-5.0 & 17 20 55.1909 & -35 45 03.7907 & 9.369\\ 
-5.4 & 17 20 55.1894 & -35 45 03.7472 & 25.136\\ 
-5.7 & 17 20 55.1889 & -35 45 03.7349 & 53.891\\ 
-6.0 & 17 20 55.1888 & -35 45 03.7344 & 75.067\\ 
-6.3 & 17 20 55.1886 & -35 45 03.7251 & 78.148\\ 
-6.7 & 17 20 55.1887 & -35 45 03.7191 & 47.466\\ 
-7.0 & 17 20 55.1896 & -35 45 03.7044 & 28.648\\ 
-7.3 & 17 20 55.1903 & -35 45 03.7147 & 38.029\\ 
-7.7 & 17 20 55.1906 & -35 45 03.7250 & 56.377\\ 
-8.0 & 17 20 55.1905 & -35 45 03.7176 & 53.954\\ 
-8.3 & 17 20 55.1900 & -35 45 03.7093 & 43.538\\ 
-8.6 & 17 20 55.1900 & -35 45 03.6913 & 40.949\\ 
-9.0 & 17 20 55.1904 & -35 45 03.6811 & 35.198\\ 
-9.3 & 17 20 55.1904 & -35 45 03.6656 & 24.791\\ 
-9.6 & 17 20 55.1903 & -35 45 03.6716 & 18.717\\ 
-10.0 & 17 20 55.1907 & -35 45 03.6974 & 17.582\\ 
-10.3 & 17 20 55.1912 & -35 45 03.7294 & 19.375\\ 
-10.6 & 17 20 55.1909 & -35 45 03.7398 & 23.868\\ 
-11.0 & 17 20 55.1901 & -35 45 03.7236 & 35.739\\ 
-11.3 & 17 20 55.1896 & -35 45 03.7096 & 54.072\\ 
-11.6 & 17 20 55.1894 & -35 45 03.6996 & 60.958\\ 
-11.9 & 17 20 55.1896 & -35 45 03.7001 & 56.060\\ 
-12.3 & 17 20 55.1897 & -35 45 03.7151 & 49.063\\ 
-12.6 & 17 20 55.1897 & -35 45 03.7203 & 42.350\\ 
-12.9 & 17 20 55.1902 & -35 45 03.7192 & 31.009\\ 
-13.3 & 17 20 55.1917 & -35 45 03.7315 & 20.392\\ 
-13.6 & 17 20 55.1935 & -35 45 03.7775 & 20.630\\ 
-13.9 & 17 20 55.1951 & -35 45 03.7924 & 21.565\\ 
-14.2 & 17 20 55.1961 & -35 45 03.7852 & 28.003\\ 
-14.6 & 17 20 55.1957 & -35 45 03.7856 & 39.724\\ 
-14.9 & 17 20 55.1947 & -35 45 03.7788 & 48.197\\ 
-15.2 & 17 20 55.1937 & -35 45 03.7819 & 45.933\\ 
-15.6 & 17 20 55.1941 & -35 45 03.7893 & 30.484\\ 
-15.9 & 17 20 55.1946 & -35 45 03.8037 & 19.581\\ 
-16.2 & 17 20 55.1944 & -35 45 03.8229 & 11.492\\ 
-16.5 & 17 20 55.1939 & -35 45 03.8483 & 5.953\\ 
-16.9 & 17 20 55.1937 & -35 45 03.8592 & 4.037\\ 
-17.2 & 17 20 55.1945 & -35 45 03.8557 & 4.049\\ 
-17.5 & 17 20 55.1943 & -35 45 03.8889 & 4.885\\ 
-17.9 & 17 20 55.1931 & -35 45 03.9351 & 5.400\\ 
-18.2 & 17 20 55.1929 & -35 45 03.9393 & 8.133\\ 
-18.5 & 17 20 55.1927 & -35 45 03.9447 & 12.453\\ 
-18.9 & 17 20 55.1927 & -35 45 03.9142 & 10.795\\ 
-19.2 & 17 20 55.1931 & -35 45 03.8412 & 10.113\\ 
-19.5 & 17 20 55.1931 & -35 45 03.8174 & 14.345\\ 
-19.8 & 17 20 55.1935 & -35 45 03.8128 & 15.384\\ 
-20.2 & 17 20 55.1937 & -35 45 03.8166 & 15.221\\ 
-20.5 & 17 20 55.1942 & -35 45 03.8205 & 21.661\\ 
-20.8 & 17 20 55.1943 & -35 45 03.8267 & 29.474\\ 
-21.2 & 17 20 55.1944 & -35 45 03.8372 & 26.080\\ 
-21.5 & 17 20 55.1943 & -35 45 03.8502 & 17.125\\ 
-21.8 & 17 20 55.1943 & -35 45 03.8580 & 13.898\\ 
-22.1 & 17 20 55.1946 & -35 45 03.8462 & 10.798\\ 
-22.5 & 17 20 55.1948 & -35 45 03.8333 & 9.661\\ 
-22.8 & 17 20 55.1954 & -35 45 03.8195 & 12.385\\ 
-23.1 & 17 20 55.1960 & -35 45 03.8100 & 19.850\\ 
-23.5 & 17 20 55.1962 & -35 45 03.8062 & 26.292\\ 
-23.8 & 17 20 55.1962 & -35 45 03.8127 & 26.225\\ 
-24.1 & 17 20 55.1961 & -35 45 03.8147 & 22.945\\ 
-24.4 & 17 20 55.1961 & -35 45 03.8109 & 20.757\\ 
-24.8 & 17 20 55.1963 & -35 45 03.8060 & 20.093\\ 
-25.1 & 17 20 55.1965 & -35 45 03.8017 & 19.849\\ 
-25.4 & 17 20 55.1967 & -35 45 03.7930 & 20.821\\ 
-25.8 & 17 20 55.1967 & -35 45 03.7911 & 18.421\\ 
-26.1 & 17 20 55.1966 & -35 45 03.7881 & 14.058\\ 
-26.4 & 17 20 55.1963 & -35 45 03.8018 & 13.170\\ 
-26.8 & 17 20 55.1965 & -35 45 03.7979 & 16.190\\ 
\multicolumn{4}{c}{Group C9}\\
+5.2 & 17 20 55.2287 & -35 45 03.6925 & 1.390\\ 
+4.9 & 17 20 55.2285 & -35 45 03.6727 & 6.600\\ 
+4.5 & 17 20 55.2285 & -35 45 03.6666 & 21.816\\ 
+4.2 & 17 20 55.2286 & -35 45 03.6675 & 50.615\\ 
+3.9 & 17 20 55.2285 & -35 45 03.6665 & 75.742\\ 
+3.5 & 17 20 55.2286 & -35 45 03.6643 & 68.601\\ 
+3.2 & 17 20 55.2285 & -35 45 03.6668 & 38.300\\ 
+2.9 & 17 20 55.2287 & -35 45 03.6640 & 14.519\\ 
+2.5 & 17 20 55.2287 & -35 45 03.6588 & 3.474\\ 
+2.2 & 17 20 55.2241 & -35 45 03.6685 & 0.883\\ 
\multicolumn{4}{c}{Group C10}\\
-7.7 & 17 20 55.1563 & -35 45 03.8775 & 1.546\\ 
-8.0 & 17 20 55.1459 & -35 45 03.7494 & 2.538\\ 
-8.3 & 17 20 55.1431 & -35 45 03.6847 & 2.936\\ 
-8.6 & 17 20 55.1424 & -35 45 03.7049 & 3.390\\ 
-9.0 & 17 20 55.1429 & -35 45 03.7173 & 2.884\\ 
-9.3 & 17 20 55.1431 & -35 45 03.7141 & 1.793\\ 
-9.6 & 17 20 55.1436 & -35 45 03.7531 & 0.872\\ 
\multicolumn{4}{c}{Group C11}\\
-3.7 & 17 20 55.2142 & -35 45 01.9928 & 1.058\\ 
-4.0 & 17 20 55.2147 & -35 45 01.9789 & 3.470\\ 
-4.4 & 17 20 55.2145 & -35 45 01.9851 & 5.465\\ 
-4.7 & 17 20 55.2145 & -35 45 01.9814 & 4.159\\ 
-5.0 & 17 20 55.2148 & -35 45 01.9944 & 2.024\\ 
-5.4 & 17 20 55.2159 & -35 45 02.0419 & 1.174\\ 
-5.7 & 17 20 55.2189 & -35 45 01.9821 & 0.851\\ 
-6.0 & 17 20 55.2219 & -35 45 02.0605 & 0.936\\ 
-6.3 & 17 20 55.2246 & -35 45 02.0941 & 0.611\\     
\enddata
\end{deluxetable}

\begin{deluxetable}{rccrrccc}
\tablecolumns{7}
\tablewidth{0pc}
\tablecaption{44 GHz Class~I Methanol masers detected in \ngcin\ \label{methanoltable}}  
\tabletypesize{\scriptsize}
\tablehead{
\colhead{Spot} & \multicolumn{2}{c}{Peak position} & \colhead{Peak intensity}  & \colhead{$T_{\rm Brightness}$} & \multicolumn{3}{c}{Velocity (\kms\/)} \\  
\colhead{Number} & \colhead{R.A.} &  \colhead{Dec}  & \colhead{(Jy beam$^{-1}$)} & \colhead{(K)} & \colhead{$v_{\rm peak}$} & \colhead{$v_{\rm min}$} & \colhead{$v_{\rm max}$}} 
\startdata
     1 & 17 20 53.351 & -35 45 19.31   & $13.48\pm0.22$ & 16340 & -3.17 &  -3.67  & -2.67 \\ 
     2 & 17 20 53.484 & -35 45 14.90   & $ 5.19\pm0.20$ & 6290  & -5.50 &  -5.83  & -5.17 \\ 
     3 & 17 20 53.600 & -35 45 18.28   & $ 2.69\pm0.20$ & 3260  & -4.17 &  -4.34  & -3.84 \\ 
     4 & 17 20 53.789 & -35 45 13.42   & $12.82\pm0.18$ & 15540 & -1.68 &  -2.18  & +1.14 \\ 
     5 & 17 20 54.052 & -35 45 15.56   & $ 0.82\pm0.17$ & 990   & -2.18 &  -2.84  & -1.84 \\ 
     6 & 17 20 54.055 & -35 45 20.38   & $ 3.01\pm0.18$ & 3650  & -4.50 &  -5.17  & -4.17 \\ 
     7 & 17 20 54.274 & -35 45 12.11   & $ 1.10\pm0.16$ & 1330  & -7.16 &  -7.66  & -6.49 \\ 
     8 & 17 20 54.333 & -35 45 22.19   & $10.71\pm0.17$ & 12980 & -6.83 &  -7.82  & -6.16 \\ 
     9 & 17 20 54.405 & -35 45 22.57   & $ 4.39\pm0.17$ & 5320  & -8.49 &  -9.98  & -7.99 \\ 
    10 & 17 20 54.419 & -35 45 15.50   & $ 0.95\pm0.16$ & 1200  & -8.32 &  -9.32  & -8.16 \\ 
    11 & 17 20 54.457 & -35 45 22.57   & $83.28\pm0.17$ & 100900 & -7.16 &  -8.16  & -6.16 \\ 
    12 & 17 20 54.474 & -35 45 14.56   & $ 0.53\pm0.15$ & 640    & -1.51 &  -1.84  & -0.85 \\ 
    13 & 17 20 54.478 & -35 45 06.64   & $ 0.62\pm0.15$ & 750    & -1.35 &  -1.84  & -1.01 \\ 
    14 & 17 20 54.552 & -35 45 22.46   & $ 5.46\pm0.17$ & 6620   & -6.16 &  -7.66  & -5.17 \\ 
    15 & 17 20 54.662 & -35 45 43.52   & $ 5.10\pm0.37$ & 6180   & -4.83 &  -5.17  & -3.67 \\ 
    16 & 17 20 54.838 & -35 45 37.31   & $ 6.63\pm0.26$ & 8040   & -4.83 &  -5.33  & -4.67 \\ 
    17 & 17 20 54.725 & -35 45 14.22   & $18.83\pm0.15$ & 22820  & -5.33 &  -6.00  & -3.67 \\ 
    18 & 17 20 54.781 & -35 45 09.67   & $ 0.81\pm0.15$ & 980    & -2.01 &  -3.01  & -1.68 \\ 
    19 & 17 20 54.821 & -35 45 14.43   & $ 1.40\pm0.15$ & 1700   & -4.67 &  -4.83  & -4.34 \\ 
    20 & 17 20 54.848 & -35 45 36.84   & $ 6.11\pm0.25$ & 7410   & -4.34 &  -4.67  & -3.84 \\ 
    21 & 17 20 54.881 & -35 45 11.65   & $ 0.74\pm0.14$ & 900    & -1.68 &  -2.01  & -1.35 \\ 
    22 & 17 20 54.921 & -35 45 13.88   & $33.12\pm0.15$ & 40140  & -3.34 &  -3.84  & -2.84 \\ 
    23 & 17 20 54.928 & -35 45 13.60   & $12.19\pm0.15$ & 14770  & -5.17 &  -6.49  & -4.83 \\ 
    24 & 17 20 54.936 & -35 45 08.27   & $ 3.89\pm0.14$ & 4710   & -4.67 &  -5.00  & -4.34 \\ 
    25 & 17 20 54.960 & -35 45 13.94   & $78.63\pm0.15$ & 95300  & -4.34 &  -4.67  & -4.00 \\ 
    26 & 17 20 54.984 & -35 45 13.08   & $ 0.95\pm0.14$ & 1200   & -5.17 &  -5.66  & -4.83 \\ 
    27 & 17 20 54.991 & -35 45 28.87   & $ 2.64\pm0.19$ & 3200   & -4.50 &  -4.83  & -4.17 \\ 
    28 & 17 20 55.009 & -35 45 14.06   & $35.40\pm0.15$ & 42900  & -2.67 &  -3.67  & -1.84 \\ 
    29 & 17 20 55.090 & -35 45 15.54   & $ 4.01\pm0.15$ & 4860   & -2.18 &  -3.17  & -1.51 \\ 
    30 & 17 20 55.095 & -35 45 36.28   & $ 1.16\pm0.24$ & 1410   & -0.35 &  -0.68  & -0.02 \\ 
    31 & 17 20 55.109 & -35 45 36.88   & $ 1.86\pm0.25$ & 2250   & -7.16 &  -7.49  & -6.66 \\ 
    32 & 17 20 55.165 & -35 45 37.53   & $ 6.05\pm0.26$ & 7330   & -3.67 &  -4.00  & -2.84 \\ 
    33 & 17 20 55.195 & -35 45 03.81   & $ 1.02\pm0.15$ & 1240   & -0.85 &  -2.01  & -0.02 \\ 
    34 & 17 20 55.208 & -35 45 37.58   & $ 1.34\pm0.26$ & 1620   & -2.01 &  -2.34  & -1.68 \\ 
    35 & 17 20 55.264 & -35 45 36.84   & $ 4.79\pm0.25$ & 5810   & -3.84 &  -4.34  & -2.67 \\ 
    36 & 17 20 55.251 & -35 45 30.18   & $18.37\pm0.19$ & 22260  & -0.68 &  -1.51  & -0.35 \\ 
    37 & 17 20 55.300 & -35 45 30.15   & $ 1.43\pm0.19$ & 1730   & -1.18 &  -1.68  & -1.01 \\ 
    38 & 17 20 55.328 & -35 45 01.76   & $ 1.42\pm0.15$ & 1720   & -1.84 &  -2.18  & -1.18 \\ 
    39 & 17 20 55.329 & -35 45 05.55   & $ 0.60\pm0.15$ & 730    & -1.18 &  -1.51  & -0.85 \\ 
    40 & 17 20 55.367 & -35 45 05.61   & $ 1.37\pm0.15$ & 1660   & -3.34 &  -4.00  & -2.84 \\ 
    41 & 17 20 55.392 & -35 45 05.16   & $ 1.02\pm0.15$ & 1240   & -1.01 &  -1.51  & -0.68 \\ 
    42 & 17 20 55.440 & -35 45 09.78   & $11.19\pm0.14$ & 13560  & -4.67 &  -5.33  & -4.50 \\ 
    43 & 17 20 55.466 & -35 45 06.62   & $ 2.83\pm0.15$ & 3430   & -1.35 &  -2.01  & -0.85 \\ 
    44 & 17 20 55.471 & -35 45 00.86   & $ 1.61\pm0.15$ & 1950   & +2.64 &  +1.64  & +4.80 \\ 
    45 & 17 20 55.513 & -35 45 06.25   & $ 1.38\pm0.15$ & 1670   & -4.50 &  -4.83  & -4.34 \\ 
    46 & 17 20 55.515 & -35 45 07.64   & $ 1.90\pm0.15$ & 2300   & -5.00 &  -5.50  & -4.67 \\ 
    47 & 17 20 55.538 & -35 45 00.62   & $ 1.63\pm0.15$ & 1980   & -4.17 &  -4.34  & -3.84 \\ 
    48 & 17 20 55.540 & -35 45 09.05   & $ 0.74\pm0.14$ & 900    & -5.33 &  -5.50  & -5.17 \\ 
    49 & 17 20 55.558 & -35 45 26.72   & $ 3.32\pm0.18$ & 4020   & -3.34 &  -3.67  & -3.01 \\ 
    50 & 17 20 55.581 & -35 45 02.53   & $ 0.96\pm0.15$ & 1200   & -2.01 &  -3.01  & -0.68 \\ 
    51 & 17 20 55.610 & -35 45 29.45   & $ 5.35\pm0.19$ & 6480   & -2.84 &  -3.17  & -2.51 \\ 
    52 & 17 20 55.651 & -35 45 31.23   & $ 4.48\pm0.21$ & 5430   & -4.34 &  -4.50  & -4.17 \\ 
    53 & 17 20 55.652 & -35 45 01.68   & $ 1.68\pm0.15$ & 2040   & -1.18 &  -2.18  & -0.02 \\ 
    54 & 17 20 55.656 & -35 45 10.62   & $ 2.86\pm0.15$ & 3470   & -5.83 &  -6.49  & -4.67 \\ 
    55 & 17 20 55.677 & -35 45 00.39   & $296.42\pm0.16$& 359260 & -2.67 &  -3.17  & -1.51 \\ 
    56 & 17 20 55.683 & -35 45 02.88   & $ 1.02\pm0.15$ & 1240   & -7.16 &  -7.66  & -6.66 \\ 
    57 & 17 20 55.708 & -35 45 11.17   & $ 1.42\pm0.15$ & 1720   & -3.51 &  -3.67  & -3.17 \\ 
    58 & 17 20 55.771 & -35 45 27.75   & $268.96\pm0.19$& 325980 & -4.50 &  -5.17  & -3.84 \\ 
    59 & 17 20 55.810 & -35 45 31.07   & $ 2.12\pm0.21$ & 2570   & -5.00 &  -5.66  & -4.83 \\ 
    60 & 17 20 55.859 & -35 45 31.74   & $ 3.45\pm0.21$ & 4180   & -5.33 &  -6.00  & -4.83 \\ 
    61 & 17 20 55.859 & -35 45 28.27   & $ 7.12\pm0.19$ & 8629   & -3.84 &  -4.34  & -3.51 \\ 
    62 & 17 20 56.004 & -35 45 33.04   & $ 4.53\pm0.23$ & 5490   & -3.84 &  -4.50  & -3.17 \\ 
    63 & 17 20 56.004 & -35 45 23.42   & $ 1.98\pm0.18$ & 2400   & -3.34 &  -3.67  & -3.17 \\ 
    64 & 17 20 56.095 & -35 45 22.41   & $ 4.12\pm0.17$ & 4990   & -4.67 &  -5.00  & -4.00 \\ 
    65 & 17 20 56.128 & -35 45 23.50   & $28.48\pm0.18$ & 34520  & -5.00 &  -5.66  & -4.50 \\ 
    66 & 17 20 56.164 & -35 45 22.80   & $ 5.23\pm0.18$ & 6340   & -7.66 &  -8.65  & -7.32 \\ 
    67 & 17 20 56.185 & -35 45 23.85   & $10.17\pm0.18$ & 12330  & -7.82 &  -8.82  & -6.49 \\ 
    68 & 17 20 56.196 & -35 45 32.22   & $ 9.45\pm0.23$ & 11450  & -4.17 &  -4.50  & -3.67 \\ 
    69 & 17 20 55.264 & -35 45 36.84   & $ 4.80\pm0.25$ & 5820   & -5.33 &  -5.66  & -5.00 \\ 
    70 & 17 20 56.285 & -35 45 19.20   & $ 2.14\pm0.17$ & 2590   & -3.51 &  -4.34  & -3.17 \\ 
    71 & 17 20 56.296 & -35 45 29.73   & $43.26\pm0.22$ & 52430  & -5.00 &  -5.83  & -4.50 \\ 
    72 & 17 20 56.342 & -35 45 20.41   & $ 1.74\pm0.18$ & 2110   & -3.51 &  -4.17  & -3.17 \\ 
    73 & 17 20 56.307 & -35 45 26.75   & $ 0.90\pm0.20$ & 1100   & -2.18 &  -2.51  & -2.01 \\ 
    74 & 17 20 56.341 & -35 45 30.68   & $28.14\pm0.23$ & 34110  & -4.00 &  -4.50  & -3.17 \\ 
    75 & 17 20 56.414 & -35 45 31.72   & $171.47\pm0.24$& 207820 & -3.34 &  -3.67  & -3.01 \\ 
    76 & 17 20 56.423 & -35 45 29.69   & $ 9.90\pm0.23$ & 12000  & -4.00 &  -6.00  & -3.67 \\ 
    77 & 17 20 56.488 & -35 45 30.49   & $ 5.38\pm0.24$ & 6520   & -2.51 &  -3.67  & -2.01 \\ 
    78 & 17 20 56.708 & -35 45 32.61   & $13.34\pm0.28$ & 16170  & -2.51 &  -3.01  & -1.18 \\ 
    79 & 17 20 57.028 & -35 45 35.92   & $ 2.02\pm0.37$ & 2450   & -5.33 &  -5.66  & -4.34 \\ 
    80 & 17 20 56.903 & -35 45 24.93   & $ 2.06\pm0.24$ & 2500   & -3.51 &  -3.84  & -3.17 \\ 
    81 & 17 20 57.070 & -35 45 26.27   & $ 2.47\pm0.26$ & 2990   & -3.67 &  -4.17  & -3.17 \\ 
    82 & 17 20 56.859 & -35 45 28.03   & $ 3.59\pm0.25$ & 4350   & -5.33 &  -5.66  & -4.83 \\ 
    83 & 17 20 57.004 & -35 45 23.18   & $ 2.07\pm0.24$ & 2510   & -5.17 &  -5.33  & -3.67 \\ 
\enddata
\end{deluxetable} 

\clearpage


\begin{deluxetable}{cclcc}
\tablecolumns{5}
\tablewidth{0pc}
\tablecaption{1.3 millimeter spectral lines detected toward SMA1\label{sma1lines}}  
\tabletypesize{\scriptsize}
\tablehead{ 
\colhead{Species} & \colhead{Transition} & \colhead{Frequency} & 
\colhead{$E_{\rm lower}$ (cm$^{-1}$)\tablenotemark{a}} & 
\colhead{$T_{\rm Brightness}$(K)\tablenotemark{b}}
} 
\startdata
H$_2$S               & $2_{2,0}-2_{1,1}$           & 216.710435 & 51.1 & 13 \\
\methylformate-E     & $18_{4,15}-17_{4,14}$       & 216.830112 & 66.2 & 6.2 \\
\methylformate-A     & $18_{2,16}-17_{2,15}$       & 216.838846 & 66.2 & 4.8 \\
\acrylonitrile       & $23_{2,22}-22_{,2,21}$      & 216.936717 & 85.9 & 0.8 \\
\methanol-E          & $5_{+1,4}-4_{+2,2}$     & 216.945600 & 31.6 & 11 \\
\methylformate-E     & $20_{1,20}-19_{1,19}$       & 216.964789 & 70.3 & 10 \\
\methylformate-A     & $20_{1,20}-19_{1,19}$       & 216.965991 & 70.3 &  \\ [-1ex]
\methylformate-E     & $20_{2,19}-19_{0,19}$       & 216.966247 & 70.3 & \raisebox{1.5ex}{ $\bigr\}$ 13} \\    
\methylformate-A     & $20_{0,20}-19_{0,19}$       & 216.967334 & 70.3 & 11 \\
Unidentified         & \nodata                     & 217.020    & \nodata & 2.3 \\
SiO                  & $5-4$                       & 217.104980 & 14.5 & 2.8 \\
\dimethylether       & $36_{0,36,1}-36_{1,35,1}$   & 217.168120 & 436.6 & 1.2 \\
\dimethylether       & $22_{4,19,2}-22_{3,20,2}$   & 217.189667 & 168.9 & \\[-1ex]
\dimethylether       & $22_{4,19,3}-22_{3,20,3}$   & 217.189667 & 168.9 & \raisebox{1.5ex}{$\bigr\}$ 2.9} \\ 
\dimethylether       & $22_{4,19,1}-22_{3,20,1}$   & 217.191424 & 168.9 & 4.8 \\
\dimethylether       & $22_{4,19,0}-22_{3,20,0}$   & 217.193170 & 168.9 & 3.6 \\
DCN                  & $3-2$\tablenotemark{c}      & 217.238630 &  7.2  & 5.7 \\
\methanol-A$^-$ ($v_t=1$)          & $6_{1,5}-7_{2,6}$     & 217.299202 & 252.6 & 8.4 \\
Unidentified         & \nodata & 217.312 & \nodata & 2.2 \\
\thirteenmethanol-A$^+$    & $10_{2,8}-9_{3,7}$    & 217.399550 & 105.6 & 2.4 \\
Unidentified         & \nodata & 217.418 & \nodata & 1.9 \\
\methanol-A$^+$ ($v_t=1$)         & $15_{6,10}-16_{5,11}$ & 217.642859 & 511.0 & 3.3 \\
\methanol-E          & $20_{+1,19}-20_{+0,20}$ & 217.886390 & 346.1 & 5.0 \\
Unidentified         & \nodata                     & 217.959    & \nodata & 1.9 \\
OC$^{13}$S           & 18--17                      & 218.198998 & 61.9 & 3.7 \\
\formaldehyde        & $3_{0,3}-2_{0,2}$           & 218.222192 &  7.3 & 8.4 \\
\methylformate-E     & $17_{1,16}-16_{1,15}$       & 218.280835 & 62.0 & 4.4 \\ 
\methylformate-A     & $17_{3,14}-16_{3,13}$       & 218.297831 & 62.0 & 4.9 \\
\hcccn               & $24-23$\tablenotemark{d}    & 218.324723 & 83.8 & 21 \\
\ethylcyanide        & $24_{3,21}-23_{3,20}$       & 218.389970 & 90.0 & 7.7 \\
\methanol-E          & $4_{+2,2}-3_{+1,2}$     & 218.440040 & 24.3 & 23 \\
\acrylonitrile       & $23_{5,19}-22_{5,18}$       & 218.451298 & 117.7 & 2.7 \\
\acrylonitrile       & $23_{5,18}-22_{5,17}$       & 218.452357 & 117.7 & 2.5 \\
\formamide           & $10_{1,9}-9_{1,8}$          & 218.459653 & 35.0 & 3.4 \\
\formaldehyde        & $3_{2,2}-2_{2,1}$           & 218.475632 & 40.4 & 7.7 \\
\dimethylether       & $23_{3,21,1}-23_{2,22,1}$   & 218.492412	& 176.1 & 4.7 \\
\acrylonitrile       & $23_{4,20}-22_{4,19}$       & 218.573646	& 104.2 & 1.7 \\
\acrylonitrile       & $23_{3,21}-22_{3,20}$       & 218.585072	&  93.7 & 1.6 \\
%
CN & $2_{0,2,3}-1_{0,1,2}$ & 226.659558 & 3.8 & -3.4 \\
CN & $2_{0,2,1}-1_{0,1,1}$ & 226.663693 & 3.8 & -0.9 \\
CN & $2_{0,2,2}-1_{0,1,1}$ & 226.679311 & 3.8 & -4.9 \\
\methylformate-E & $20_{2,18}-19_{2,18}$ & 226.713060 & 76.0 & 5.8\\
\methylformate-A & $20_{2,19}-19_{2,18}$ & 226.718688 & 76.0 & 5.3 \\
\methylformate-E & $20_{3,18}-19_{3,17}$ & 226.773130 & 76.0 & 5.2 \\
\methylformate-A & $20_{1,19}-19_{1,18}$ & 226.778707 & 76.0 & 6.0 \\
CN & $2_{0,3,3}-1_{0,2,2}$ & 226.874191 & 3.8 &  \\ [-1ex]
CN & $2_{0,3,4}-1_{0,2,3}$ & 226.874781 & 3.8 & \raisebox{1.5ex}{$\bigr\}$ -6.9} \\
CN & $2_{0,3,2}-1_{0,2,1}$ & 226.875896 & 3.8 & -6.1 \\
CN & $2_{0,3,2}-1_{0,2,2}$ & 226.887420 & 3.8 & -2.2 \\
CN & $2_{0,3,3}-1_{0,2,3}$ & 226.892128 & 3.8 & -1.5 \\
\methylformate-E & $19_{4,16}-18_{4,15}$ & 227.019490 & 73.5 & 4.8 \\
\methylformate-A & $19_{2,17}-18_{2,16}$ & 227.028070 & 73.4 & 5.0 \\
\methanol-E & $21_{+1,20}-21_{+0,21}$ & 227.094601 & 379.6 & 6.9 \\
Unidentified & \nodata & 227.126 & \nodata & 2.2 \\
Unidentified & \nodata & 227.170 & \nodata & 2.0 \\
Unidentified & \nodata & 227.221 & \nodata & 1.8 \\
\methanol-E & $12_{-1,12}-11_{+2,9}$ & 227.229600 & 122.0 & 2.2 \\
g-\ethanol & $13_{3,10,1}-12_{3,9,1}$ & 227.294752 & 95.7 & 1.8 \\
$^{13}$C$^{34}$S & $5-4$ & 227.300506 & 15.16 & 3.9 \\
\hcccn & $25-24$\tablenotemark{d} & 227.418905 & 91.03 & 17 \\
\methylformate-E & $21_{1,21}-20_{1,20}$ & 227.560955 & 77.5 & 12 \\
\methylformate-E & $21_{2,20}-20_{2,19}$ & 227.561753 & 77.5 & \\ [-1ex]
\methylformate-A & $21_{1,21}-20_{1,20}$ & 227.561944 & 77.5 & \raisebox{1.5ex}{$\bigr\}$ 12} \\
\methylformate-A & $21_{0,21}-20_{0,20}$ & 227.562740 & 77.5 & 10 \\
t-\ethanol & $18_{5,13,2}-18_{4,14,2}$ & 227.606079 & 114.3 & \\ [-1ex]
\formamide & $11_{0,11}-10_{0,10}$     & 227.606176 & 38.5 &  \raisebox{1.5ex}{1.5}\\
\dimethylether & $26_{5,21,0}-26_{4,22,0}$ & 227.647849 & 239.7 & \\
\dimethylether & $26_{5,21,1}-26_{4,22,1}$ & 227.647984 & 239.7 & $\bigr\}$ 4.5\\ 
\dimethylether & $26_{5,21,2}-26_{4,22,2}$ & 227.648120 & 239.7 & \\
\dimethylether & $24_{3,22,2}-24_{2,23,2}$ & 227.654386 & 190.9 & \\ [-1ex]
\dimethylether & $24_{3,22,3}-24_{2,23,3}$ & 227.654386 & 190.9 & \raisebox{1.5ex}{$\bigr\}$ 4.6} \\
\dimethylether & $24_{3,22,1}-24_{2,23,1}$ & 227.657025 & 190.9 & 4.9 \\
\dimethylether & $24_{3,22,0}-24_{2,23,0}$ & 227.659664 & 190.9 & 2.0 \\
\ethylcyanide & $25_{3,22}-24_{3,21}$ & 227.780972 & 97.3 & 6.1\\   
\methanol-A$^+$ & $16_{1,16}-15_{2,13}$ & 227.814650 & 219.8 & 9.5\\
g-\ethanol & $13_{1,12,1}-12_{1,11,1}$ & 227.891911 & 89.7 & 0.9 \\
\acrylonitrile & $24_{6,19}-23_{6,18}$ & 227.906683 & 141.5 &  \\[-1ex]
\acrylonitrile & $24_{6,18}-23_{6,17}$ & 227.906709 & 141.5 & \raisebox{1.5ex}{$\bigr\}$ 1.6}\\
Unidentified & \nodata & 227.942 & \nodata & 1.9 \\
\acrylonitrile & $24_{5,20}-23_{5,19}$ & 227.966032 & 125.0 & 1.8 \\
\acrylonitrile & $24_{5,19}-23_{5,18}$ & 227.967589 & 125.0 & 1.8 \\
\hcccn\ (v$_7$=1e) & $25-24$ & 227.977277 & 314.3 & 3.4 \\
Unidentified & \nodata & 228.015 & \nodata & 2.4 \\
g-\ethanol & $13_{3,10,0}-12_{3,9,0}$ & 228.029050 & 92.4 & 1.2 \\
\acrylonitrile & $24_{3,22}-23_{3,21}$ & 228.090537 & 101.0 & 1.2 \\
\acrylonitrile & $24_{4,21}-23_{4,20}$ & 228.104614 & 111.5 & 1.3 \\
Unidentified & \nodata & 228.152 & \nodata & 4.2 \\
\acrylonitrile & $24_{4,20}-23_{4,19}$ & 228.160305 & 111.5 & 2.6 \\
Unidentified & \nodata & 228.210 & \nodata & 3.4 \\
Unidentified & \nodata & 228.223 & \nodata & -1.4\\
\hcccn\ (v$_7$=1f) & $25-24$ & 228.303174 & 314.4 & 5.6 \\
\ethylcyanide & $25_{2,23}-24_{2,22}$ & 228.483136 & 94.3 & 8.8 \\
Unidentified & \nodata & 228.533 & \nodata & 1.9 \\
Unidentified & \nodata & 228.555 & \nodata & 2.2 \\
\enddata
\tablenotetext{a}{To convert to $E_l$ (K) multiply by 1.438 K*cm.}
\tablenotetext{b}{Referred to the beamsize of $2\farcs 27\times
  1\farcs 42$ in USB, and $2\farcs 40 \times 1\farcs 56$ in LSB}
\tablenotetext{c}{Comprised of five components}
\tablenotetext{d}{Comprised of three components}
\end{deluxetable}

\begin{deluxetable}{lcccc}
\tablecolumns{5}
\tablewidth{0pc}
\tablecaption{Derived 1.3 millimeter Continuum Properties\label{mass}}  
\tabletypesize{\scriptsize}
\tablehead{ 
\colhead{Source} & \colhead{$T_{dust}$} &
\colhead{$M_{gas}$} & 
\colhead{$N_{H_2}\times 10^{23}$ } & \colhead{$n_{H_2}\times 10^{7}$} \\
 & \colhead{(K)} & \colhead{(M$_\odot$)}  & \colhead{(\ct\/)} & \colhead{(\cc\/)}
} 
\startdata
SMA1 & 155 - 165  & 16 - 14  & 7.7 - 6.8  & 1.3 - 1.2  \\
SMA2 & 135 - 155  &  7 -  6  & 6.4 - 5.7  & 1.6 - 1.4  \\
SMA3 & 20 - 50    & 74 - 23  & 40 - 12    & 7 - 2      \\
SMA4 & 20 - 50    & 52 - 16  & 30 - 9     & 6 - 2      \\
SMA5 & 20 - 50    & 28 -  9  & 18 - 6     & 4 - 1      \\
SMA6 & 25 - 50    & 32 - 13  & 58 - 17    & 16 - 5     \\
SMA7 & 20 - 50    & 22 -  7  &  8 - 3     & 1 - 0.4    \\
\enddata
\end{deluxetable}

\end{document}